# Emotional Brain State Classification on fMRI Data Using Deep Residual and Convolutional Networks

Maxime Tchibozo[b,c], Donggeun Kim[b], Zijing Wang[c], Xiaofu He[a,b,c,*]

[a] Department of Psychiatry, Vagelos College of Physicians and Surgeons, Columbia University, New York, NY, United States

[b] New York State Psychiatric Institute, New York, NY, United States

[c] Data Science Institute, Columbia University, New York, NY, United States

* **Corresponding author**: Xiaofu He, Ph.D., Columbia University and the New York State Psychiatric Institute, 1051 Riverside Drive, Unit 43, New York, NY 10032, United States. Phone: 1-646 774 5875, Email: xh2170@cumc.columbia.edu

**Abstract:**

The goal of emotional brain state classification on functional MRI (fMRI) data is to recognize brain activity patterns related to specific emotion tasks performed by subjects during an experiment. Distinguishing emotional brain states from other brain states using fMRI data has proven to be challenging due to two factors: a difficulty to generate fast yet accurate predictions in short time frames, and a difficulty to extract emotion features which generalize to unseen subjects.

To address these challenges, we conducted an experiment in which 22 subjects viewed pictures designed to stimulate either negative, neutral or rest emotional responses while their brain activity was measured using fMRI. We then developed two distinct Convolution-based approaches to decode emotional brain states using only spatial information from single, minimally pre-processed (slice timing and realignment) fMRI volumes. In our first approach, we trained a 1D Convolutional Network (84.9% accuracy; chance level 33%) to classify 3 emotion conditions using One-way Analysis of Variance (ANOVA) voxel selection combined with hyperalignment. In our second approach, we trained a 3D ResNet-50 model (78.0% accuracy; chance level 50%) to classify 2 emotion conditions from single 3D fMRI volumes directly.

Our Convolutional and Residual classifiers successfully learned group-level emotion features and could decode emotion conditions from fMRI volumes in milliseconds. These approaches could potentially be used in brain computer interfaces and real-time fMRI neurofeedback research.

## 1. Introduction

Emotional brain state classification of functional MRI (fMRI) data aims to recognize activity patterns in the brain regions specifically activated during an emotion task (Saarimäki et al., 2016). Brain state classification is often referenced as cognitive state classification or brain decoding (Zhang et al., 2021), and has proven to be a useful tool for real-time neuro-feedback research (Christopher deCharms, 2008; Cox et al., 1995). It has also been applied in Brain-Computer Interfaces (BCI) to successfully assist individuals with motor disabilities in interacting with their environment (Mahmood et al., 2019; Shanechi, 2019; Willett et al., 2021).

Researchers have successfully designed accurate brain state classification methods which leverage Graph Convolutional Networks (GCN) and imitate brain connectivity topologies (Zhang et al., 2021). In the case of human emotions however, brain activity patterns involve specialized brain regions (Krain et al., 2009; Krylova et al., 2021), and emotional brain state classification methods using fMRI have proven to be challenging to implement in practice. Firstly, existing methods for brain state classification of fMRI images are usually subject-specific and their extracted features are not always generalizable to groups of new subjects (G. Chen et al., 2021; de Vos et al., 2020; O'Connor et al., 2021). Secondly, existing methods often rely on time series inputs or require time-consuming pre-processing steps which can make fast online inferences on individual fMRI images infeasible, limiting their potential use in real-time neurofeedback experiments (H. Li & Fan, 2019).

While the classification of brain states from fMRI data was traditionally performed using classical Machine Learning algorithms (Mokhtari & Hossein-Zadeh, 2013; Ryali et al., 2010) and Multi-Voxel Pattern Analysis-based classifiers (MVPA) (Bush et al., 2018; Haxby, 2012; Haynes, 2015; LaConte et al., 2007), Deep Learning methods have steadily gained ground in Neuroimaging. Convolutional Neural Networks (CNN) have successfully been used to diagnose disorders such as Alzheimer's disease, Dementia, and Schizophrenia using MRI data (Bron et al., 2020; Gallay et al., 2020; Larsen et al., 2020; H. Li et al., 2019). Additionally, 3D CNN (Maturana & Scherer, 2015) have been used to extract patterns from 3D fMRI images in brain tissue segmentation experiments (Huo et al., 2019; Kleesiek et al., 2016) and in brain state classification experiments combining widely different conditions such as motor, auditory and memory-related tasks (Vu et al., 2020; Wang et al., 2020). Using Human Connectome Project fMRI data (Van Essen et al., 2013), researchers successfully trained Convolutional brain state classification models which generalized across cognitive domains. However, a model's ability to classify tasks across domains which activate different brain regions (e.g., motor, auditory and visual tasks) is not necessarily indicative of an ability to distinguish different emotion tasks which activate similar brain regions (Vu et al., 2020).

In this paper, we propose two novel methods to address the generalizability and real-time classification challenges presented by emotional brain state classification on fMRI images. In the first method, we coupled a 1D CNN with an Analysis of Variance (ANOVA) voxel selection, Multi-Voxel Pattern Analysis (MVPA), and hyperalignment (Haxby et al., 2011) for an emotional brain state classification of individual fMRI images which generalizes to unseen subjects. In the second method, we trained a spatial 3D adaptation of the ResNet-50 classifier (K. He et al., 2016) which classifies emotional brain states from individual 3D fMRI images in milliseconds ($< 1\ TR$ i.e., Repetition Time) with minimal pre-processing.

## 2. Materials and Methods

### 2.1. Subjects

The fMRI dataset used for both experiments was acquired from a study (X. He, 2022) involving a group of consenting subjects with a history of Major Depressive Disorder (MDD) (10 subjects) and a group of Healthy Control subjects (HC) (12 subjects). In total, there were 22 subjects and the two groups were matched for age, sex, ethnicity, and handedness. The study was approved by both the New York State Psychiatric Institute and Columbia University College of Physicians & Surgeons ethics committees.

### 2.2. fMRI Data Acquisition

Blood Oxygen Level-Dependent (BOLD) contrast fMRI data was measured with a GE Discovery MR750 3.0 Tesla scanner using gradient-echo T2*-weighted echo-planar imaging pulse (EPI) ($TR = 2000$ msec, $TE = 25$ msec, $flip\ angle = 77°$). Each scanned fMRI volume consisted of 45 contiguous axial slices with 3 mm slice thickness, $3 \times 3$ mm in-plane resolution (matrix size $64 \times 64$). Five dummy scans were used to allow the fMRI signal to reach a steady state (X. He, 2022).

### 2.3. Emotion Task

Subjects passively viewed either negative or neutral images which were selected from the International Affective Pictures System (IAPS) (Bradley & Lang, 2017) and displayed in continuous 20-second blocks. Negative and neutral emotion-inducing images were labeled negative (negative brain state) and neutral (neutral brain state) respectively, and were alternated with resting blocks where a single cross-hair was shown (rest brain state) (Fig.1). There are therefore 3 conditions in the emotional brain state classification task: view cross-hair (rest state), view neutral stimuli (neutral state), and view negative stimuli (negative state).

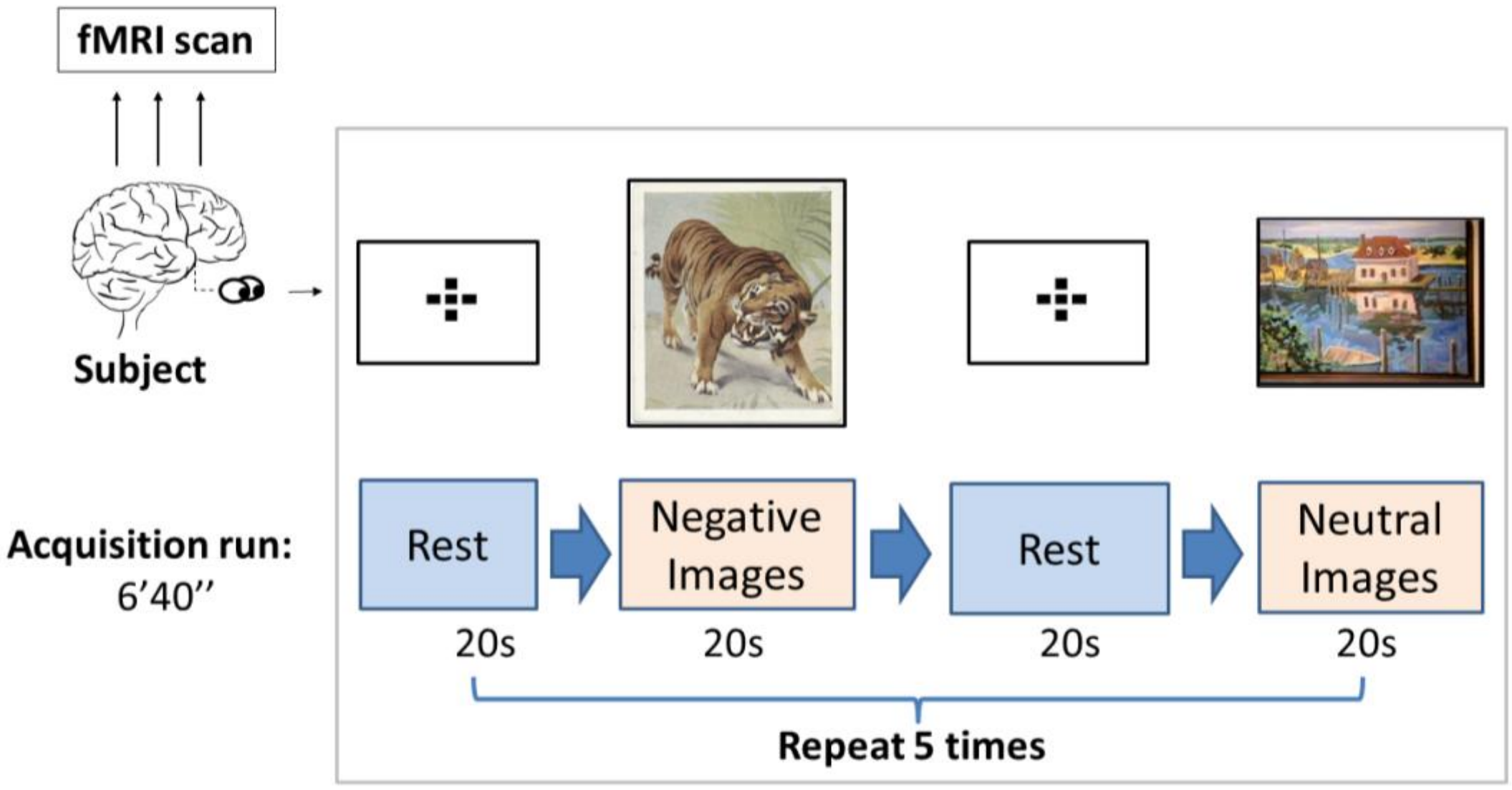

Figure 1. Experimental protocol for the emotion task (surrogate emotion images used for presentation purposes, courtesy of the New York Public Library Digital Collections).

### 2.4. FMRI data pre-processing

During fMRI data acquisition, subjects are prone to head movements which cause misalignments between scanned fMRI volumes over time. We used SPM12 (Friston et al., 2007) to correct the subject motion and to convert the fMRI data to a Python-interpretable format. SPM12 data pre-processing steps were limited to slice timing correction and head movement correction (local subject space realignment, i.e., spatially aligning all fMRI volumes to the fMRI volume of the first scan). This minimal pre-processing differs from traditional fMRI data experiments, which usually include additional co-registration, segmentation, normalization, or smoothing steps.

### 2.5. Feature selection and classification

We designed two experiments to perform emotional brain state classification on fMRI data.

### 2.5.1. For Experiment A

When applied to emotional brain state classification of fMRI images, CNN extract complex 3D brain activity patterns while using fewer learnable parameters than other types of Artificial Neural Networks (LeCun et al., 2015). In experiment A, we first applied a whole-brain filtering mask to the 3D fMRI data to extract only the brain tissue voxels. To encode the 3D fMRI voxels into a format interpretable by a 1D CNN classifier, we used a One-Way ANOVA feature selection and then applied hyperalignment to project the fMRI data of each subject into a common representational space (Haxby et al., 2011).

**ANOVA Feature Selection**

ANOVA computes F-scores for each voxel in a 3D image based on class-specific labels (class $c = negative, neutral\ or\ rest$) with the F-score defined as the ratio of the between-class and within-class variances. We applied ANOVA to identify the $m$ voxels which consistently contained the most predictive label information across all images in the training dataset.

**Hyperalignment**

The pre-hyperalignment dataset for this experiment can be represented as a set of subject matrices $\boldsymbol{S_j}$, each representing the fMRI time series data of a subject $j$ in its respective local brain space. Each subject matrix $\boldsymbol{S_j}$ was constructed by stacking $n_j$ samples (or time points) which contained $m_j$ $(m_j = m)$ ANOVA-selected fMRI voxels. For this experiment, we aggregated data from several fMRI runs to obtain a fixed number of time points for each subject $(n_j = n)$.

Hyperalignment searched for an orthogonal transformation matrix $\boldsymbol{R_j} \in \mathbb{R}^{n_j, n_j}$ which applied rotations and/or reflections to the transpose of each subject matrix $\boldsymbol{S_j}$ and maximized the correlation of the transformed data over all subjects simultaneously (Yousefnezhad & Zhang, 2017). This can be formulated as the following procrustean optimization problem (Haxby et al., 2011, 2020):

$$\boldsymbol{R} = \underset{\Omega}{argmin} \sum_j \left\| \boldsymbol{S_j^t} \Omega_j - \boldsymbol{T^t} \right\|$$

$$\text{subject to } \Omega^t \Omega = \boldsymbol{I}$$

Here, the common representational space between-subject matrix $\boldsymbol{T} \in \mathbb{R}^{m,n}$ represented the averaged subject matrices from all training subjects projected into the target common representational space. $\boldsymbol{T}$ and $\boldsymbol{R}$ were both computed algorithmically using the PyMVPA software (Hanke et al., 2009). The general hyperalignment procedure was shown in Figure 2.

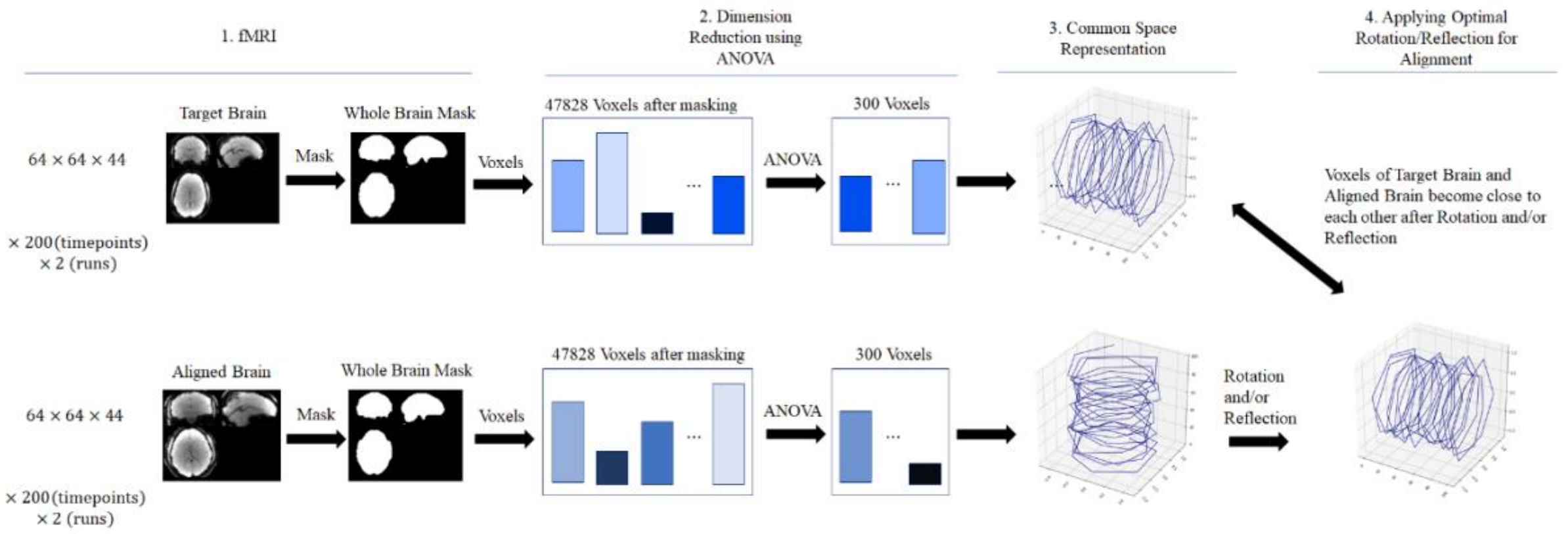

Figure 2. Hyperalignment searches for the transformation which orthogonalizes each subject's selected voxels and maximizes cross-subject correlations (simplified illustration for presentation purposes).

Leave-One-Subject-Out Cross-Validation (LOOCV) was implemented to evaluate the model's generalization performance, and the $m$ selected voxel features were specific to each cross-validation fold. The orthogonal solution matrix $\boldsymbol{R}$ was specific to each cross-validation fold and was computed using training data only. We projected both the training subject data and test subject data into the common representational space using the same matrix $\boldsymbol{R}$ to avoid data leakage during cross-validation.

**1D Convolutional Neural Network**

After the hyperalignment of the fMRI data, the extracted 1D feature vectors ($m$ hyperaligned voxels) were input into successive convolution layers, using convolution kernels of fixed window size to extract patterns. These patterns (feature maps) were then vectorized into a fully connected 1D vector and passed through successive fully connected layers with Rectified Linear Unit activation functions (ReLU). The final layer was connected to the output categories, which, in this experiment, predicted the rest, neutral and negative brain states.

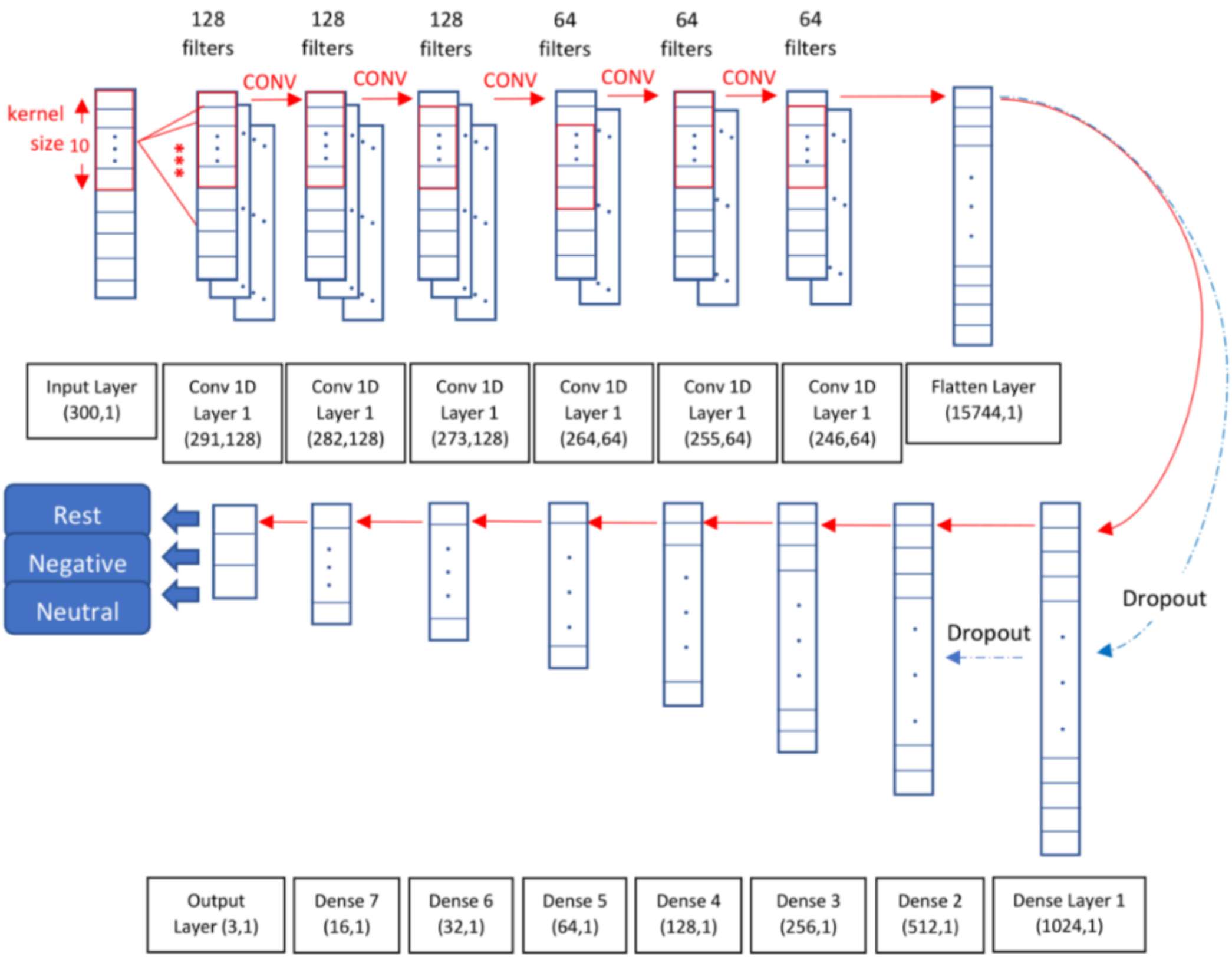

Figure 3. Architecture of the ANOVA + MVPA + 1D CNN brain state classifier.

The architecture (Fig.3) of this classifier was designed to extract features complex enough to capture the spatial patterns contained in the hyperaligned vectors, while simultaneously limiting the overfitting drawback common to many Deep Artificial Neural Networks.

### 2.5.2. For Experiment B

Residual Neural Networks (ResNets) extend CNNs by combining them with shortcut connections (skip-connections) between residual blocks which can skip groups of layers. This simplifies the training of models with deep architectures and alleviates vanishing gradient issues (Hochreiter, 1998). While ResNets are almost exclusively used for 2D image analysis, H. Chen et al., (2018) have proposed a 3D adaptation of the ResNet for brain segmentation, using 3D spatial convolutional kernels grouped inside residual block layers which combined different image processing methods (convolutions, ReLU activations and batch normalizations). In our second experiment, we directly input the SPM12 minimally pre-processed (slice timing and realignment) 3D fMRI data into a 3D adaptation of the ResNet-50 classifier (Jihong Ju, 2017) to extract patterns for emotional brain state classification. The convolution blocks extracted feature maps from the fMRI data, with feature maps from the final convolution group then vectorized into a 1D vector and fully connected to the output categories to classify brain states. This architecture (Fig. 4) replicated the original ResNet-50 from K. He et al. (2016) with 3D convolutional kernels instead of 2D kernels.

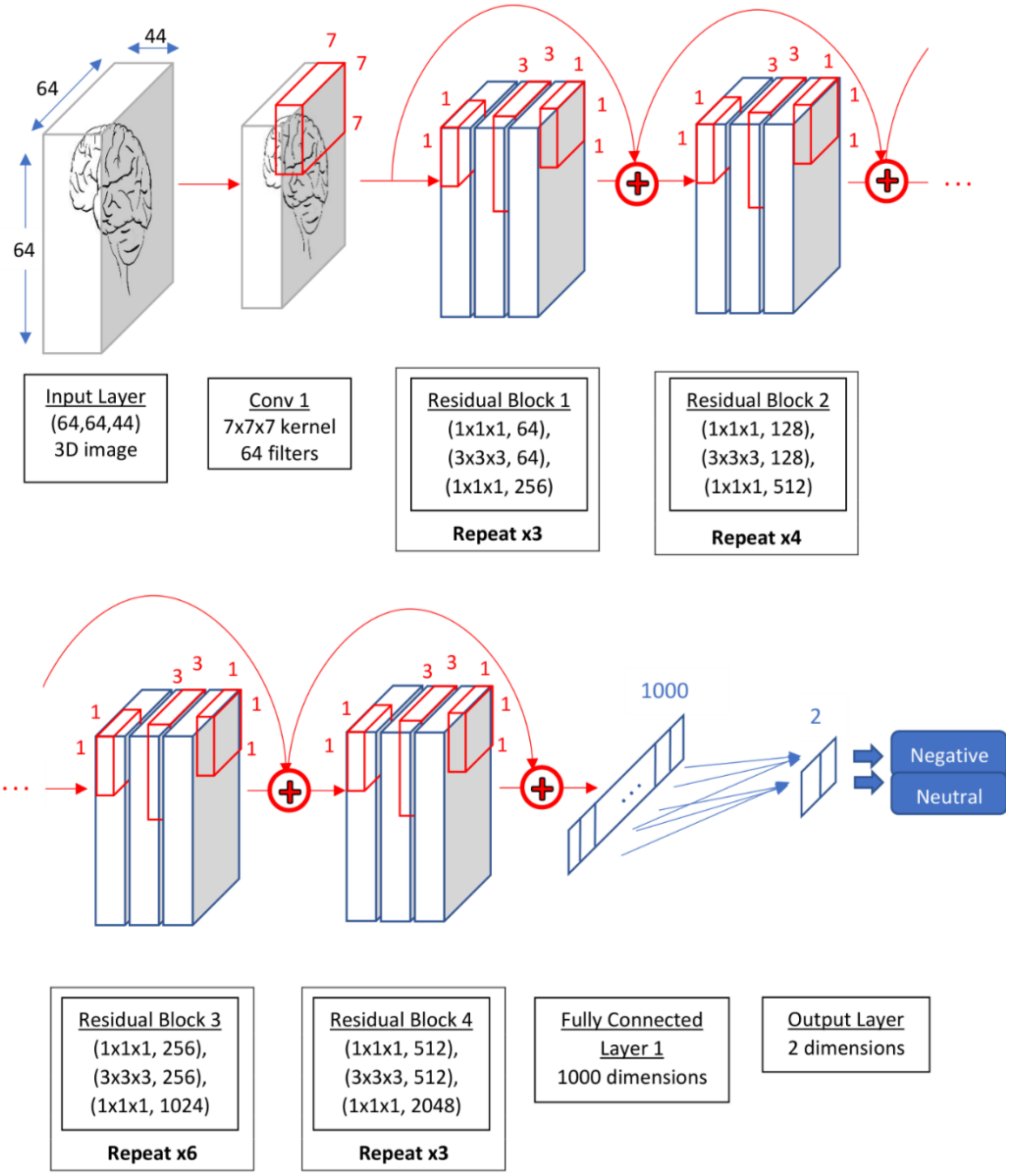

Figure 4. Architecture of the 3D ResNet-50 brain state classifier.

## 2.6. Model Training

For both experiments, model performance on the emotional brain state classification task was measured with the SoftMax Cross-Entropy loss function, and both loss functions were minimized through Adam Stochastic Optimization (Kingma & Ba, 2015).

In Experiment A, we used a stepwise decreasing learning rate schedule, with the learning rate initialized at $lr_A^{(0)}$ for the first epoch and iteratively reduced by a factor $d_A$ whenever the loss function did not improve for $P_A$ consecutive epochs (patience parameter). The count of consecutive non-improving epochs was reset after each learning rate decrease. This dynamic approach reduced the likelihood of the optimizer being trapped in an unsuitable local minimum. We introduced Dropout layers (Srivastava et al., 2014), Batch Normalization layers (Ioffe & Szegedy, 2015), and L2 Norm regularization (Krogh & Hertz, 1992) with parameter $\lambda_A$ in the model architecture to reduce overfitting. Accuracy was measured as the proportion of correct rest vs. neutral vs. negative brain state predictions on the 1D hyperaligned test vectors. In Experiment B, we used a stepwise decreasing learning rate schedule similar to Experiment A, but with the learning rate initialized at $lr_B^{(0)}$, the scaling factor set to $d_B$ and the patience parameter equal to $P_B$ . We used an L2 Norm regularization with parameter $\lambda_B$ to reduce overfitting. Accuracy was measured as the proportion of correct neutral vs. negative emotional brain state predictions on the minimally pre-processed 3D fMRI test images.

### 2.7. Training Strategy

Experiments A and B used two different approaches to prepare the fMRI data and evaluate performance. For cross-subject classification in Experiment A, accuracy was measured using $K$-fold Leave-One-Subject-Out Cross-Validation (LOOCV). Training data consisted of hyperaligned voxels from $K - 1$ subjects, with testing data consisting of hyperaligned voxels from the left-out subject $K$. Classification accuracy was measured over all $K$ left-out subjects separately and averaged. To train a model capable of classifying minimally pre-processed 3D fMRI images directly in Experiment B, we randomly shuffled voxel images from all subjects together and selected 80% to create the combined training and validation datasets, leaving aside the remaining 20% to create the testing dataset. The training and validation dataset was then split at random, with 80% of its fMRI images used for training, and the remaining 20% for validation.

We compared model performance in both experiments to Linear and Kernelized Support Vector Machines (SVM), Random Forests (RF) (Breiman, 2001), Gradient Boosting Trees (XGB) (Friedman, 2002), Linear Discriminant Analysis (LDA), and Multilayer Perceptron (MLP) classifiers.

## 3. Results

### 3.1. Subject Data

#### 3.1.1. For Experiment A

To select relevant features and reduce data dimensionality ahead of classification in Experiment A, we applied an ANOVA feature selection and hyperalignment to the minimally pre-processed fMRI data. To ensure consistency of the hyperalignment process, we restricted the Experiment A dataset to subjects that had an identical number of runs, regardless of their treatment assignment. While the original dataset contained fMRI images from $22$ subjects ($10$ subjects with a history of MDD with age: 20.1 ± 1.1; $12$ control subjects with age: 19.2 ± 0.9), the subset of data used for Experiment A contained fMRI images from the $K = 11$ subjects from the combined groups whose brain activity was measured on exactly $2$ runs, with exactly $200$ real-time emotional brain states measured for each run. In total, we used $n = 2 \times 200 = 400$ fMRI images per subject, generating a total of 4400 fMRI images. Before hyperalignment, we applied a group-level ANOVA feature selection to extract $m = 300$ important voxel features from each fMRI image data point. The decision to select $300$ important voxels was made after a careful evaluation of the influence of the number of selected features $m$ on overfitting and underfitting of the 1D CNN classifier. Each 1D hyperaligned important voxel vector was assigned an emotional brain state label (0: rest state, 1: neutral state, 2: negative state). Training sets initially contained $4000$ emotional brain state fMRI images each, with the $400$ remaining fMRI images – all drawn from the same left-out subject – used for validation. This dataset was initially imbalanced, with 50% of fMRI images corresponding to the rest state, 25% to the neutral state, and 25% to the negative state. To ensure the correct interpretability of our performance metrics, we artificially augmented the minority classes (negative state, neutral state) using Bootstrap resampling with replacement (Efron & Tibshirani, 1986), ensuring that all three classes contained an equal amount of fMRI images ($600$ emotional brain states per subject). This approach limited the likelihood of the classifier overly predicting the majority brain state (rest state) and allowed for a direct comparison with random chance.

#### 3.1.2. For Experiment B

In the second experiment, there was no additional transformation or feature selection of the minimally pre-processed (slice timing and realignment) fMRI data. Each fMRI data point consisted of a $64 \times 64 \times 44$-voxel image, as output by the SPM12 software. In Experiment A, data from subjects with 1 or 3 runs had been discarded to ensure each LOOCV test set contained the same amount of information. For Experiment B on the other hand, we used every available fMRI image from 39 separate runs, originating from all 22 subjects who had between 1 and 3 runs each. In this experiment, due to memory-intensive provisions of 3D fMRI images, we restricted the classifier to predict only the negative and neutral classes, using a total of 3048 fMRI 3D fMRI images with their associated labels (0: neutral state, 1: negative state). The proportion of fMRI images was balanced across both classes. In total, we used 1950 3D fMRI images for training, 488 3D fMRI images for validation, and 610 3D fMRI images for testing. We trained 10 ResNet-50 classifiers using bootstrap resampling over the same training, testing, and validation datasets to obtain confidence intervals on the performance of this classifier.

### 3.2. Model Parameters

Model architectures and hyper-parameters were selected by grid-searching to minimize overfitting between the training and validation loss functions. In Experiment A, a vector of 15744 features was extracted from hyperaligned voxels by 6 consecutive convolution layers and was input into 7 consecutive fully connected layers. Convolution layers contained either 64 or 128 convolution filters depending on their depth within the model's architecture and each convolution filter had a window size of 10. The learning rate was initialized at $lr_A^{(0)} = 4.5 \times 10^{-3}$, the learning rate decay factor was set to $d_A = 0.2$ and the patience parameter was set to $P_A = 50$ epochs. The L2 Norm regularization parameter was set to $\lambda_A = 5 \times 10^{-2}$ and dropout rates were set to either 0.25 or 0.5 depending on the layer. In Experiment B, a vector of 1000 features was extracted from 3D fMRI data by multiple convolution groups and was input into a single fully connected layer. The learning rate was initialized at $lr_B^{(0)} = 5 \times 10^{-4}$, the learning rate decay factor was set to $d_B = 0.3$, and the patience parameter was set to $P_B = 30$ epochs. The L2 Norm regularization parameter was set to $\lambda_B = 2 \times 10^{-4}$.

### 3.3. Classification Performance

#### 3.3.1. For Experiment A

The model converged, as shown by Figure 5, and model performance was evaluated below. This model classified a batch of 600 1D voxel vectors in 0.771 ± 0.08 sec, for an average prediction time of 1.29 msec per 1D voxel vector.

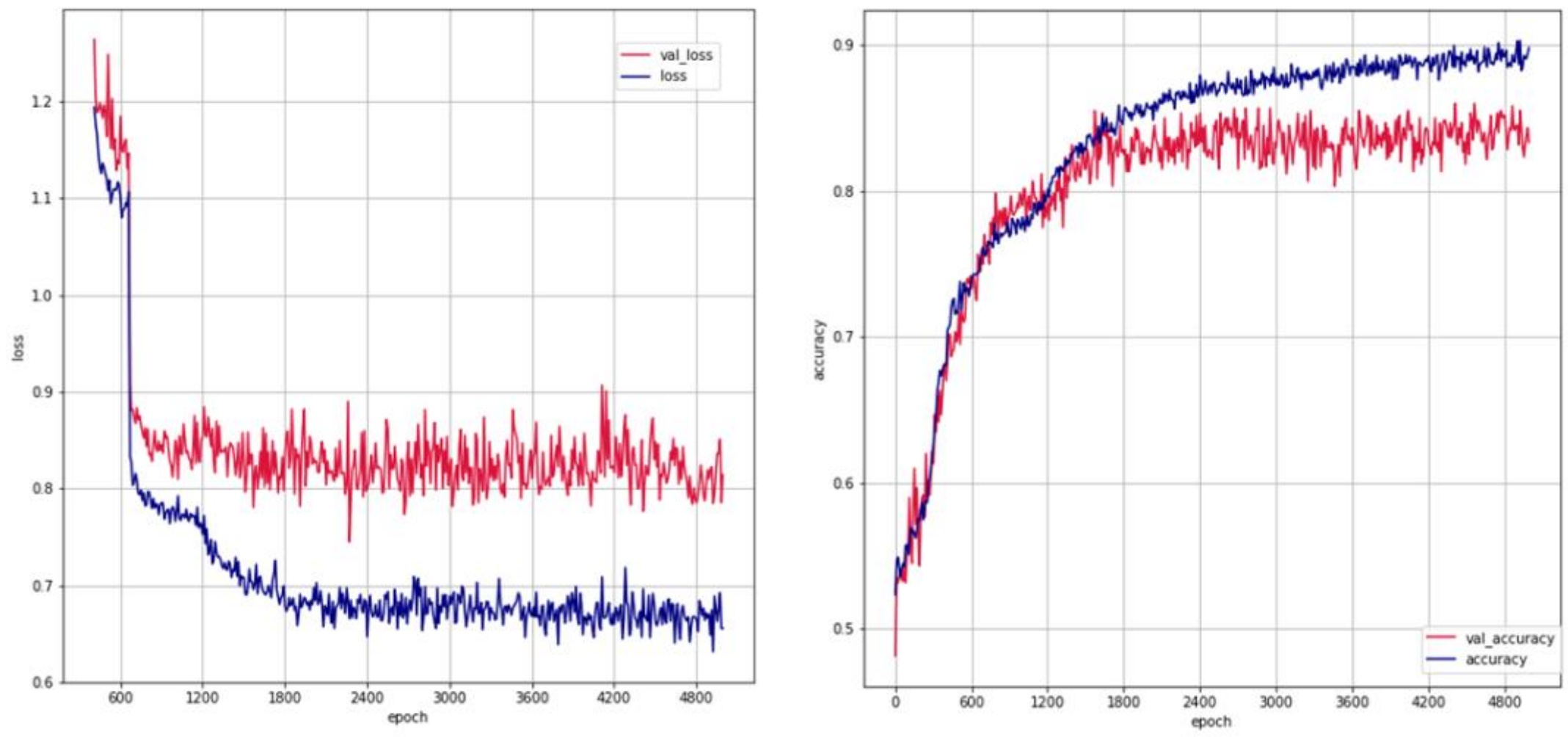

(a) (b)

Figure 5. (a) Training Loss (blue) and validation loss (red) as functions of epoch (feature extraction + 1D CNN), training on all subjects except subject #7, validation on subject #7. (b) Training accuracy (blue) and validation accuracy (red) as functions of epoch (feature extraction + 1D CNN), training on all subjects except subject #7, validation on subject #7.

**Accuracy**

| Subject (Fold #) | 1D CNN (+MVPA) | SVM (Linear) | SVM (RBF) | RF | XGB | LDA | MLP |
|---|---|---|---|---|---|---|---|
| 0 | 0.868 | 0.763 | 0.767 | 0.723 | 0.718 | 0.752 | 0.735 |
| 1 | 0.853 | 0.763 | 0.830 | 0.765 | 0.760 | 0.735 | 0.768 |
| 2 | 0.857 | 0.753 | 0.777 | 0.750 | 0.737 | 0.735 | 0.740 |
| 3 | 0.820 | 0.767 | 0.767 | 0.697 | 0.733 | 0.772 | 0.747 |
| 4 | 0.747 | 0.715 | 0.707 | 0.687 | 0.700 | 0.705 | 0.672 |
| 5 | 0.873 | 0.758 | 0.822 | 0.753 | 0.765 | 0.778 | 0.742 |
| 6 | 0.850 | 0.715 | 0.799 | 0.757 | 0.753 | 0.735 | 0.743 |
| 7 | 0.868 | 0.773 | 0.832 | 0.743 | 0.743 | 0.790 | 0.763 |
| 8 | 0.870 | 0.747 | 0.752 | 0.712 | 0.778 | 0.738 | 0.778 |
| 9 | 0.863 | 0.750 | 0.767 | 0.728 | 0.722 | 0.752 | 0.745 |
| 10 | 0.870 | 0.715 | 0.838 | 0.820 | 0.815 | 0.742 | 0.760 |

Table 1. Leave-One-Subject-Out accuracy of the 1D CNN classifier in comparison to classical Machine Learning algorithms for each LOOCV test fold subject.

Experiment A combined One-Way ANOVA feature selection and hyperalignment, and used LOOCV to measure accuracy. The mean LOOCV accuracy (± standard deviation) from the 1D CNN in the first experiment was 84.9% (± 3.5%) for the 1D CNN, making it at least 6% more accurate than classifiers such as SVM (Linear kernel 74.7% ± 2.1%; RBF kernel 78.7% ± 3.9%), Random Forests (RF) (73.9% ± 3.5%), Gradient Boosting (XGB) (74.8% ± 3.0%), Linear Discriminant Analysis (LDA) (74.9% ± 2.3%) and Multi-Layer Perceptrons (MLP) (74.5% ± 2.6%) (Hanson et al., 2004) (Table 1). The 1D CNN we developed for this experiment significantly outperformed other statistical and Machine Learning classification methods.

**Confusion Matrix**

The confusion matrix shows the classifier's performance on each emotion task. A perfect classifier would have a diagonal confusion matrix, with $200$ predictions for each diagonal element, and 0 predictions elsewhere. Table 2 shows the confusion matrix of the 1D CNN classifier averaged over all $11$ cross-validation folds.

| | Rest (Predicted) | Neutral (Predicted) | Negative (Predicted) |
|---|---|---|---|
| Rest (True) | 182.7 | 9.7 | 7.5 |
| Neutral (True) | 12.6 | 164.8 | 22.5 |
| Negative (True) | 9.7 | 28.4 | 161.9 |

Table 2. Confusion Matrix of the 1D CNN classifier averaged over all $11$ Cross-Validation test folds ($600$ labels per test fold)

**Precision, Recall, balanced F1-score, and Receiver Operating Characteristic**

We evaluated the following performance metrics: Precision, Recall (also known as Sensitivity), balanced F1-score, and Receiver Operator Characteristic (ROC) over each cross-validation fold (Scheinost et al., 2019). For a given class $c$ ($c = negative, neutral \ or \ rest$), Precision measures the proportion of correctly predicted images from class $c$ in the dataset out of all images predicted as belonging to class $c$ by the classifier. A classifier that does not produce false positives would have a Precision of 1. For a given class $c$ ($c = negative, neutral\ or\ rest$), Recall measures the proportion of correctly predicted images from class $c$ in the dataset out of all images in the dataset which truly belong to class $c$. A classifier that does not produce false negatives would have a Recall of 1. The balanced F1-score is the harmonic mean between Precision and Recall and is robust to imbalanced class distributions. The ROC curve measures a classifier's ability to reliably discriminate between classes by simultaneously evaluating true positive and false positive rates as functions of the classifier's prediction confidence. Table 3 shows the precision, recall, and balanced F1-scores, while Figure 6 shows the ROC curve for Experiment A.

| | Precision | Recall | F1-score |
|---|---|---|---|
| Class 0: rest | 0.892 ± 0.04 | 0.914 ± 0.06 | 0.902 ± 0.04 |
| Class 1: neutral | 0.820 ± 0.07 | 0.824 ± 0.05 | 0.819 ± 0.04 |
| Class 2: negative | 0.849 ± 0.06 | 0.810 ± 0.08 | 0.825 ± 0.04 |

Table 3. Precision, Recall, and F1-score averaged over all 11 Cross-Validation test folds (200 labels per class and per subject)

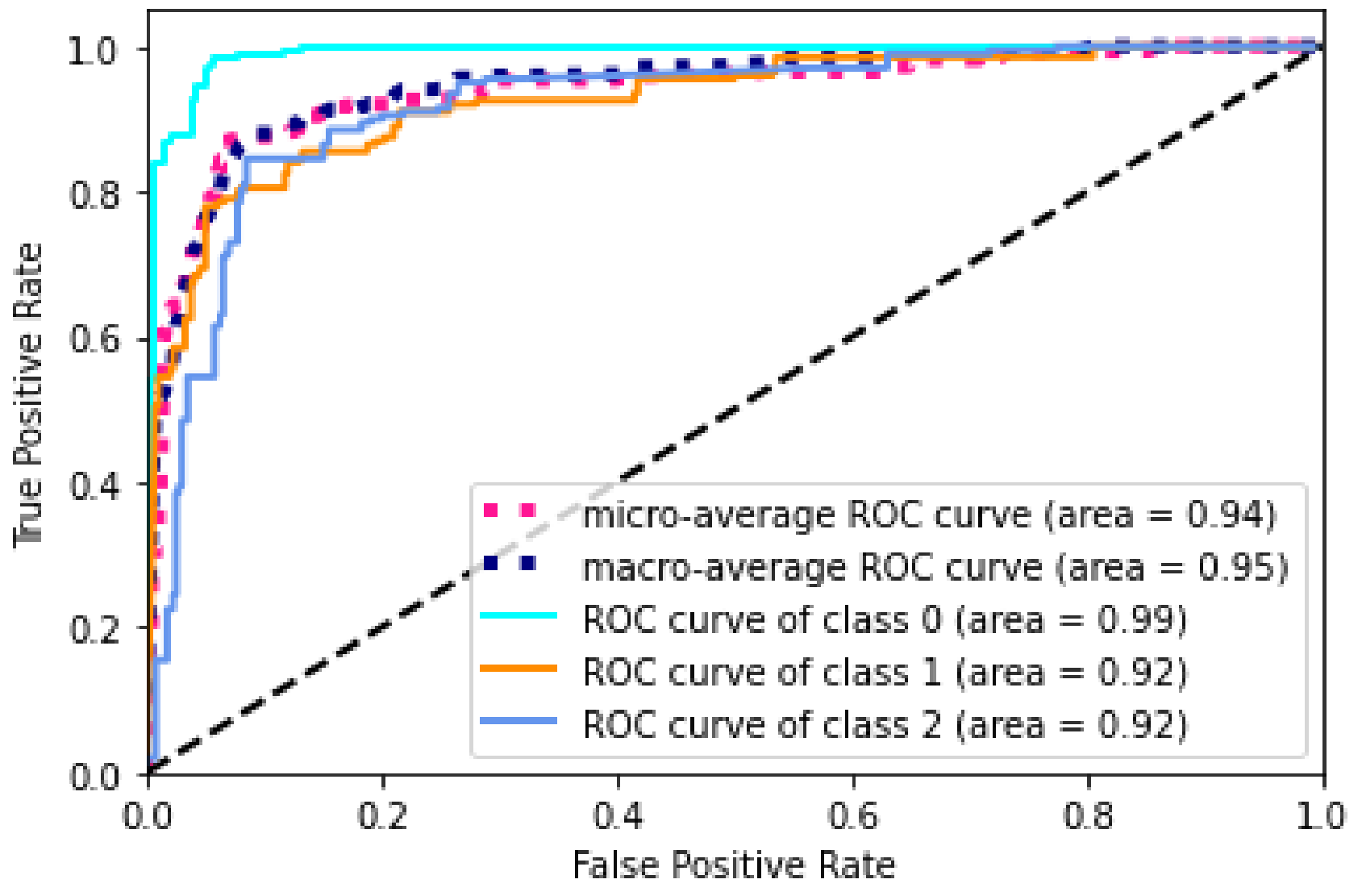

Figure 6. Multi-Class test set Receiver Operating Characteristic (ROC) and Area Under the test set ROC Curve (area), training on all subjects except subject #10, validation on subject #10. Each colored solid line represents a different emotion task label.

A classifier that perfectly separates a class from all the others would yield a ROC curve equivalent to a single point in the upper left corner of the plot, whereas a random chance classifier would yield a ROC curve equivalent to a diagonal line from bottom left to top right corners of the plot. ROC curves for all three classes in Experiment A were near the (0, 1) point in the ROC coordinate space (Fig. 6). The Rest class (Fig. 6, class 0) yielded an Area Under the ROC curve (AUROC) of 0.99 when cross-validating on images from subject #11. The neutral (Fig. 6, class 1) and negative (Fig. 6, class 2) classes both yielded an Area Under the ROC curve of 0.92 when cross-validating on fMRI images from subject #11.

### 3.3.2. For Experiment B

The ResNet-50 model from Experiment B converged with some overfitting on the training data, as illustrated by the generalization gap between the training and validation metrics in Figure 7 (a) and (b). In this experiment, similar overfitting was present for other Deep Learning methods as well as SVM and Random Forests. This classifier performed brain states inference on individual 3D fMRI images directly, and classified a batch of 1950 3D fMRI images in 2.687 sec, for an average prediction time of 1.38 msec per 3D fMRI image. The performance of the classifier was evaluated below.

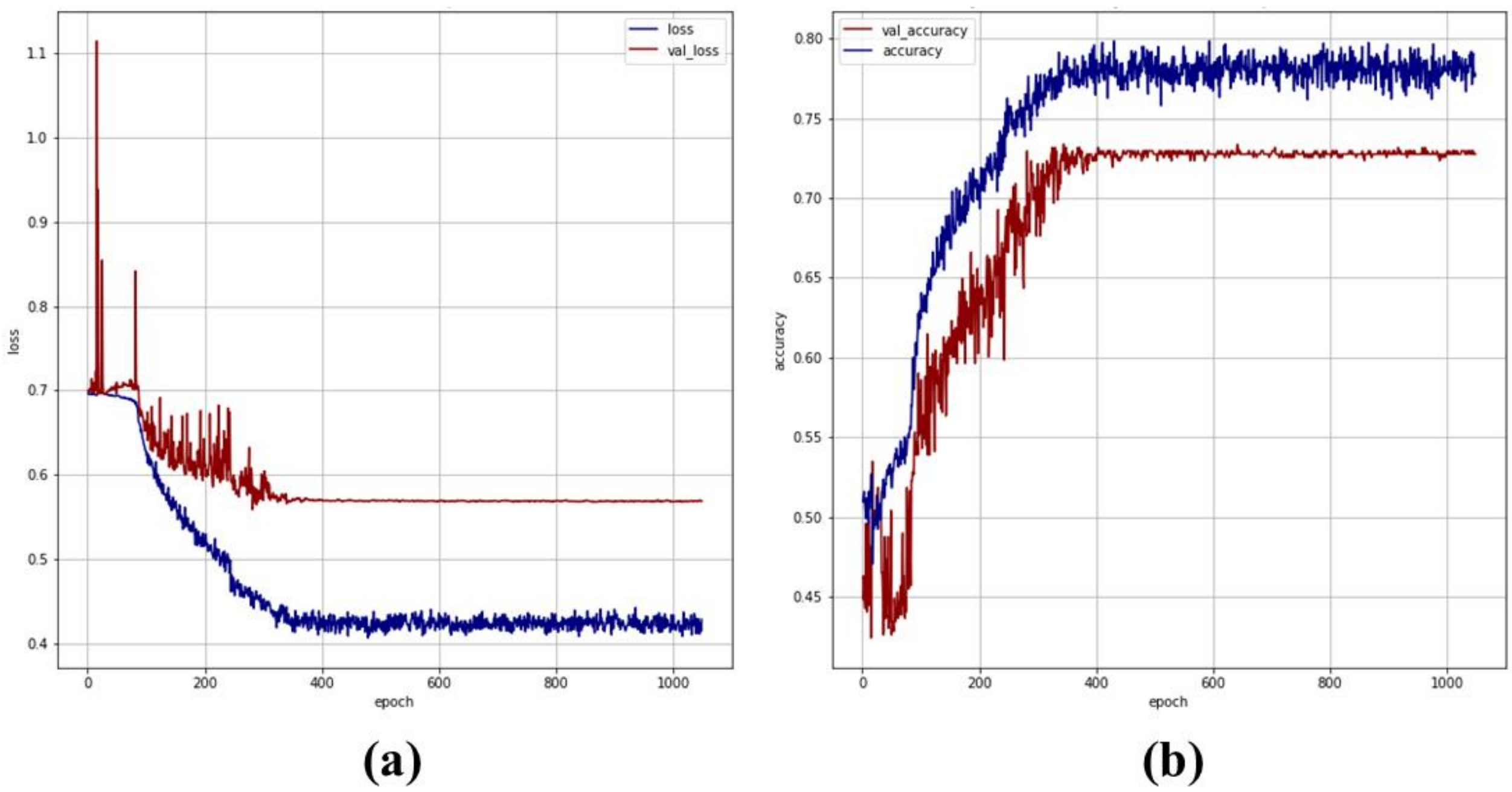

Figure 7. (a) Training loss (blue) and validation loss (red) as functions of epoch (3D ResNet-50). (b) Training accuracy (blue) and validation accuracy (red) as functions of epoch (3D ResNet-50).

**Accuracy**

Accuracy was measured using random training, validation, and test splits, and the 3D ResNet-50 model achieved a test accuracy of 78.0% (± 4.4%), surpassed by Random Forests (RF) (86.7% ± 0.9%), but outperforming Linear SVM (76.6 ± 1.8%), Multi-Layer Perceptrons (MLP) (51.2 ± 2.9%) and RBF-kernel SVM (45.8 ± 2.4%) (Table 4). Several Machine Learning methods were incapable of performing emotional brain state classification on the memory-intensive 3D fMRI data from Experiment B (Gradient Boosting, Linear Discriminant Analysis).

| Training Instance # | ResNet-50 | SVM (Linear) | SVM (RBF) | RF | MLP |
|---|---|---|---|---|---|
| 0 | 0.754 | 0.754 | 0.441 | 0.857 | 0.562 |
| 1 | 0.803 | 0.766 | 0.457 | 0.870 | 0.479 |
| 2 | 0.820 | 0.754 | 0.449 | 0.861 | 0.500 |
| 3 | 0.754 | 0.752 | 0.489 | 0.864 | 0.505 |
| 4 | 0.790 | 0.775 | 0.439 | 0.877 | 0.503 |
| 5 | 0.770 | 0.807 | 0.413 | 0.854 | 0.530 |
| 6 | 0.770 | 0.769 | 0.497 | 0.879 | 0.528 |
| 7 | 0.738 | 0.764 | 0.456 | 0.869 | 0.464 |
| 8 | 0.885 | 0.734 | 0.482 | 0.856 | 0.549 |
| 9 | 0.721 | 0.780 | 0.457 | 0.880 | 0.505 |

Table 4. Accuracy of the ResNet-50 classifier in comparison to other classical Machine Learning algorithms over the test dataset (610 samples). We repeated the training and testing processes 10 times during evaluation.

**Confusion Matrix**

| | Neutral (Predicted) | Negative (Predicted) |
|---|---|---|
| Neutral (True) | 250 | 53 |
| Negative (True) | 59 | 248 |

Table 5. Confusion Matrix of the ResNet-50 classifier over the test set.

The classifier achieved a test accuracy of 81.6% over the original test dataset (Table 5), in line with the confidence range of 78.0% (± 4.4%) established when training the ResNet-50 classifier 10 times (Table 4). The false positive (FP: 59) and false negative (FN: 53) rates in the confusion matrix were similar for both neutral and negative emotion tasks (Table 5).

**Precision, Recall, F1-score, and ROC**

Table 6 shows the precision, recall, and balanced F1-score of the Experiment B ResNet-50 classifier, while Figure 8 shows its ROC curve.

| | Precision | Recall | F1-score |
|---|---|---|---|
| Class 0: neutral | 0.809 | 0.825 | 0.817 |
| Class 1: negative | 0.824 | 0.808 | 0.816 |

Table 6. Precision, Recall, and F1-score over the test set (support of 303 labels for class 0, 307 labels for class 1).

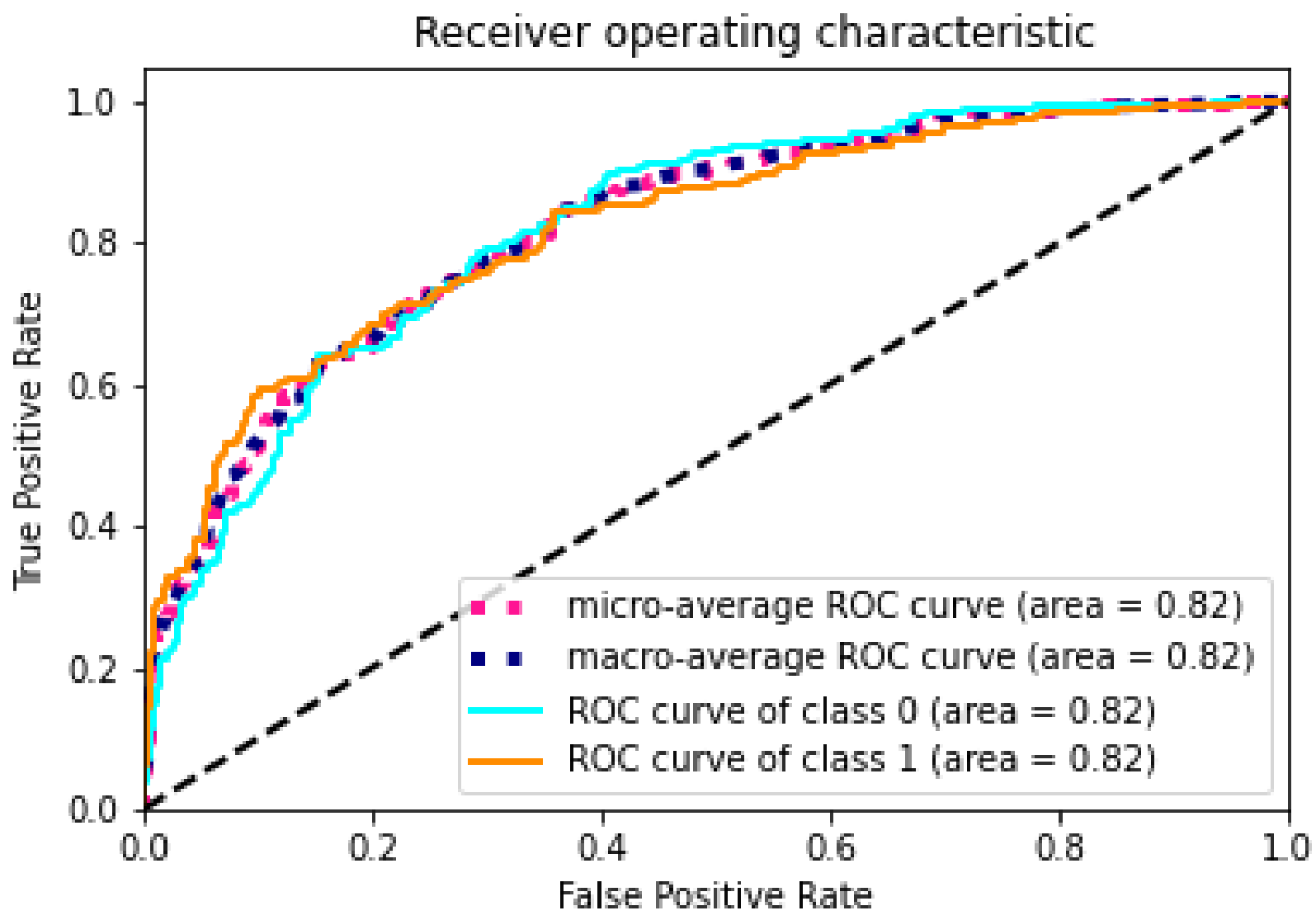

Figure 8. Receiver Operating Characteristic (ROC) and Area Under the ROC Curve (area) on the test set. Each colored solid line represents a different emotion task label.

Classification over all classes in Experiment B was better than random chance, as the ROC curves for both classes were above the diagonal line drawn from bottom left (0, 0) to the top right (1, 1) points in the ROC coordinate space (Fig. 8). The neutral (Fig. 8, class 0) and negative (Fig. 8, class 1) classes both yielded an Area Under the ROC curve (AUROC) of 0.82 when evaluating the original test set.

## 4. Discussion

### Summary

The first approach detailed in this paper (ANOVA + MVPA + 1D CNN) addressed the challenge of extracting generalizable group-level spatial-only features, and defined a method to accurately decode emotional brain states from $64 \times 64 \times 44$-dimension minimally pre-processed (slice timing, realignment) fMRI images. The dimensionality of data was reduced to 300 hyperaligned voxels – a considerable compression for fMRI image purposes. The 1D CNN we developed (detailed in Section 2.5.1) achieved an average LOOCV test accuracy of 84.9% (± 3.5%) on the balanced dataset, which contained an equal amount of rest, neutral and negative states (random chance test accuracy: 33.3%). This classifier was at least 6% more accurate than comparable methods and showed relatively limited overfitting. The second approach (ResNet-50) alleviated the need to manually extract complex features altogether and specifically addressed the challenge of performing fast emotional brain state classifications from individual 3D fMRI images. In this experiment, fMRI data was high-dimensional and minimally pre-processed, and the 3D ResNet-50 model we developed correctly decoded brain states from individual fMRI images in 1.38 msec per image, making it suitable for real-time (< 1 TR) and online brain decoding. This model (detailed in Section 2.5.2) achieved a test accuracy of 78.0% (± 4.4%) – similar to comparable methods – on a balanced 3D fMRI image test dataset containing an equal amount of neutral and negative emotion states (random chance test accuracy: 50%). In this experiment, accuracy was measured using 80/20/20 random training, validation, and test splits, and should not be directly compared to accuracy in the first experiment (LOOCV).

**Key Findings**

In both experiments, models were capable of correctly learning features from a group of subjects and inferring emotional brain states on fMRI images from different subjects. LOOCV results for the 1D CNN were consistent across all 11 cross-validation folds, indicating that the model extracted cross-subject features which generalized to unseen subjects. While all benchmark classifiers obtained their lowest test accuracy on subject 4, the 1D CNN yielded comparatively better results (Table 1). The confusion matrix (Table 2) averaged over 11 folds showed that the model misclassified neutral brain states as negative more often than it misclassified emotional states (neutral, negative) as rest states. In the second experiment, we trained several ResNet classifiers with deeper architectures (ResNet-50, ResNet-101, ResNet-152) and different hyper-parameter configurations, and selected the best model based solely on generalization capability, not accuracy. Several models based on a ResNet-101 architecture outperformed all other classifiers including Random Forests, but this was at the cost of larger overfitting which was not desirable for our experiment. The ResNet-50 classifier selected for this paper was outperformed only by Random Forest classifiers (RF 86.7% ± 0.9%) on the Experiment B test set and achieved better results than Linear SVM (76.6% ± 1.8% with standard scaling). The comparison shown in Table 4 indicated that classification on 3D fMRI data was challenging, even for advanced classical Machine Learning algorithms such as the MLP (51.2 ± 2.9%) or RBF-kernel SVM (45.8 ± 2.4%). Notably, our machines were unable to handle the computational cost of training Linear Discriminant Analysis (LDA) and Gradient Boosting (XGB) classifiers on the $64 \times 64 \times 44$ voxel 3D fMRI dataset.

Aside from the insights drawn from addressing two key neuroimaging challenges, our experiments led us to identify three issues complicating the use of deep CNN and ResNets for emotional brain state classification. Firstly, the complexity of minimally pre-processed 3D fMRI brain images caused the classification error to decrease only after several hours of training. Deep spatial convolutional networks extracted highly complex patterns at a significant computational cost. Due to memory limitations, we discarded the rest states in Experiment B and trained 3D ResNet classifiers using only two classes (negative vs. neutral). Secondly, both CNN and ResNets were prone to overfitting and occasionally learned to recognize patterns that did not generalize to unseen test data. While overfitting is usually visible by comparing training and validation accuracies, we found the training and validation loss functions to be better indicators of overfitting. In several instances, after the training and validation accuracies had converged, validation loss increased, indicating that although the model was still correctly classifying brain states, its confidence in these predictions was decreasing. Finally, adapting CNN models to neuroimaging introduced multiple hyper-parameters (batch size, dropout rate, learning rate, regularization rate, number of epochs). Tuning parameters in this context was an arduous task as CNN and ResNet models were highly sensitive to parameter changes. For computational reasons, grid searching through all hyper-parameters simultaneously was not feasible. Ongoing efforts to improve processing unit performance should alleviate these limitations and facilitate the adoption of CNN and ResNet methods in neuroimaging.

**Future Work**

The dataset for both experiments was drawn from an emotion task study involving 22 subjects, including 10 who had a history of MDD. Due to the small sample size of the dataset, we combined fMRI data from both history of MDD and HC groups before performing emotional brain state classification with CNN and ResNets. In the future, we will analyze whether any significant differences can be found between models trained only on data from subjects with a history of MDD and models trained only on data from HC subjects in emotion task fMRI experiments. Additionally, we aim to explore larger fMRI datasets such as the Human Connectome Project data (Van Essen et al., 2013) and the ABCD study data (Casey et al., 2018) to further reduce overfitting and fully utilize the capabilities of our models. In the second experiment of this paper, we used a training/validation/test split with data from all subjects shuffled together to neutralize the possibility of the ResNet-50 model learning from the temporal correlations between different brain images. Predictions were made based on spatial voxel features alone, whereas classifiers trained on 4D (temporal + spatial) fMRI data would typically extract both spatial and temporal features (Dahan et al., 2021; Q. Li et al., 2021). Our approach allowed potential data leakage when similar brain images sampled from a single run of the same subject appeared in both the training and test sets. To directly evaluate the generalizability of extracted features for the second experiment involving ResNets, we will use LOOCV and an independent dataset for testing in the future. Finally, we will explore lighter models such as EfficientNets (Tan & Le, 2019), which are designed to reproduce similar results to ResNet while using orders of magnitude fewer parameters.

### Data and Code Availability Statement

FMRI data from subjects will not be made publicly available due to ethics and privacy concerns. The models produced for both experiments are available upon request to the corresponding author, and all code for both experiments is accessible via a public repository (Tchibozo, 2021).

## 5. Conclusions

We proposed two convolution-based approaches to address two key neuroimaging challenges: developing emotional brain state classification methods that are generalizable across out-of-sample test subjects and that can perform a fast and accurate decoding of individual minimally pre-processed 3D fMRI images. In the first experiment, we applied an ANOVA voxel extraction followed by hyperalignment to encode the 3D fMRI data into 1D vectors. These vectors were used as inputs to a CNN which extracted features for cross-subject inference using 1D convolutional kernels. In our second experiment, we directly used a set of 3D fMRI images after slice timing and realignment to train a spatial 3D ResNet-50 classifier. The 1D CNN from our first experiment convincingly outperformed other Machine Learning methods while being able to extract group-level emotional brain state features, and we demonstrated in our second experiment that ResNet models could accurately recognize emotional brain states from minimally pre-processed fMRI images in a few milliseconds (< 1 TR), satisfying a key requirement for real-time and online brain decoding. We believe these approaches to emotional brain state classification could be applied to real-time cross-subject fMRI neurofeedback research and could potentially build on existing methods (Taschereau-Dumouchel et al., 2018;

Tymofiyeva et al., 2020) to help provide non-invasive treatments for phobias and emotional dysregulations.

**Funding Sources:**

This work was supported in part by the Columbia Irving Institute Imaging Pilot Award (He X.) and the New York State Psychiatry Institute MRI Pilot Award (He X.).

Supplementary Materials:

# 1. Experiment A

## 1.1. Loss and Accuracy Plots

### 1.1.1. Subject 0

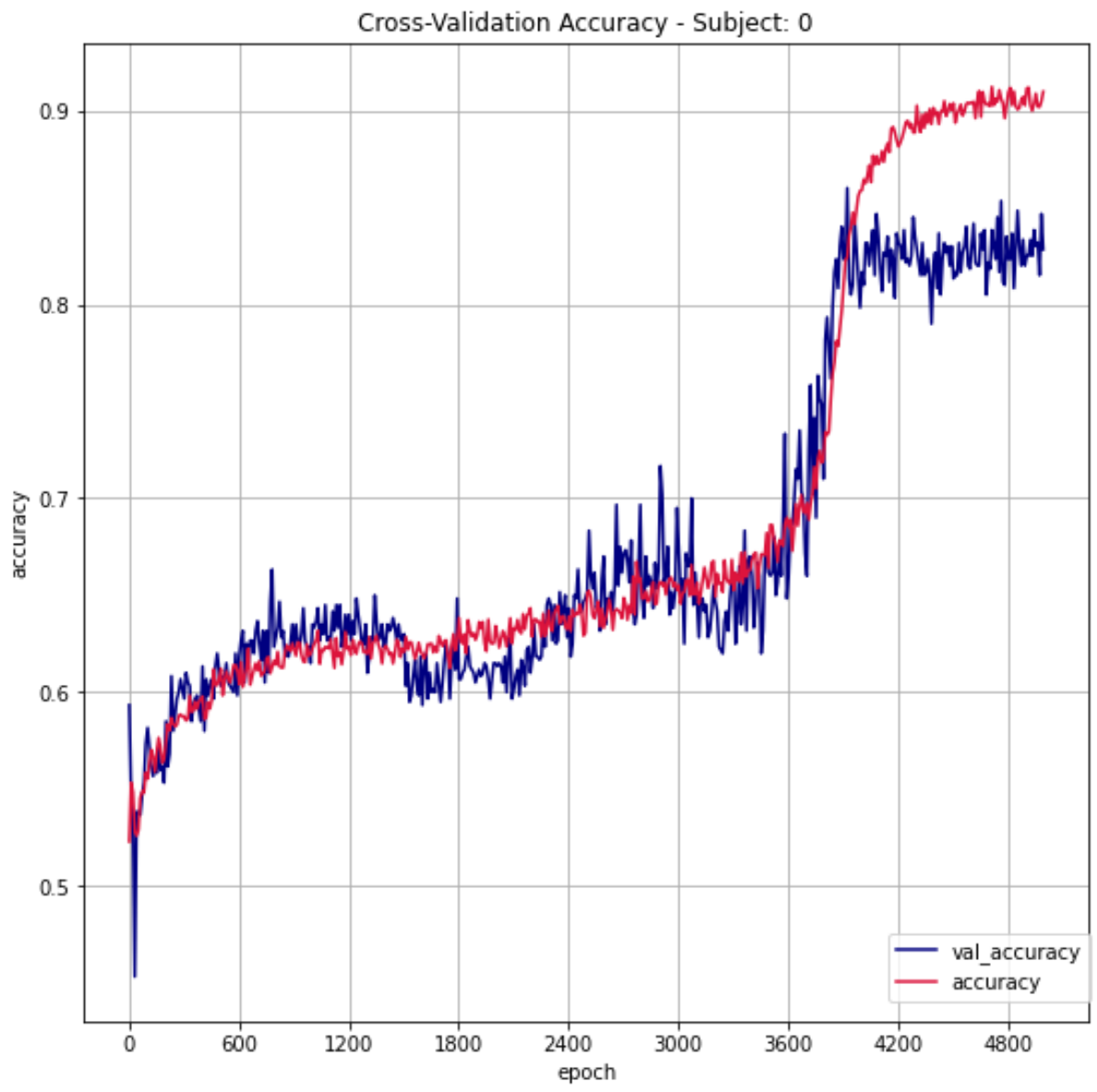

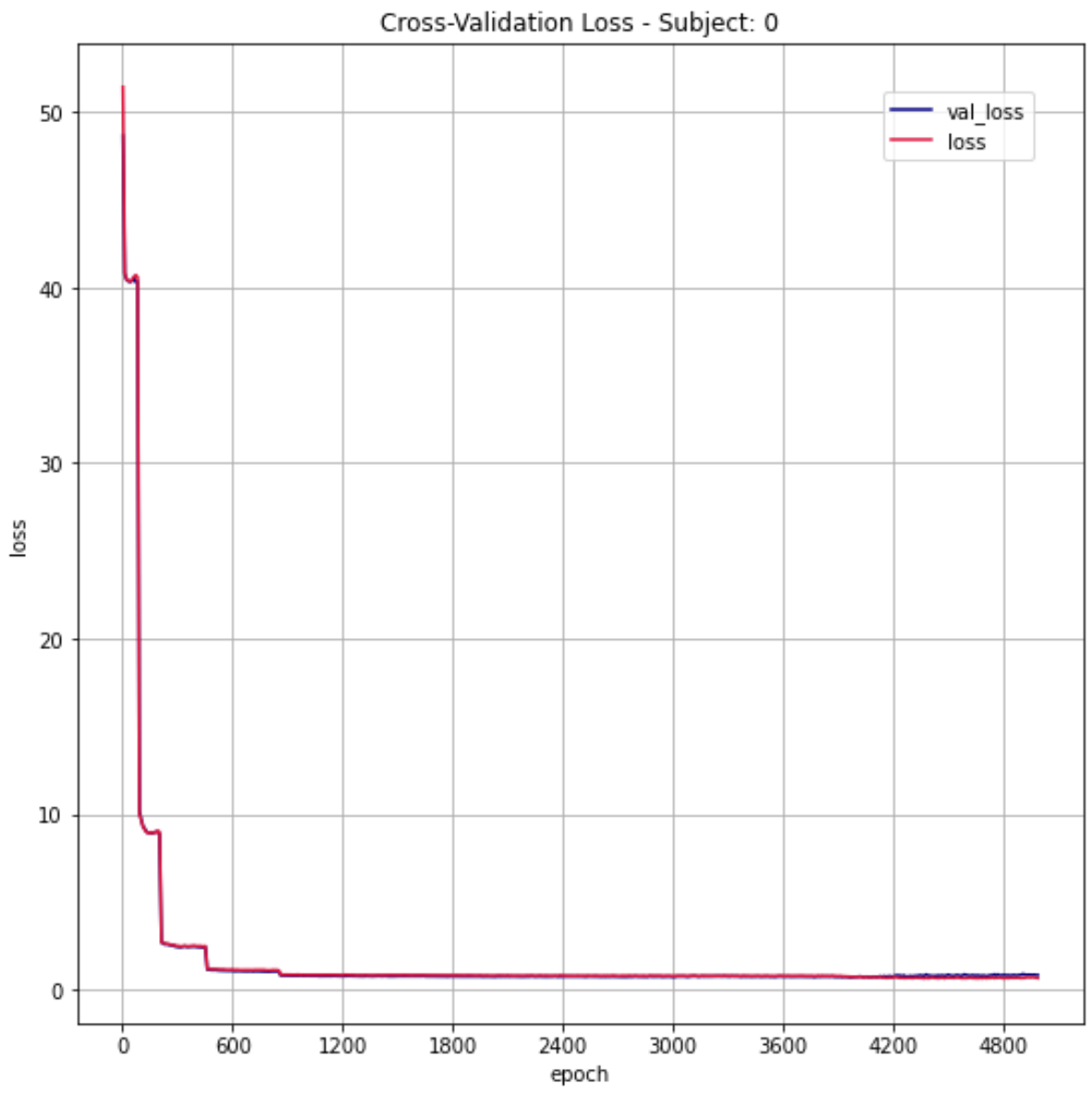

### 1.1.2. Subject 1

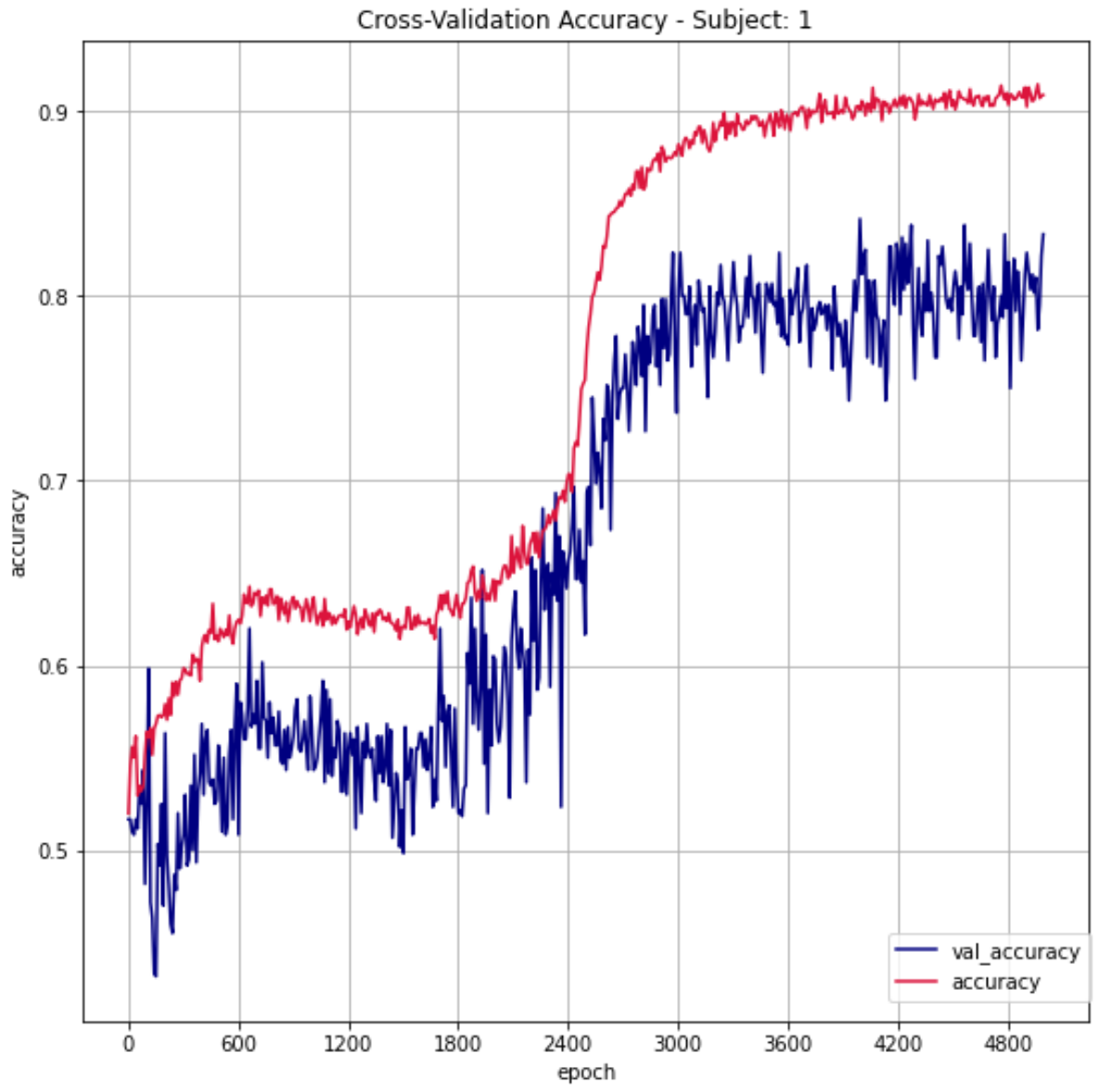

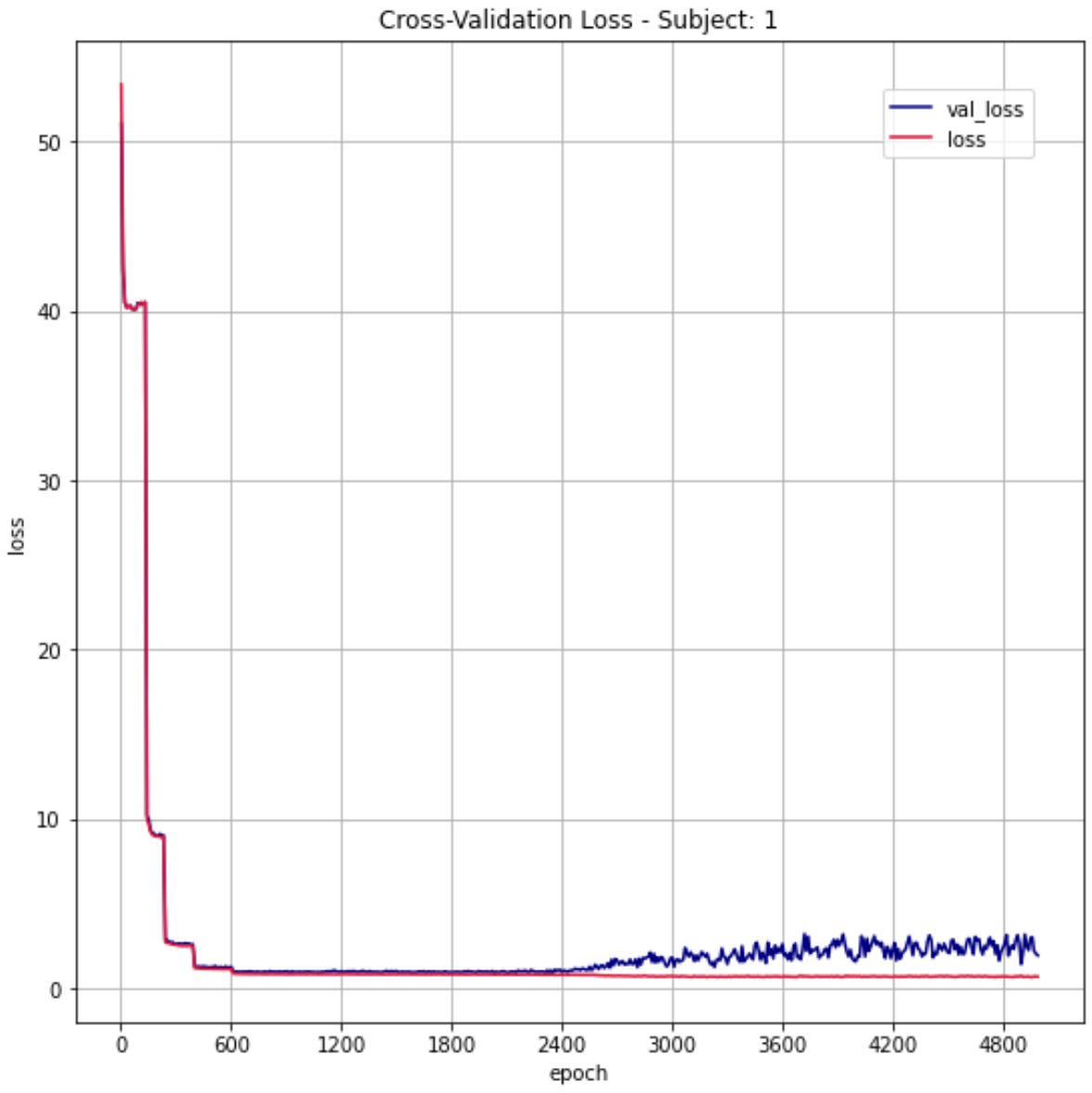

### 1.1.3. Subject 2

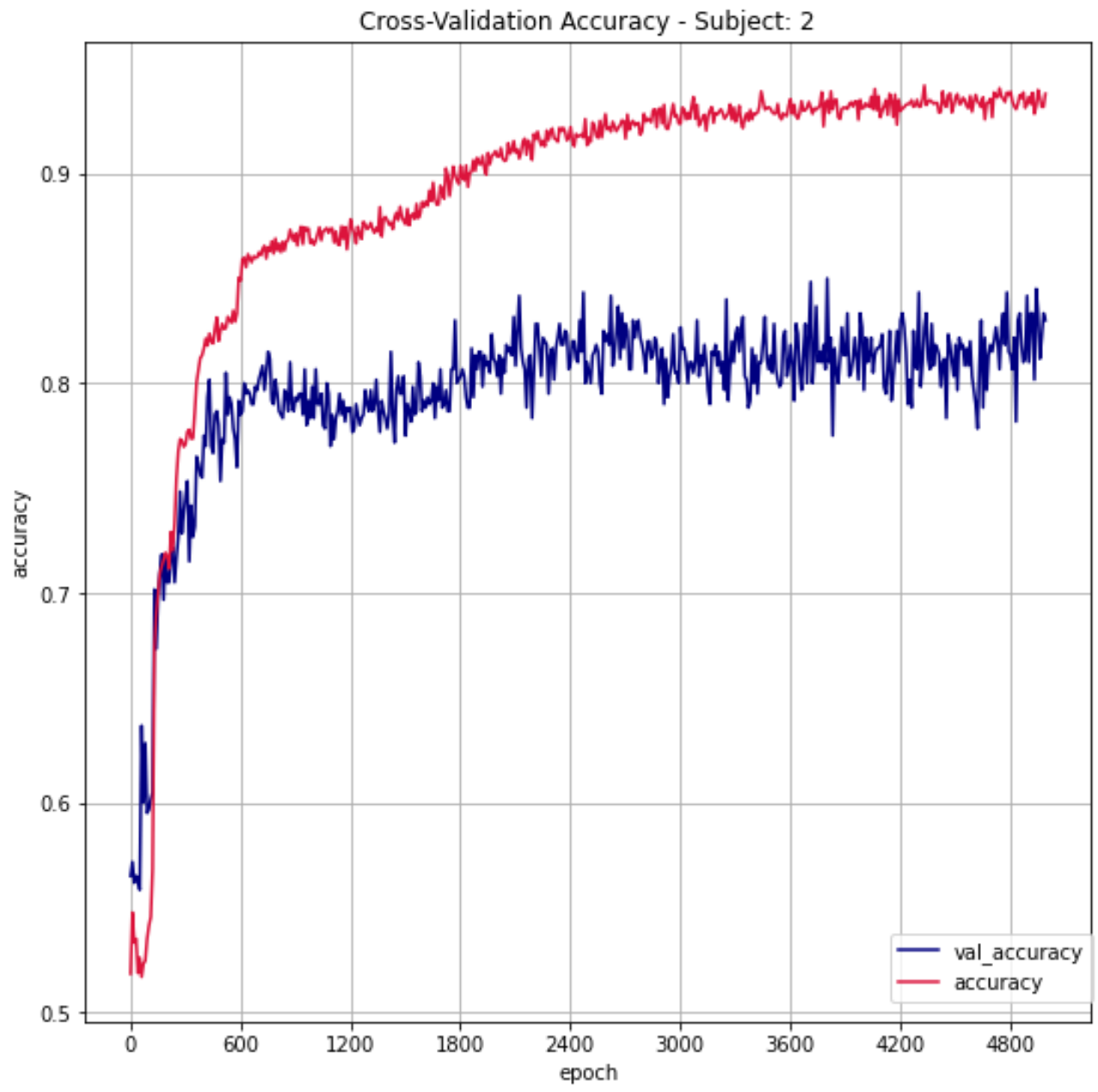

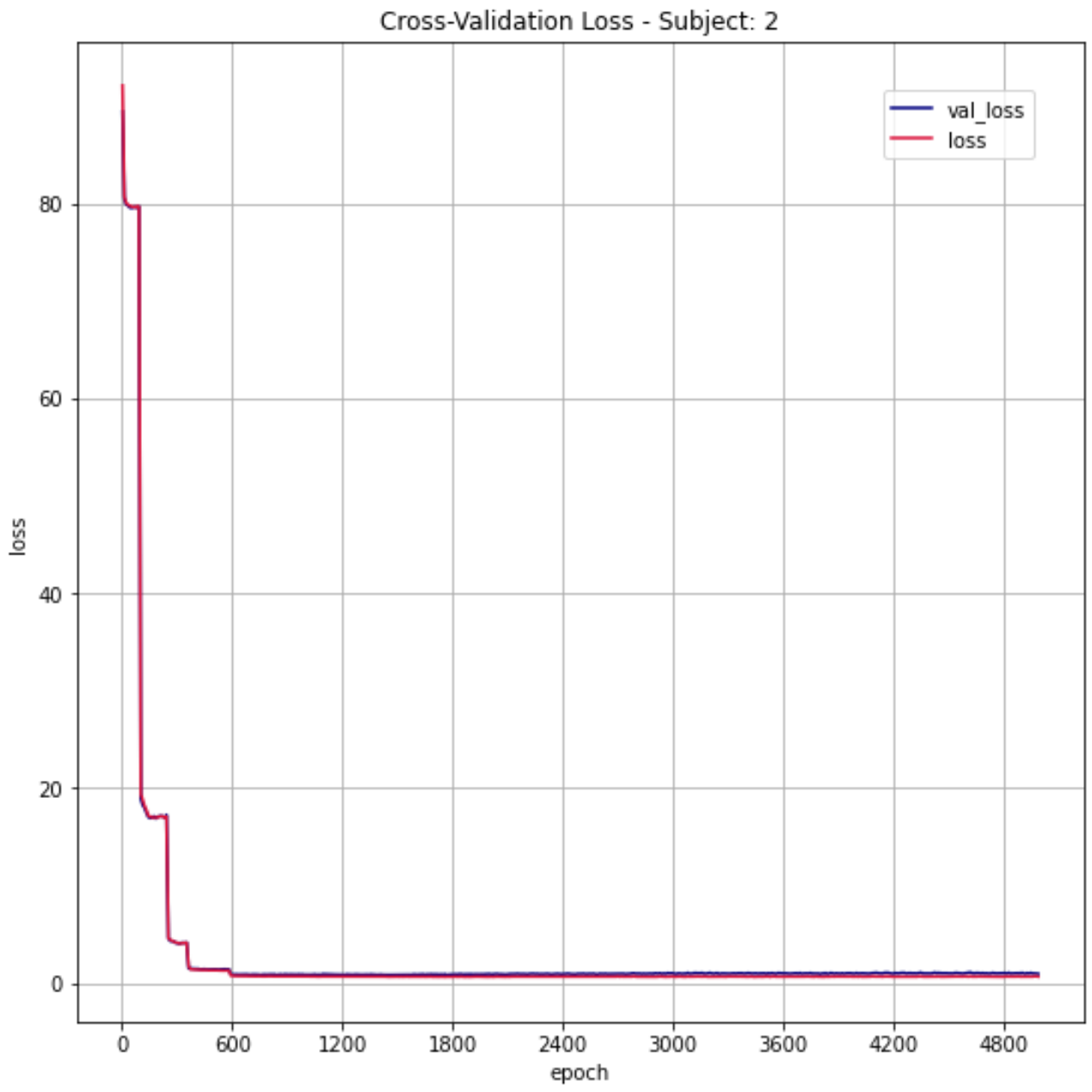

### 1.1.4. Subject 3

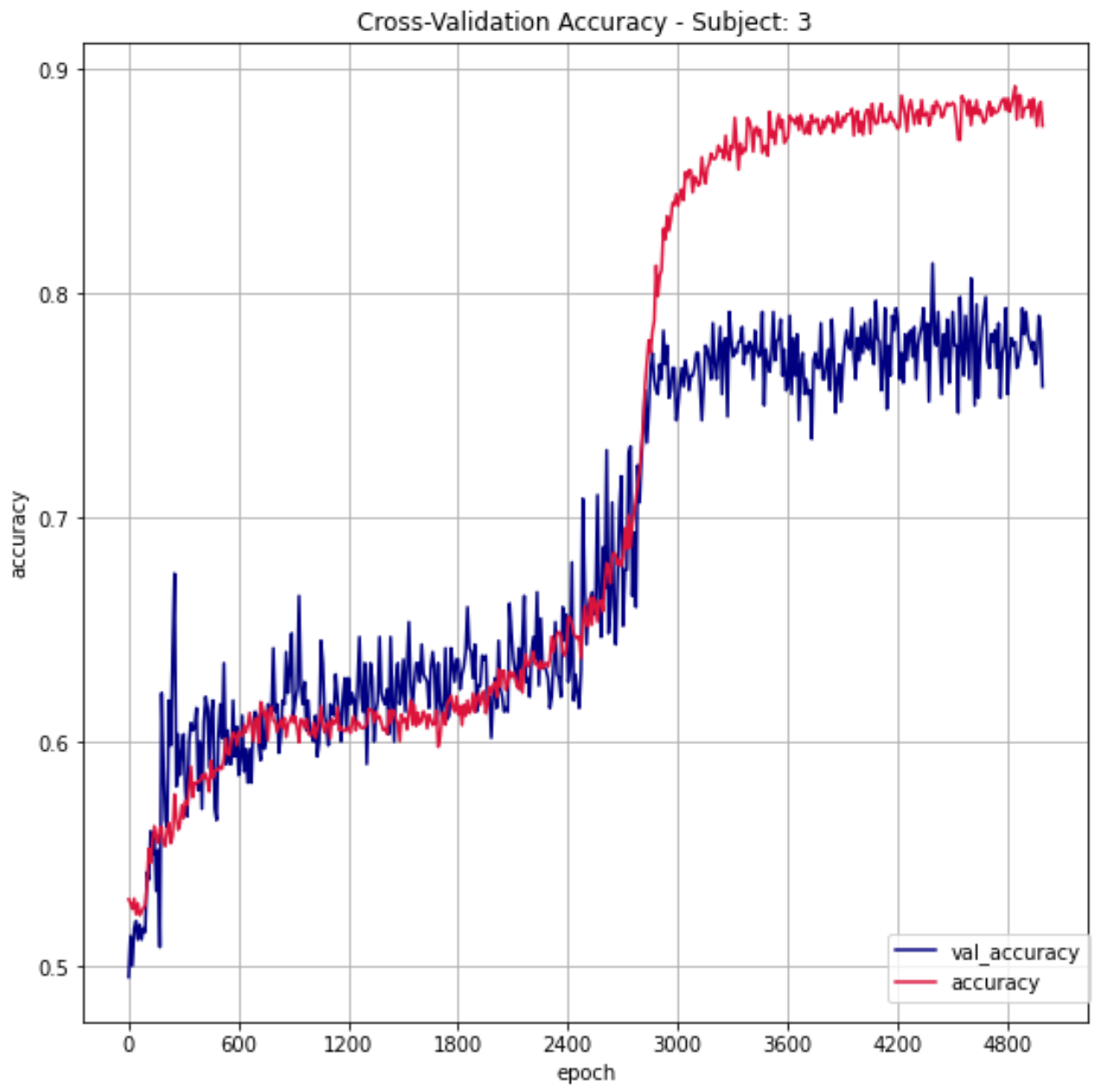

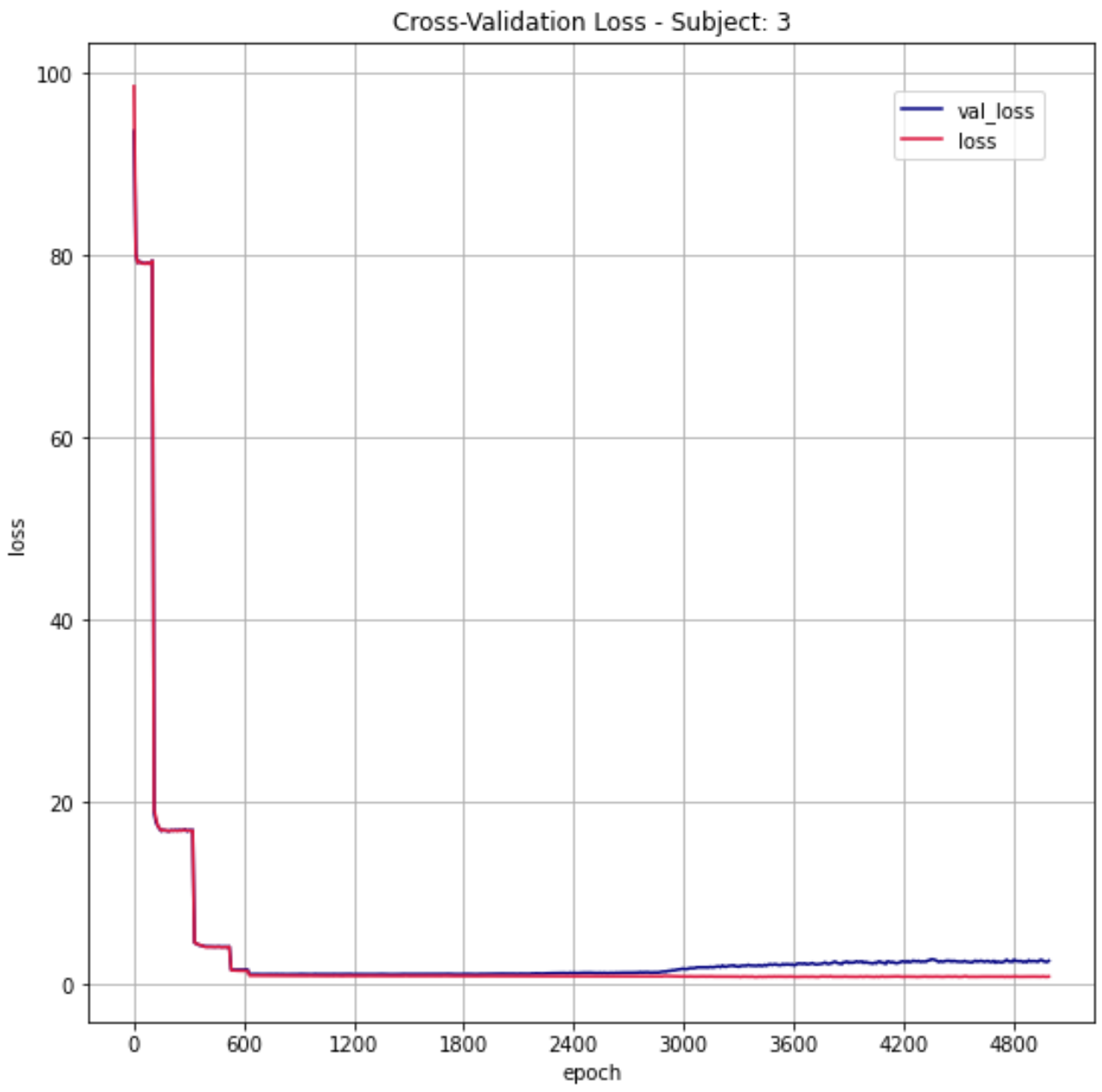

### 1.1.5. Subject 4

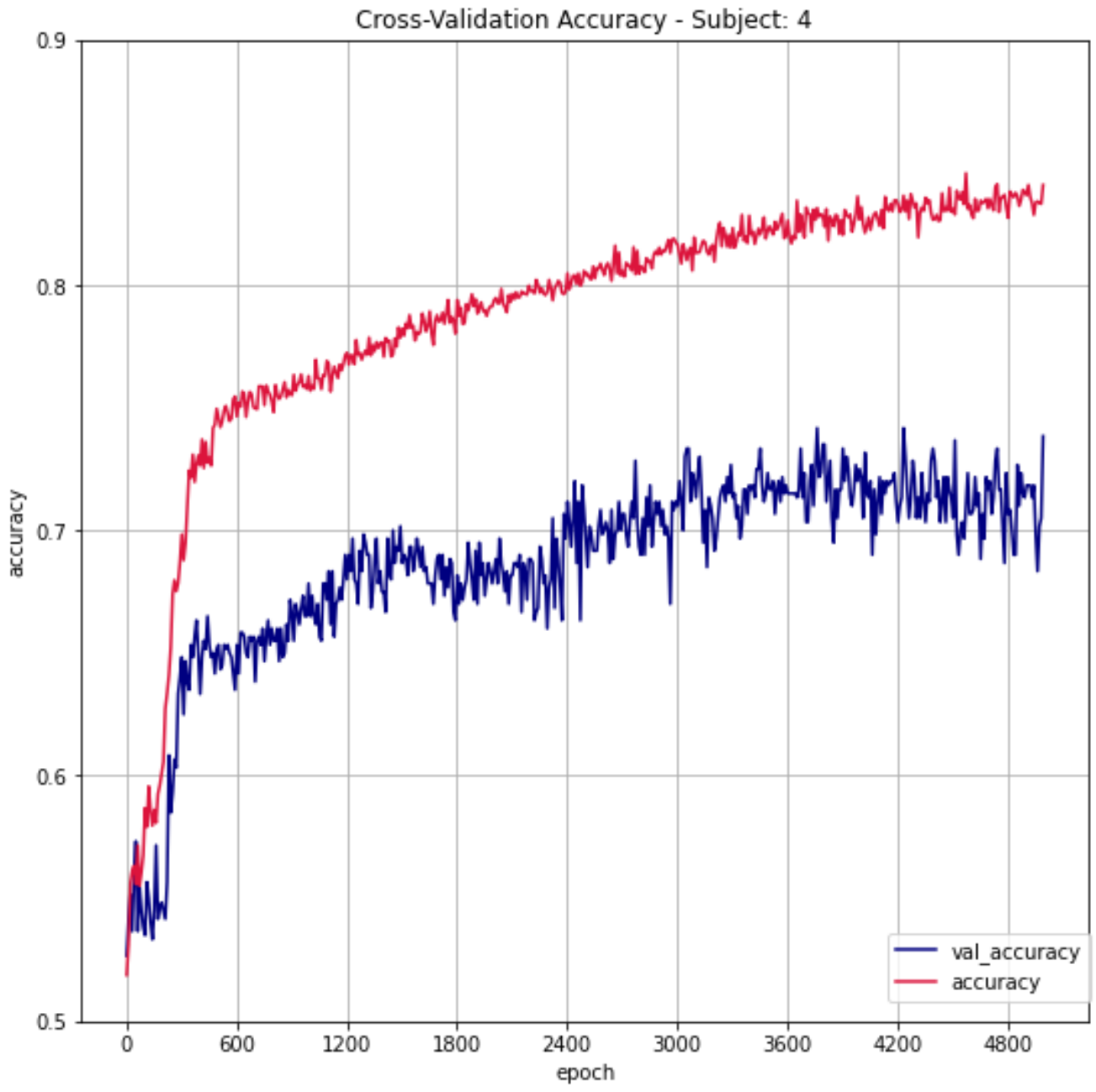

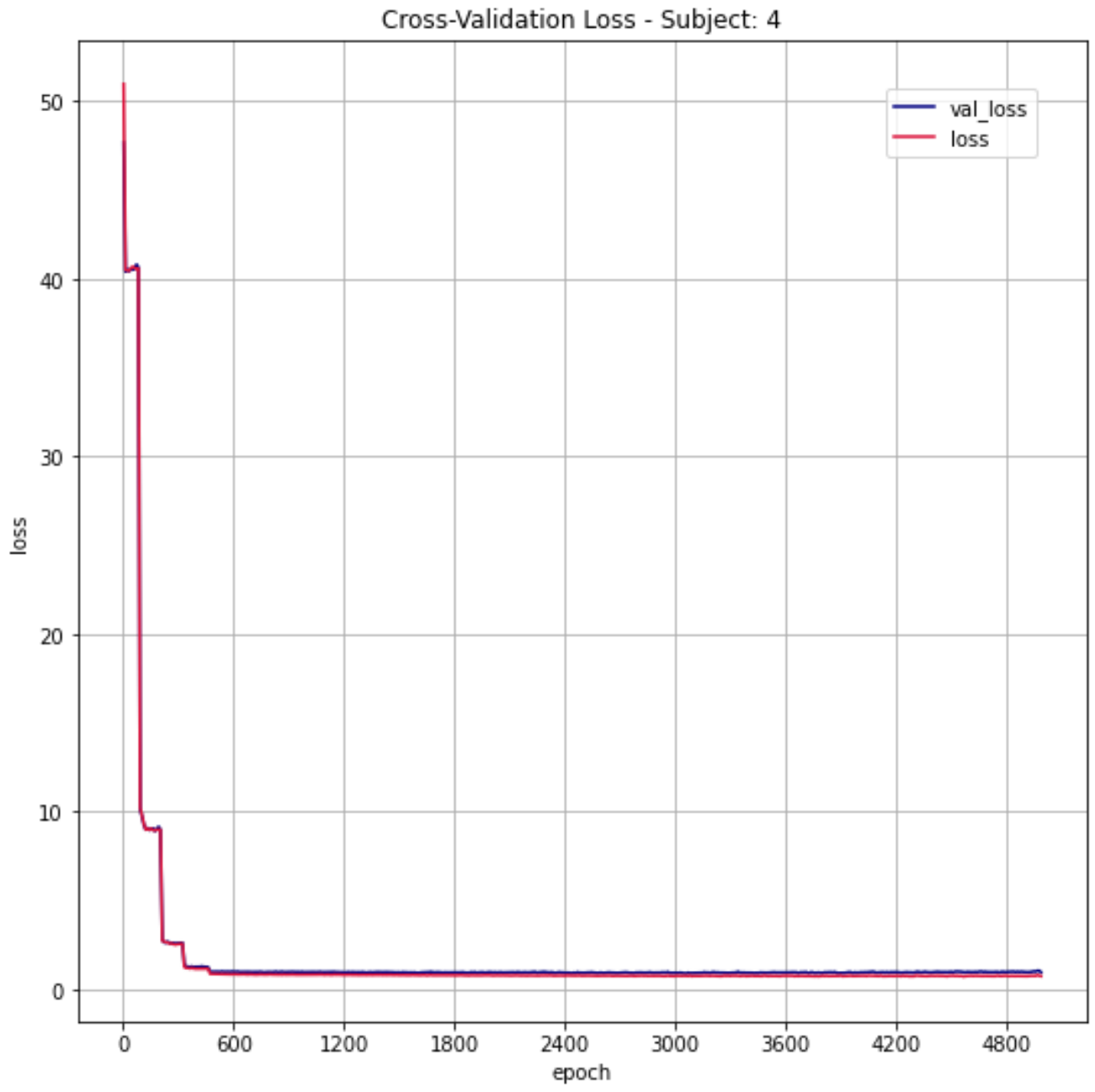

### 1.1.6. Subject 5

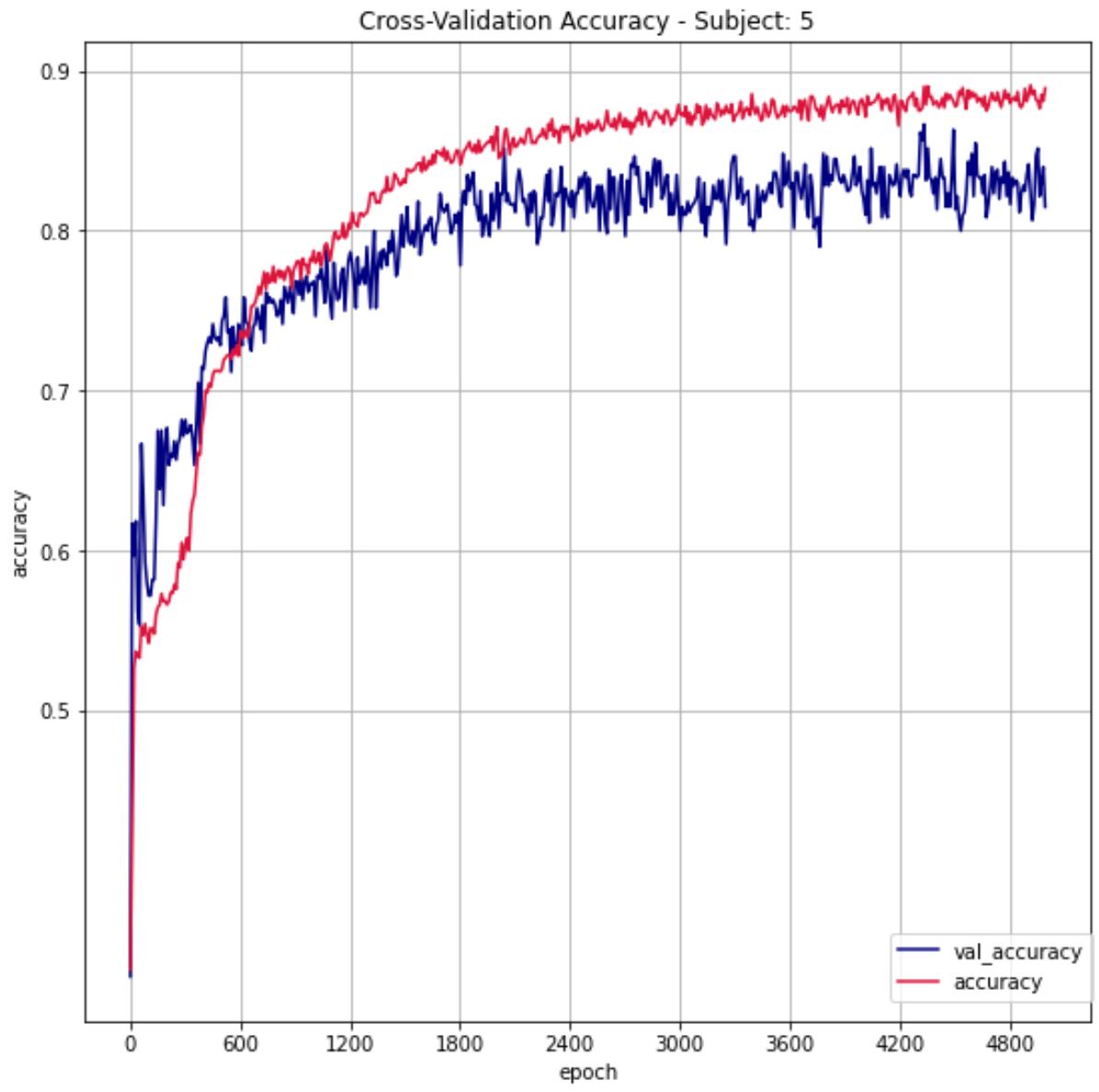

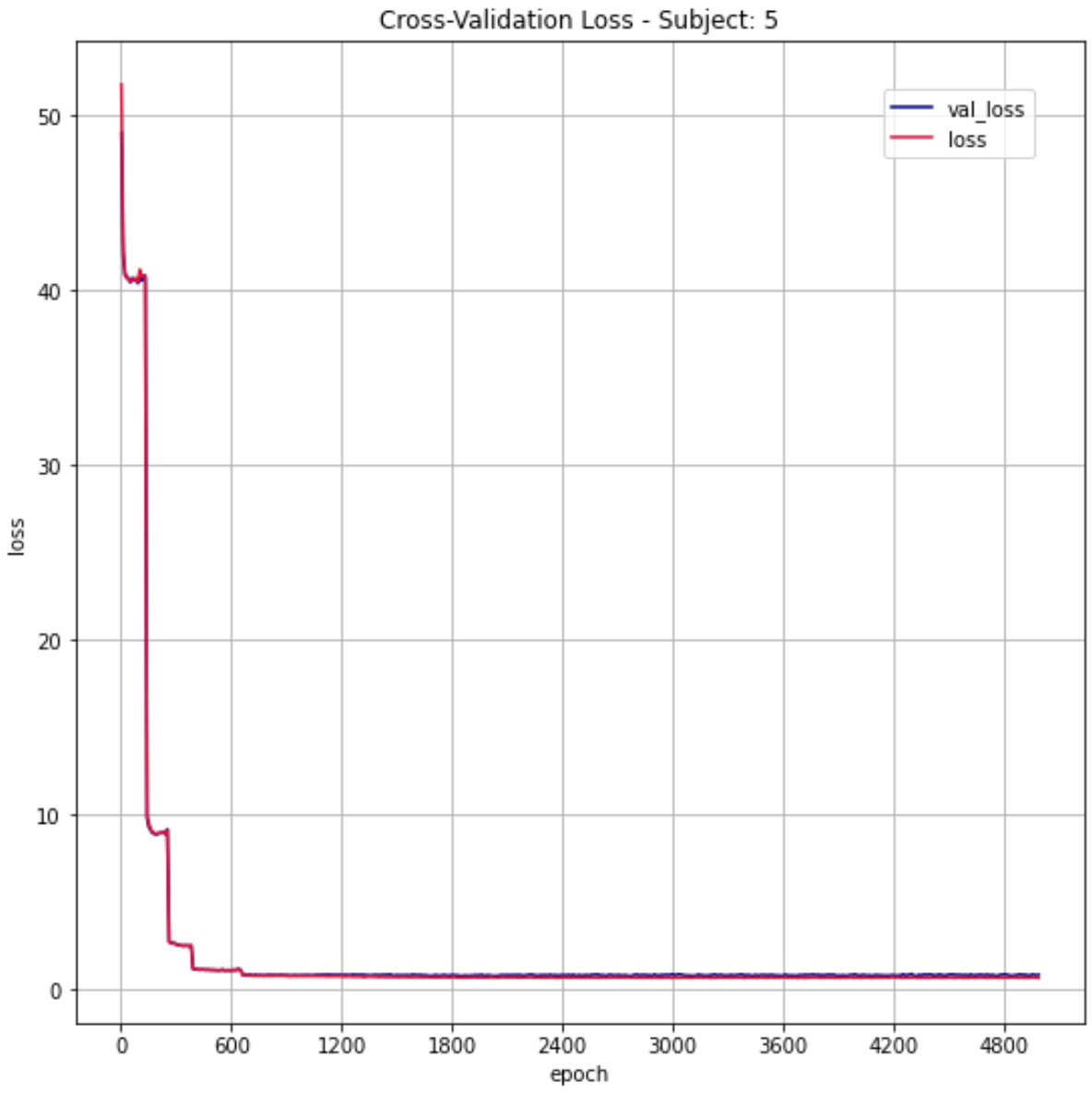

### 1.1.7. Subject 6

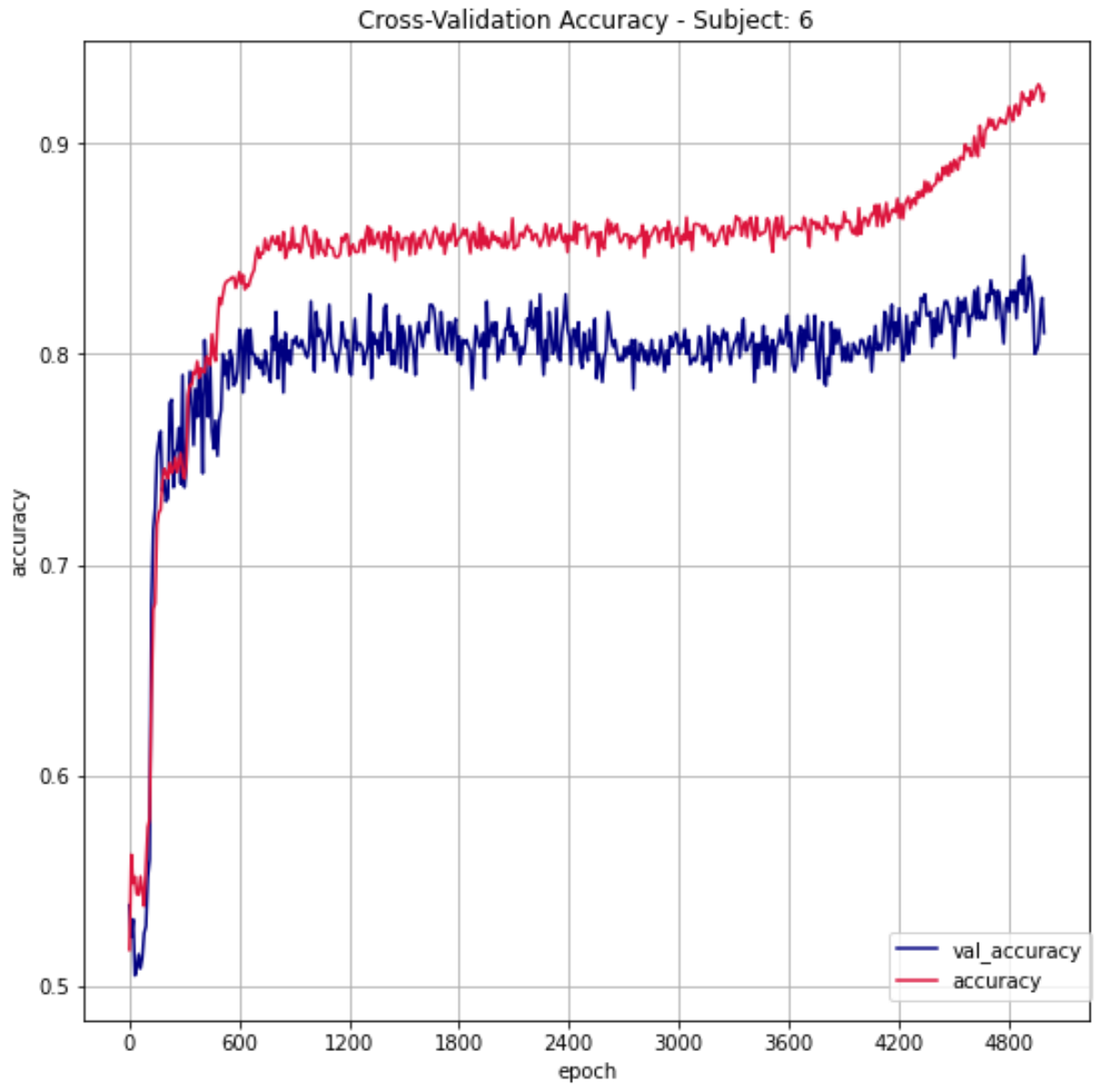

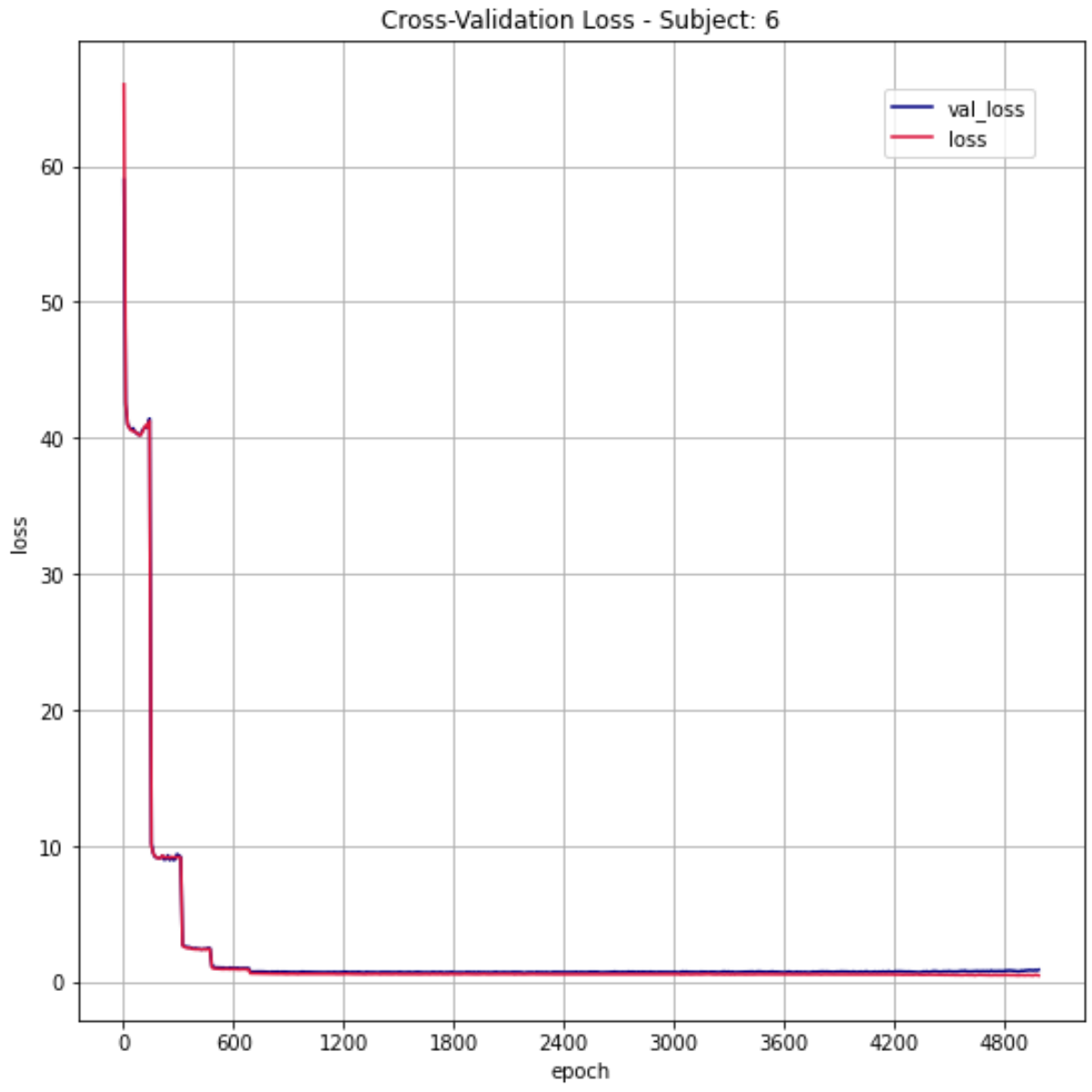

### 1.1.8. Subject 7

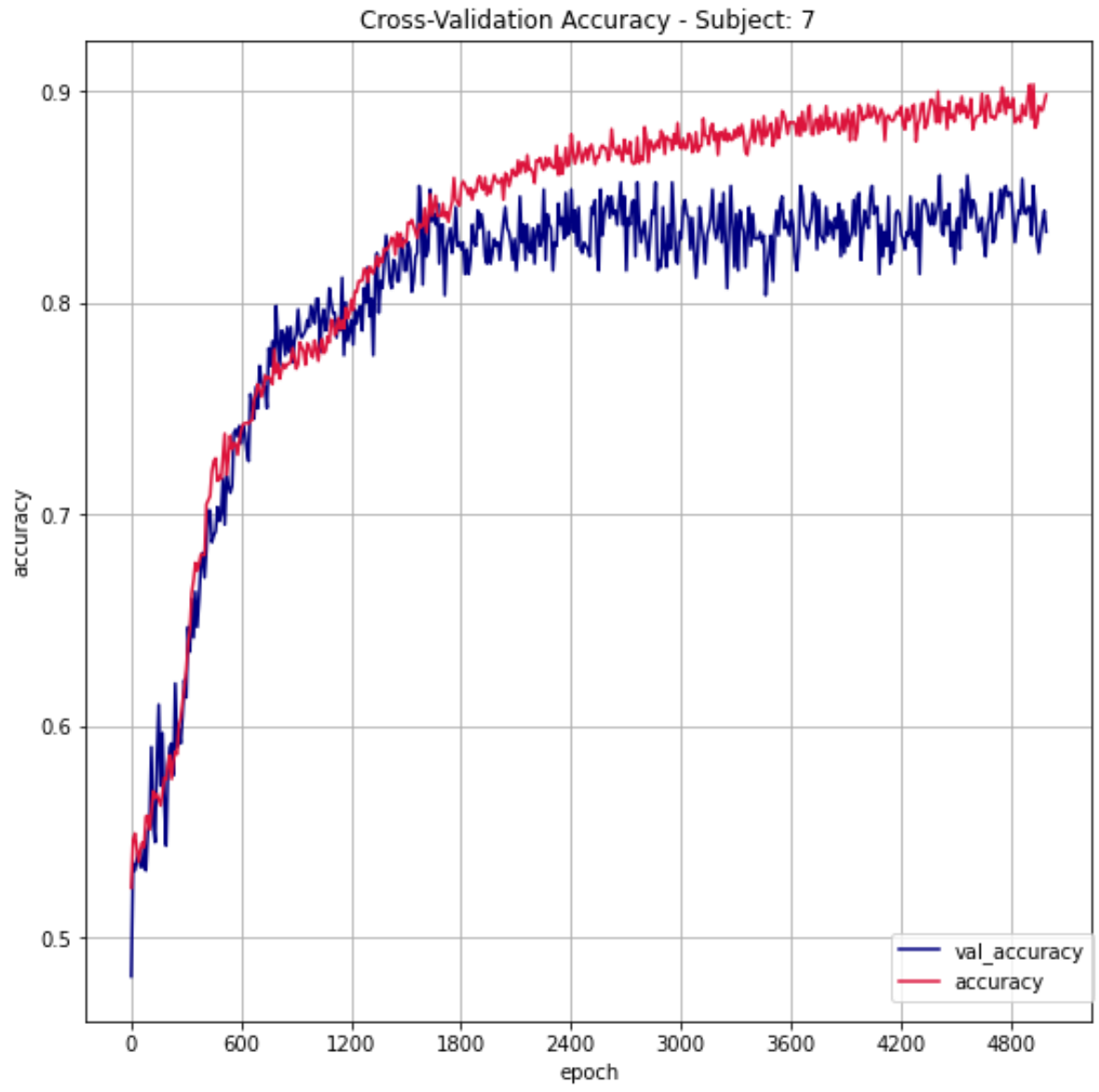

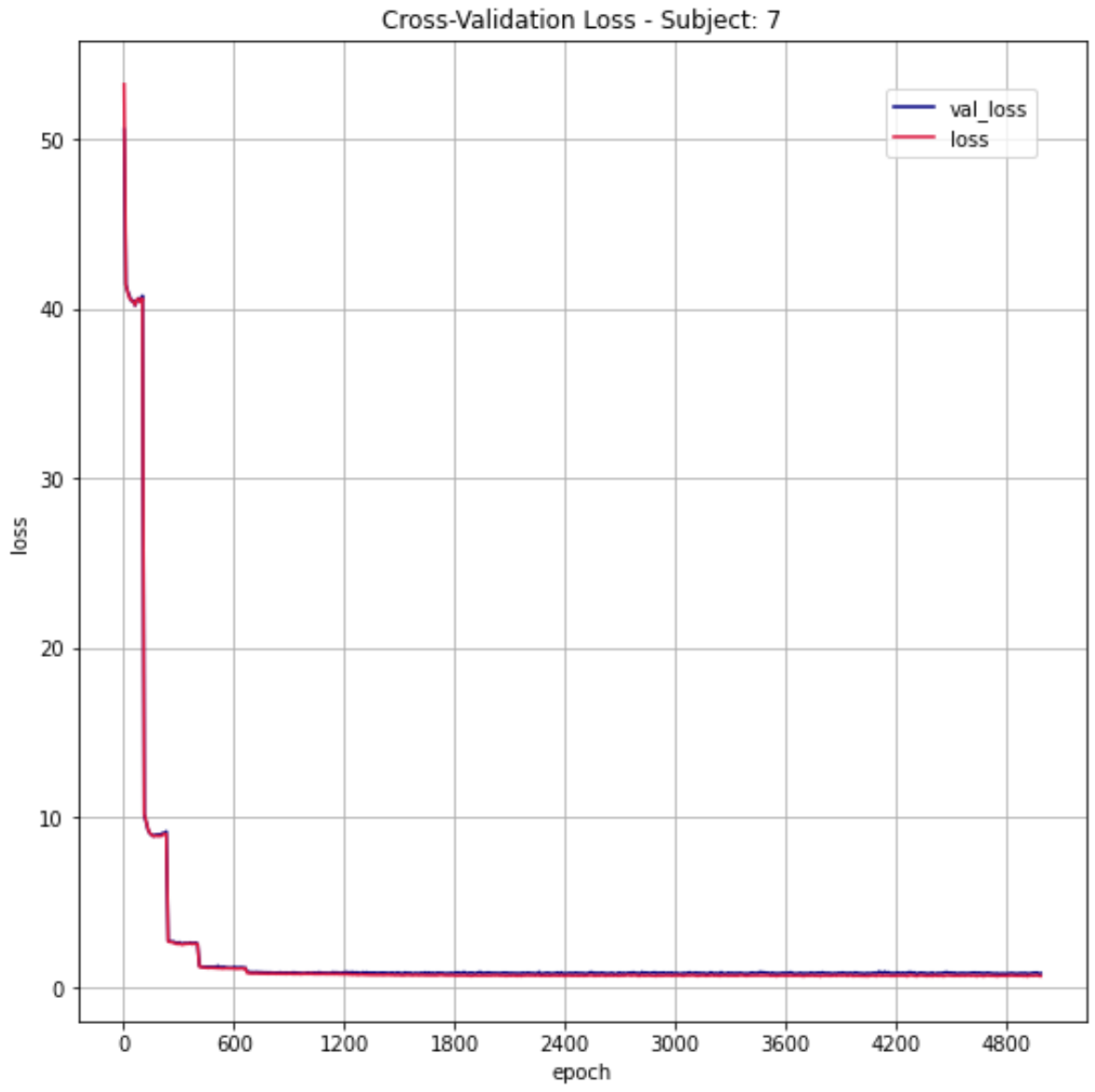

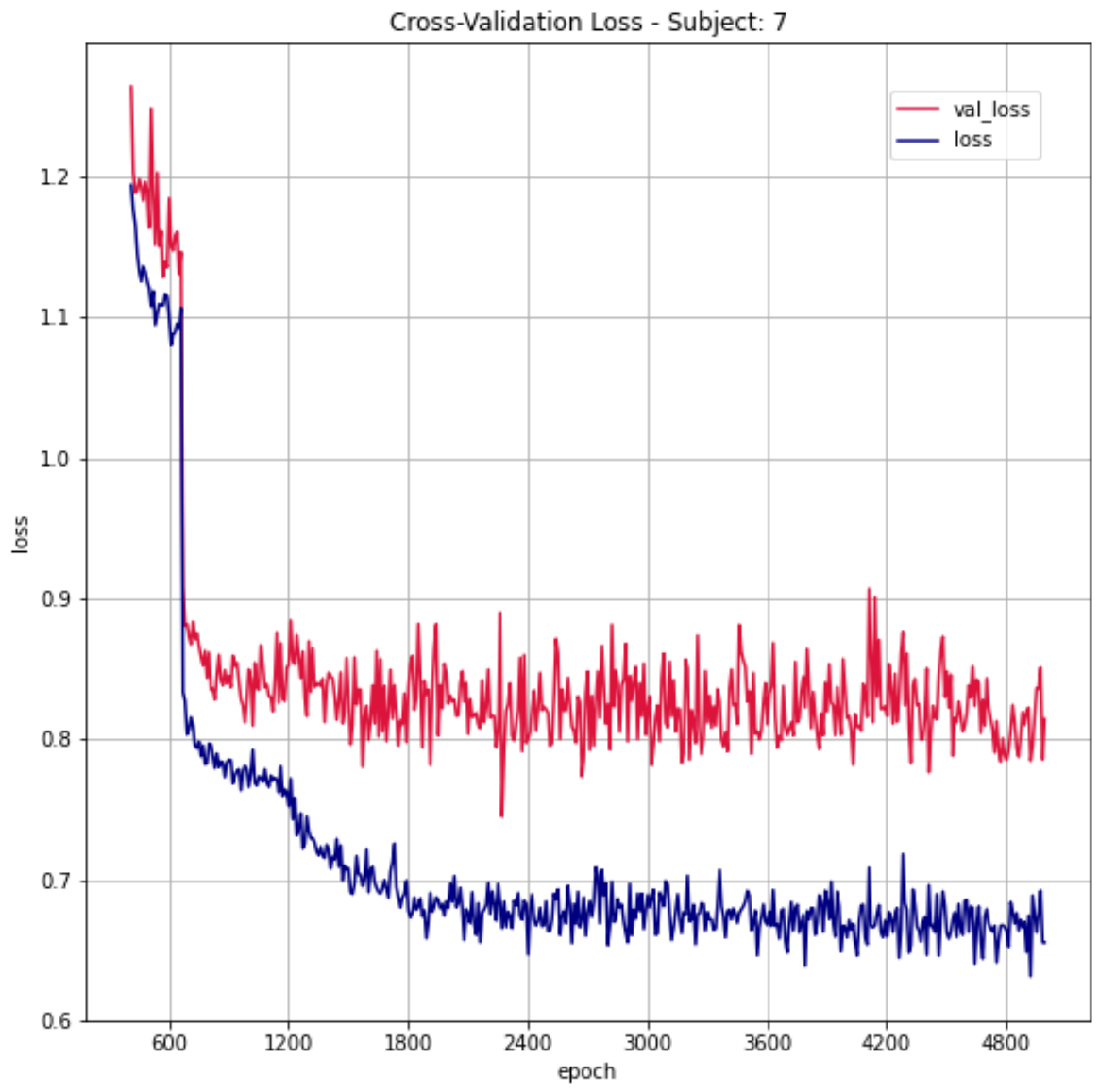

Closer look at the loss plot (epochs > 400) for subject 7

### 1.1.9. Subject 8

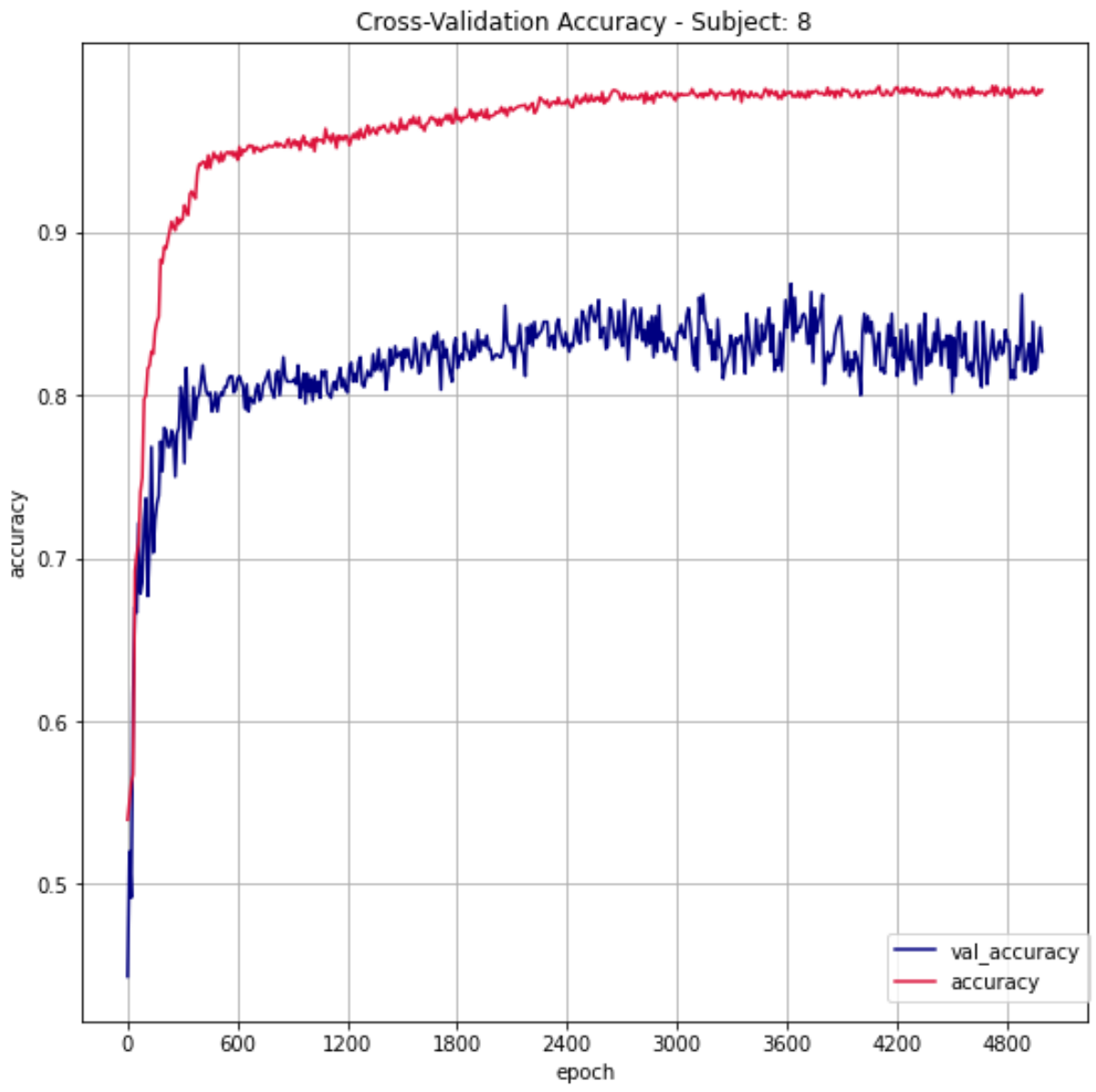

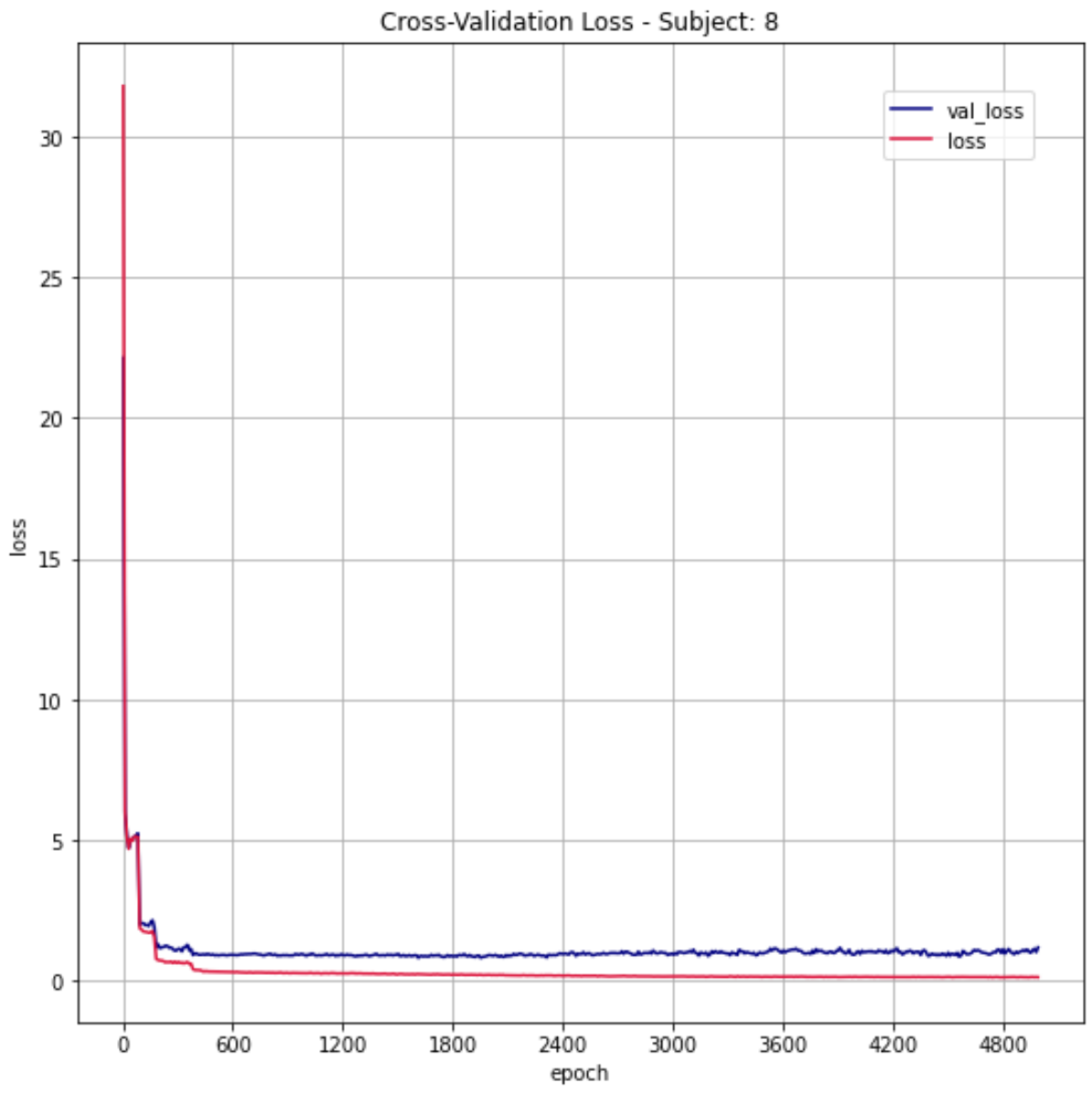

### 1.1.10. Subject 9

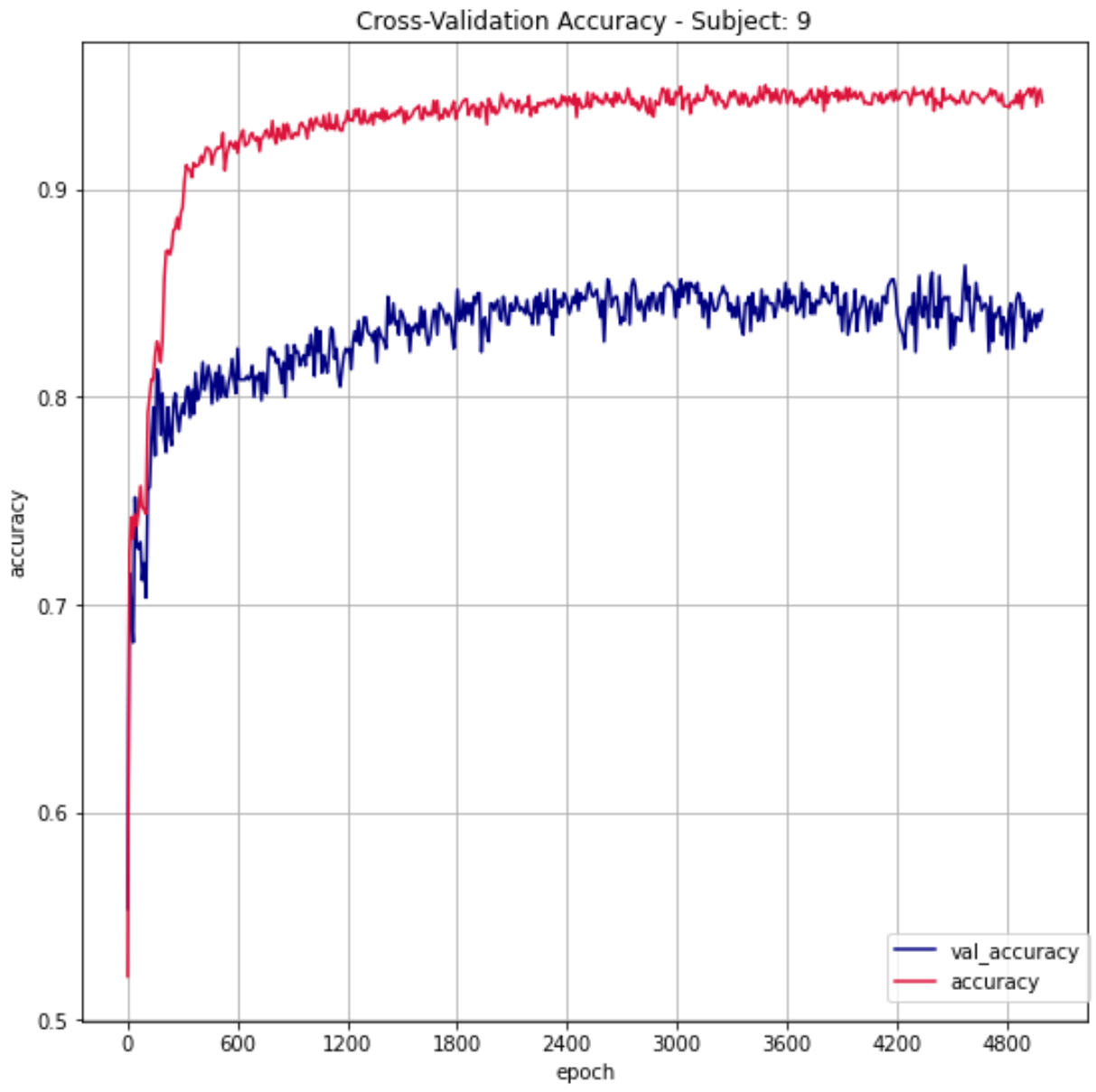

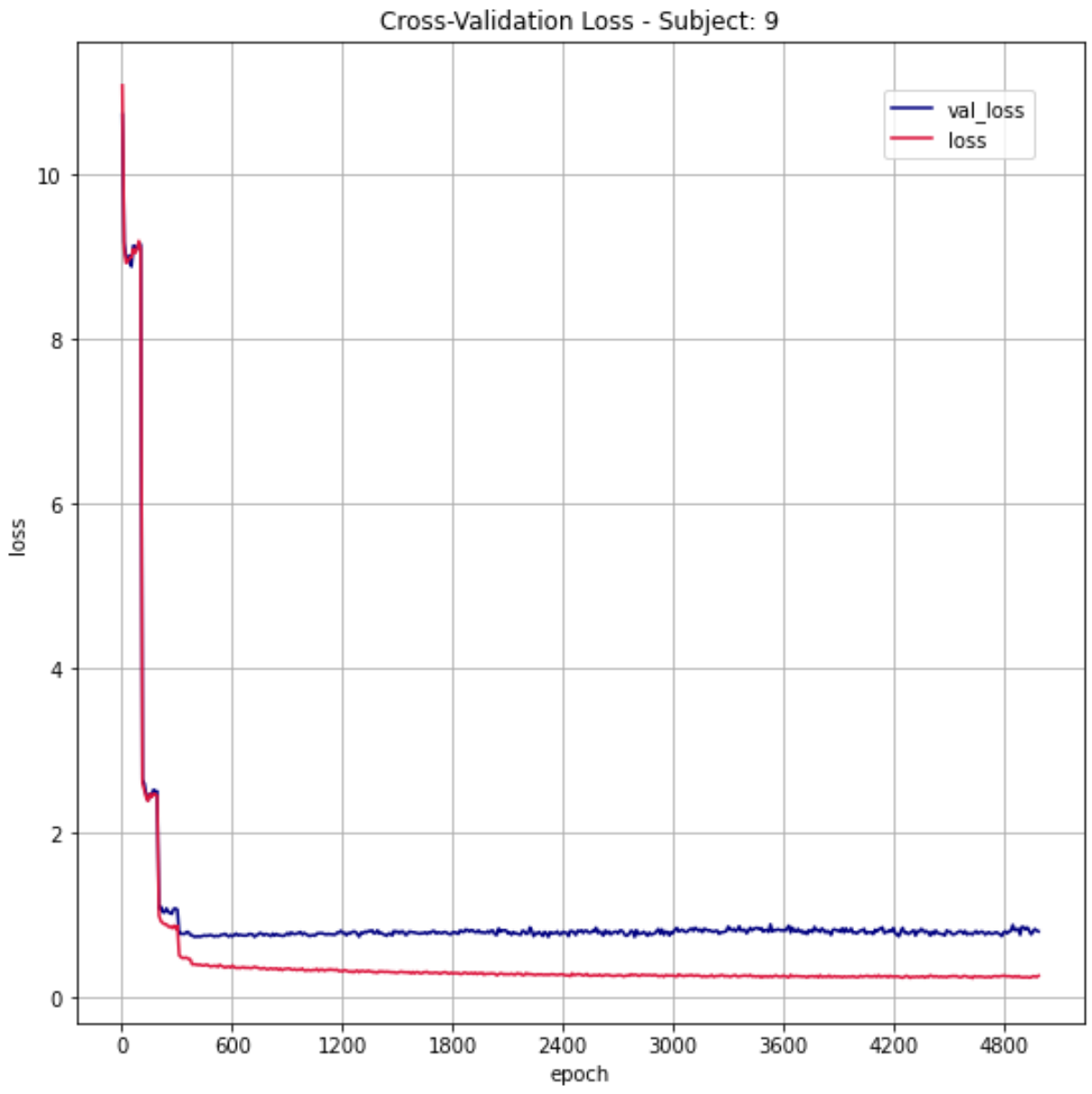

### 1.1.11. Subject 10

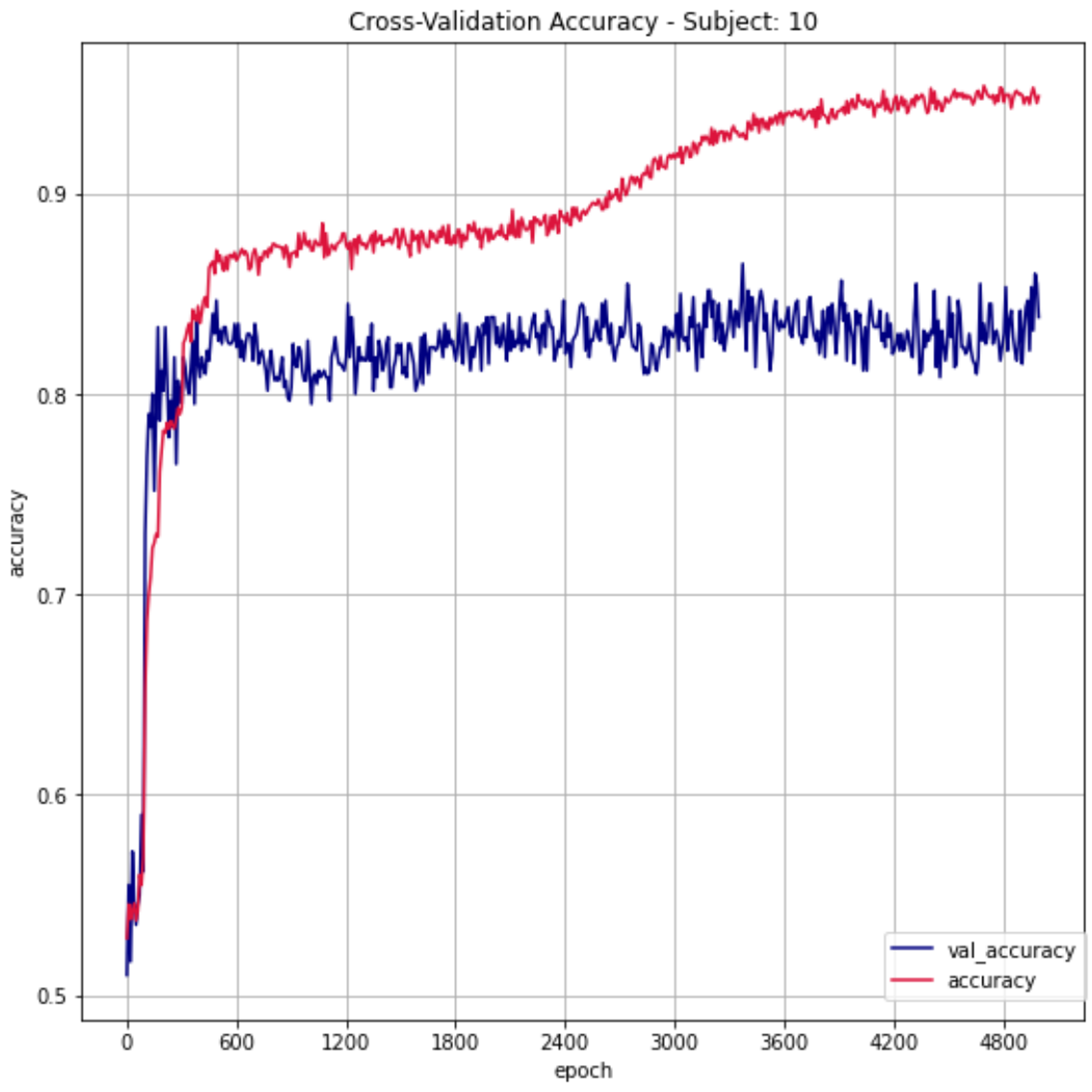

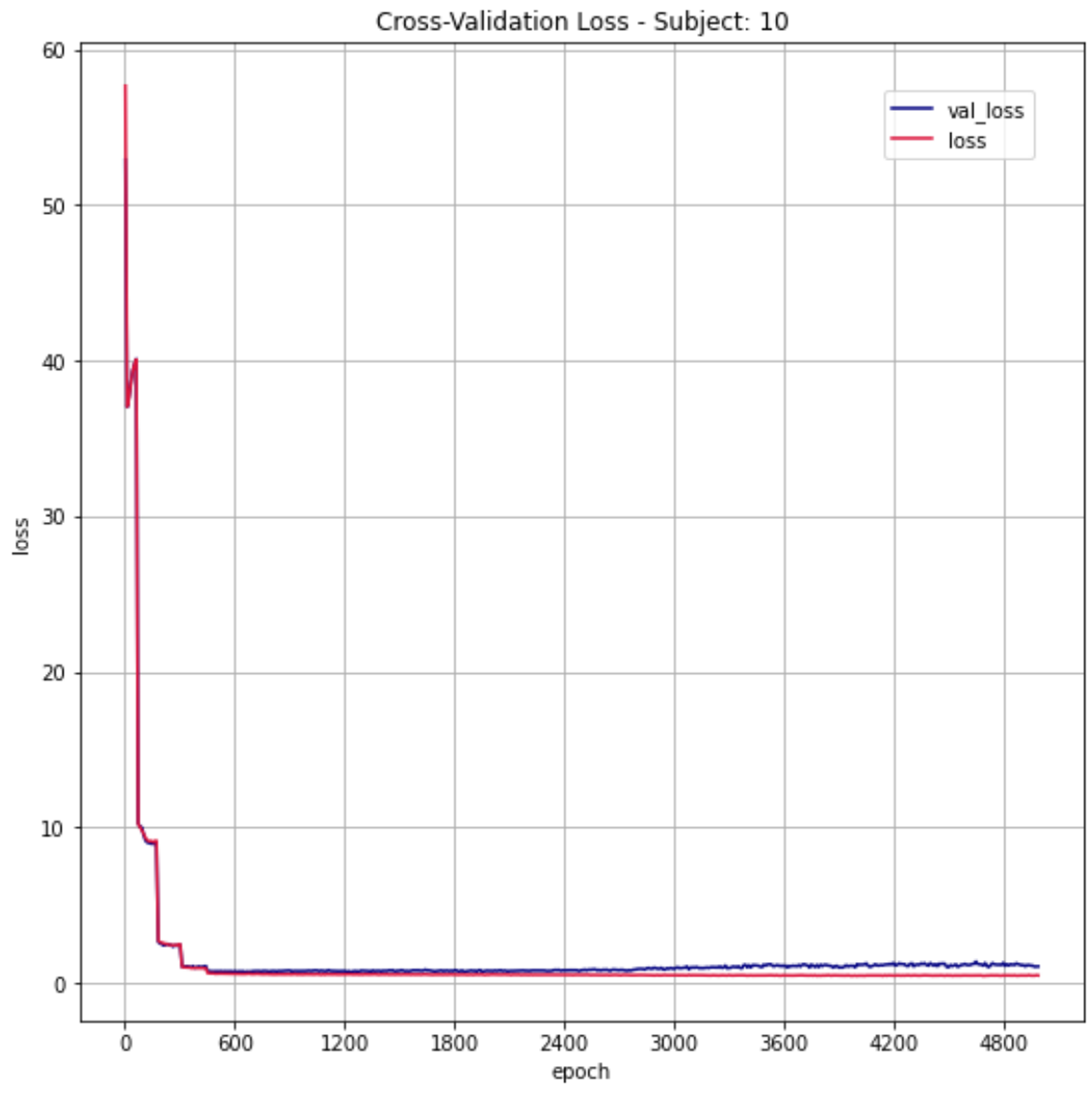

## 1.2. ROC Curves

### 1.2.1. Subject 10

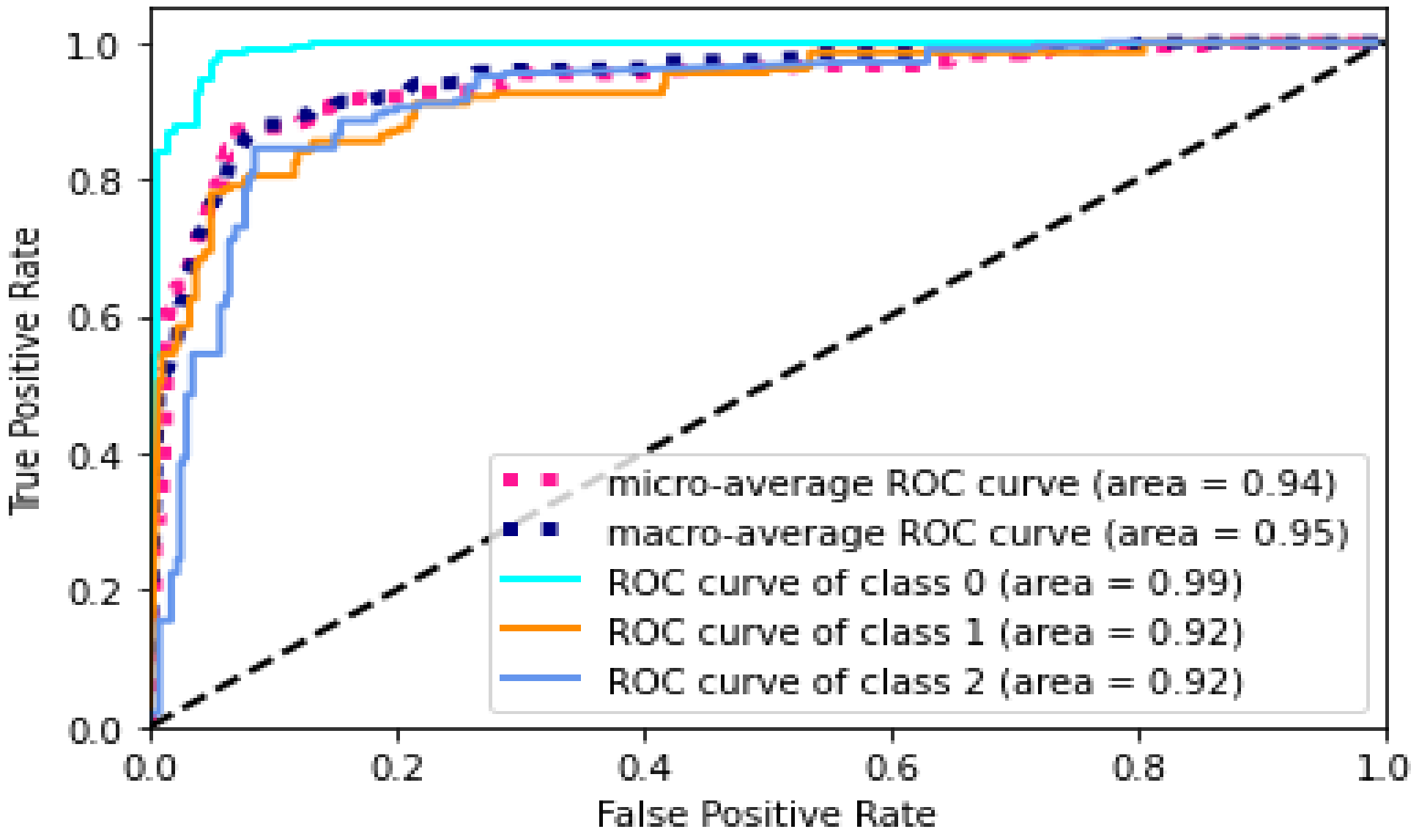

### 1.3. Confusion Matrices, Precision, Recall, F1-scores

#### 1.3.1. Subject 0

##### 1.3.1.1. Confusion Matrix

| | Rest (Predicted) | Neutral (Predicted) | Negative (Predicted) |
|---|---|---|---|
| Rest (True) | 198 | 0 | 2 |
| Neutral (True) | 3 | 161 | 36 |
| Negative (True) | 12 | 26 | 162 |

##### 1.3.1.2. Precision, Recall, F1-scores

| | Precision | Recall | F1-score |
|---|---|---|---|
| Class 0: rest | 0.930 | 0.990 | 0.959 |
| Class 1: neutral | 0.861 | 0.805 | 0.832 |
| Class 2: negative | 0.810 | 0.810 | 0.810 |

### 1.3.2. Subject 1

#### 1.3.2.1. Confusion Matrix

| | Rest (Predicted) | Neutral (Predicted) | Negative (Predicted) |
|---|---|---|---|
| Rest (True) | 184 | 16 | 0 |
| Neutral (True) | 9 | 183 | 8 |
| Negative (True) | 22 | 33 | 145 |

#### 1.3.2.2. Precision, Recall, F1-scores

| | Precision | Recall | F1-score |
|---|---|---|---|
| Class 0: rest | 0.856 | 0.920 | 0.887 |
| Class 1: neutral | 0.789 | 0.915 | 0.847 |
| Class 2: negative | 0.948 | 0.725 | 0.822 |

### 1.3.3. Subject 2

#### 1.3.3.1. Confusion Matrix

| | Rest (Predicted) | Neutral (Predicted) | Negative (Predicted) |
|---|---|---|---|
| Rest (True) | 180 | 12 | 8 |
| Neutral (True) | 8 | 176 | 16 |
| Negative (True) | 3 | 39 | 158 |

#### 1.3.3.2. Precision, Recall, F1-scores

| | Precision | Recall | F1-score |
|---|---|---|---|
| Class 0: rest | 0.942 | 0.900 | 0.921 |
| Class 1: neutral | 0.775 | 0.880 | 0.824 |
| Class 2: negative | 0.868 | 0.790 | 0.827 |

### 1.3.4. Subject 3

#### 1.3.4.1. Confusion Matrix

| | Rest (Predicted) | Neutral (Predicted) | Negative (Predicted) |
|---|---|---|---|
| Rest (True) | 156 | 7 | 37 |
| Neutral (True) | 10 | 169 | 21 |
| Negative (True) | 14 | 19 | 167 |

#### 1.3.4.2. Precision, Recall, F1-scores

| | Precision | Recall | F1-score |
|---|---|---|---|
| Class 0: rest | 0.867 | 0.780 | 0.821 |
| Class 1: neutral | 0.867 | 0.845 | 0.856 |
| Class 2: negative | 0.742 | 0.835 | 0.786 |

### 1.3.5. Subject 4

#### 1.3.5.1. Confusion Matrix

| | Rest (Predicted) | Neutral (Predicted) | Negative (Predicted) |
|---|---|---|---|
| Rest (True) | 163 | 32 | 5 |
| Neutral (True) | 19 | 162 | 19 |
| Negative (True) | 16 | 61 | 123 |

#### 1.3.5.2. Precision, Recall, F1-scores

| | Precision | Recall | F1-score |
|---|---|---|---|
| Class 0: rest | 0.823 | 0.815 | 0.819 |
| Class 1: neutral | 0.635 | 0.810 | 0.712 |
| Class 2: negative | 0.837 | 0.615 | 0.709 |

### 1.3.6. Subject 5

#### 1.3.6.1. Confusion Matrix

| | Rest (Predicted) | Neutral (Predicted) | Negative (Predicted) |
|---|---|---|---|
| Rest (True) | 192 | 8 | 0 |
| Neutral (True) | 13 | 165 | 22 |
| Negative (True) | 4 | 29 | 167 |

#### 1.3.6.2. Precision, Recall, F1-scores

| | Precision | Recall | F1-score |
|---|---|---|---|
| Class 0: rest | 0.919 | 0.960 | 0.939 |
| Class 1: neutral | 0.817 | 0.825 | 0.821 |
| Class 2: negative | 0.884 | 0.835 | 0.859 |

### 1.3.7. Subject 6

#### 1.3.7.1. Confusion Matrix

| | Rest (Predicted) | Neutral (Predicted) | Negative (Predicted) |
|---|---|---|---|
| Rest (True) | 190 | 6 | 4 |
| Neutral (True) | 30 | 153 | 17 |
| Negative (True) | 9 | 24 | 167 |

#### 1.3.7.2. Precision, Recall, F1-scores

| | Precision | Recall | F1-score |
|---|---|---|---|
| Class 0: rest | 0.830 | 0.950 | 0.886 |
| Class 1: neutral | 0.836 | 0.765 | 0.799 |
| Class 2: negative | 0.888 | 0.835 | 0.861 |

### 1.3.8. Subject 7

#### 1.3.8.1. Confusion Matrix

| | Rest (Predicted) | Neutral (Predicted) | Negative (Predicted) |
|---|---|---|---|
| Rest (True) | 182 | 13 | 6 |
| Neutral (True) | 10 | 181 | 9 |
| Negative (True) | 13 | 28 | 159 |

#### 1.3.8.2. Precision, Recall, F1-scores

| | Precision | Recall | F1-score |
|---|---|---|---|
| Class 0: rest | 0.887 | 0.905 | 0.896 |
| Class 1: neutral | 0.815 | 0.905 | 0.858 |
| Class 2: negative | 0.914 | 0.795 | 0.850 |

### 1.3.9. Subject 8

#### 1.3.9.1. Confusion Matrix

| | Rest (Predicted) | Neutral (Predicted) | Negative (Predicted) |
|---|---|---|---|
| Rest (True) | 183 | 9 | 8 |
| Neutral (True) | 20 | 152 | 28 |
| Negative (True) | 1 | 12 | 187 |

#### 1.3.9.2. Precision, Recall, F1-scores

| | Precision | Recall | F1-score |
|---|---|---|---|
| Class 0: rest | 0.897 | 0.915 | 0.906 |
| Class 1: neutral | 0.879 | 0.760 | 0.815 |
| Class 2: negative | 0.839 | 0.935 | 0.884 |

### 1.3.10. Subject 9

#### 1.3.10.1. Confusion Matrix

| | Rest (Predicted) | Neutral (Predicted) | Negative (Predicted) |
|---|---|---|---|
| Rest (True) | 186 | 4 | 10 |
| Neutral (True) | 4 | 155 | 41 |
| Negative (True) | 2 | 21 | 177 |

#### 1.3.10.2. Precision, Recall, F1-scores

| | Precision | Recall | F1-score |
|---|---|---|---|
| Class 0: rest | 0.969 | 0.930 | 0.949 |
| Class 1: neutral | 0.861 | 0.775 | 0.816 |
| Class 2: negative | 0.776 | 0.885 | 0.827 |

### 1.3.11. Subject 10

#### 1.3.11.1. Confusion Matrix

| | Rest (Predicted) | Neutral (Predicted) | Negative (Predicted) |
|---|---|---|---|
| Rest (True) | 197 | 0 | 3 |
| Neutral (True) | 13 | 156 | 31 |
| Negative (True) | 11 | 20 | 169 |

#### 1.3.11.2. Precision, Recall, F1-scores

| | Precision | Recall | F1-score |
|---|---|---|---|
| Class 0: rest | 0.891 | 0.985 | 0.936 |
| Class 1: neutral | 0.886 | 0.780 | 0.830 |
| Class 2: negative | 0.833 | 0.845 | 0.839 |

## 1.4. Result Lists

| Subject (Fold #) | 1D CNN Accuracy | 1D CNN Loss | 1D CNN AUC |
|---|---|---|---|
| 0 | 0.868 | 1.138 | 0.959 |
| 1 | 0.853 | 1.741 | 0.944 |
| 2 | 0.857 | 0.689 | 0.950 |
| 3 | 0.820 | 2.104 | 0.916 |
| 4 | 0.747 | 0.932 | 0.857 |
| 5 | 0.873 | 0.806 | 0.937 |
| 6 | 0.850 | 0.752 | 0.953 |
| 7 | 0.868 | 0.776 | 0.932 |
| 8 | 0.870 | 1.922 | 0.953 |
| 9 | 0.863 | 1.169 | 0.941 |
| 10 | 0.870 | 0.975 | 0.945 |

### 1.5. Hyperalignment Algorithm

**Preprocessing**

Original data $X \in \mathbb{R}^{m,n_1,n_2}$ where $m$ corresponds to timepoints, and the image of the scan at each timepoint in $\mathbb{R}^{n_1,n_2}$. Suppose we flatten each voxel in $\mathbb{R}^m$ and place them side-by-side such that the data becomes $\mathbb{R}^{m,n}$ where $n = n_1 + n_2$. Let's call the reorganized data as $Y \in \mathbb{R}^{m,n}$ after being normalized using Frobenius norm.

**Alignment**

The overview of alignment is as following:

- $S \in \mathbb{R}^{m,n}$ refers to collection of voxels for one subject/source. $m$ corresponds to timepoints and $n$ corresponds to number of distinct voxels.
- $T \in \mathbb{R}^{m,n}$ refers to collection of voxels for the target common space
- $\Omega \in \mathbb{R}^{m,m}$ is an orthogonal matrix that applies a combination of rotations and reflections to each voxel consistently across timepoints
- $R \in \mathbb{R}^{m,m}$ solves the orthogonal procrustean problem of the following form:

$$R = arg\min_{\Omega} ||S'\Omega - T'||$$

$$\text{subject to } \Omega'\Omega = I$$

It is straightforward to show that the solution $R$ can be obtained from the singular value decomposition (svd) of $T'S$. Suppose $T'S = U\Sigma V'$. Then, the optimal value $R$ is:

$$R = VU'$$

One remark is that both $S$ and $T$ are normalized with Frobenius norm.

The question then becomes how do we obtain the T? It is computed using level-1, level-2 and level-3 alignments.

### 1.5.1. Level-1 Alignment

Suppose there are $p$ subjects in the study such that collection of sources is of the form $S_1, S_2,\ldots,$ $S_p$ where $S_i \in \mathbb{R}^{m,n}$ for every $i$. Under this assumption, the user may specify a reference dataset $S_r$ from the collection of sources. Then the pseudo-code is as follows:

| Algorithm 1: Level-1 alignment |
| --- |
| Choose reference dataset $S_r$ from $\{S_1, S_2, \ldots, S_p\}$<br><br>for $i$ from 1 to $p$<br><br>    $U\Sigma V' = \mathrm{S}_{\mathrm{r}}' S_i$ by applying singular value decomposition<br><br>    $R_i = VU'$<br><br>    $S_i^a \leftarrow R_i' S_i$<br><br><br>    $S_r \leftarrow wS_i^a + (1 - w)S_r$ for some $w \in \mathbb{R}$<br><br>return $(R_i)_{i=1}^p, (S_i^a)_{i=1}^p, S_r$ |

### 1.5.2. Algorithm: Level-2 alignment (optional)

| Algorithm 2: Level-2 alignment (optional) |
| --- |
| Load reference dataset $S_r$ Level-1 alignment |

Specify the number of iterations, $q$

$(S_i^a)_{i=1}^p = (S_i)_{i=1}^p$ deep copy input data

for $i$ from 1 to $q$

for $j$ from 1 to $p$

$S_r^{\text{tmp}} \leftarrow \left(pS_r - S_j^a\right)/(p-1)$

$U\Sigma V' = S_r^{\text{tmp}\prime} S_j^a$ by applying singular value decomposition

$R_j = VU'$

$S_j^a \leftarrow R_j' S_j^a$

$S_r \leftarrow \frac{1}{p}\sum_{j=1}^p S_j^a$

return $\left(R_j\right)_{j=1}^p, \left(S_j^a\right)_{j=1}^p, S_r$

### 1.5.3. Level-3 alignment

We can then apply dimension-reduction techniques to $(S_i^a)_{i=1}^p$ from Level-3 alignments

| Algorithm 3: Level-3 alignment |
|---|
| Load reference dataset $S_r$ from either Level-1 or Level-2 alignments<br>for $i$ from 1 to $p$<br>$U\Sigma V' = \mathrm{S}_\mathrm{r}' S_i$ by applying singular value decomposition<br>$R_i = VU'$<br>$S_i^a \leftarrow R_i' S_i$<br>return $(R_i)_{i=1}^p, (S_i^a)_{i=1}^p, S_r$ |

## 2. Experiment B

## 2.1. Loss and Accuracy Plots

### 2.1.1. L2 Regularization: 0.0025

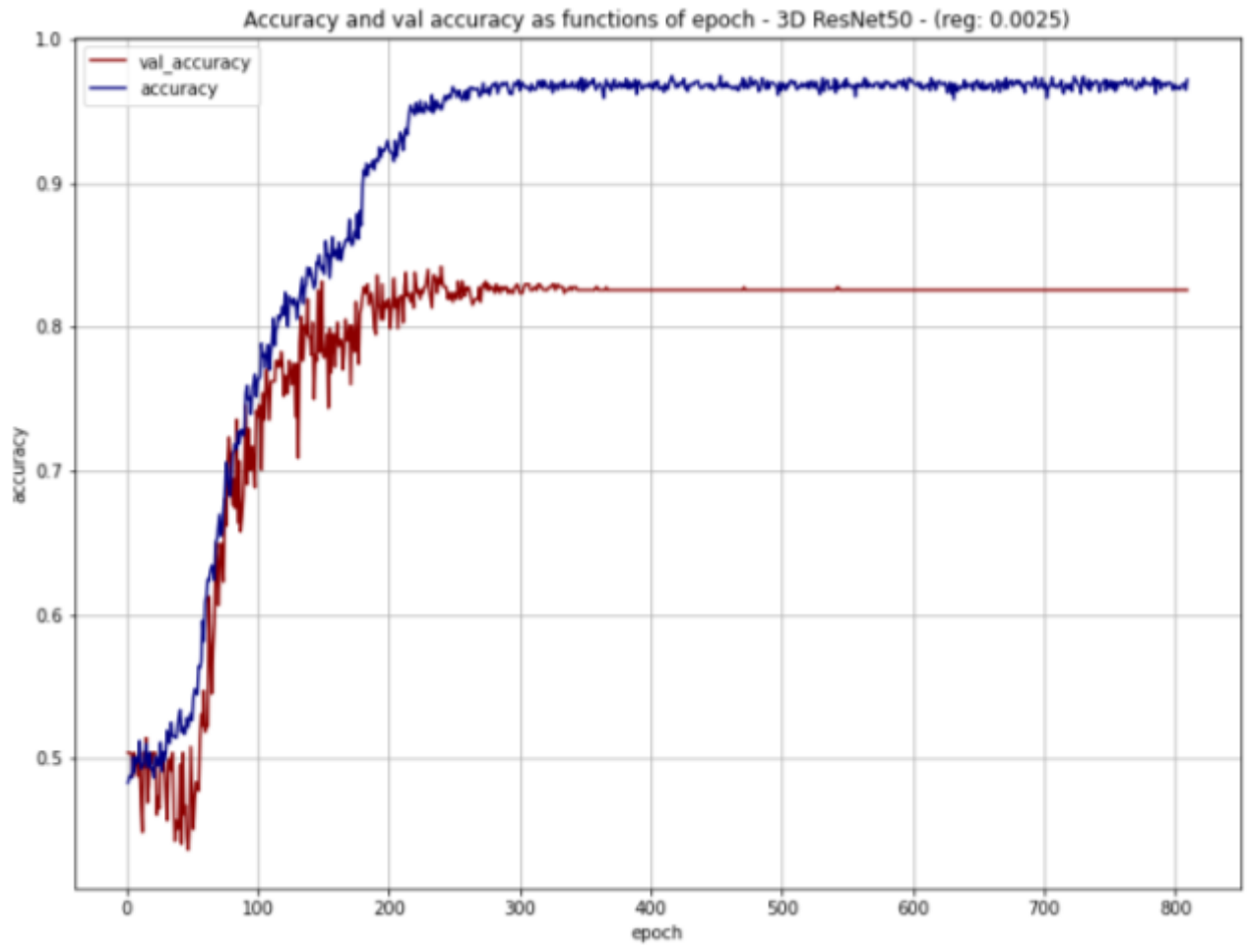

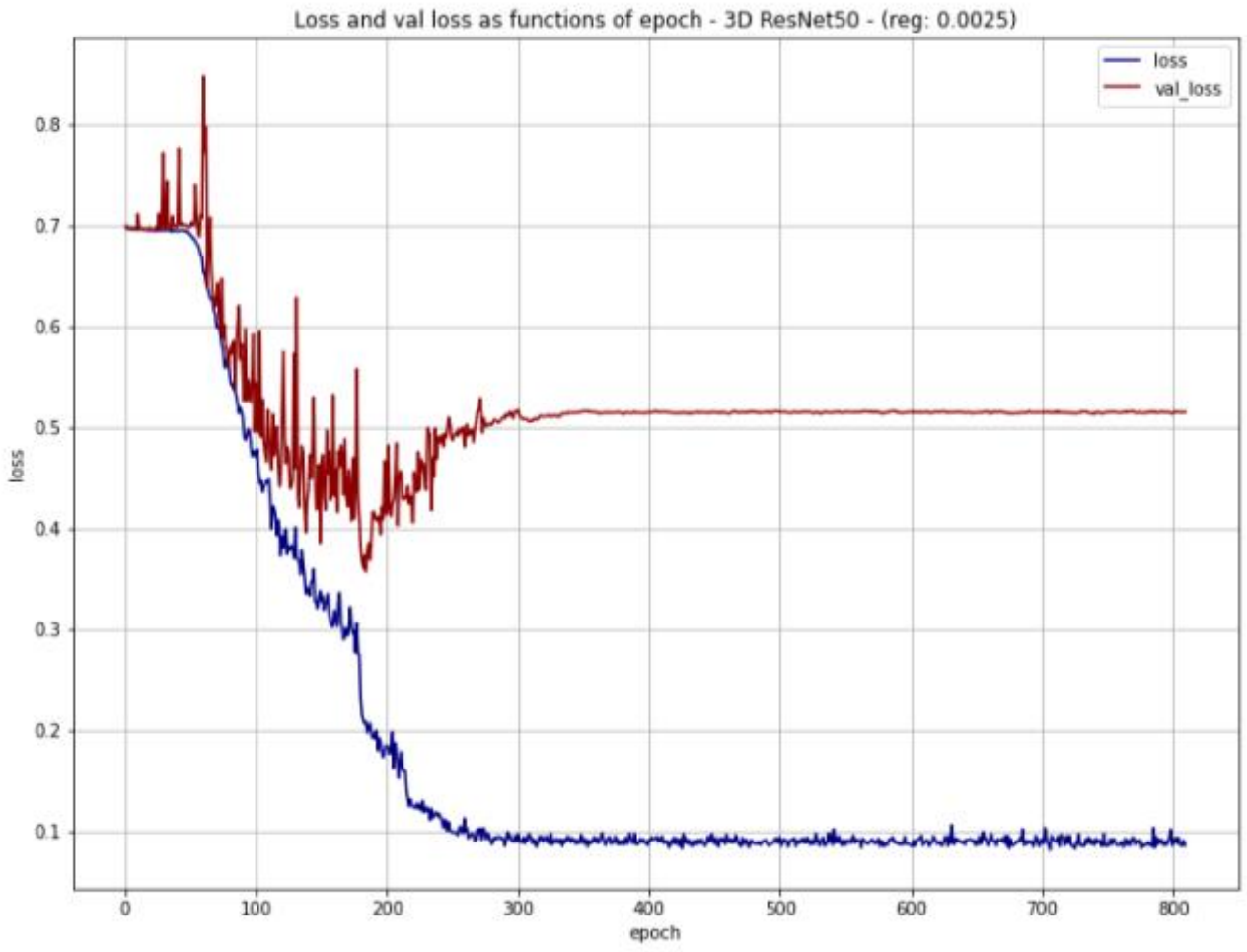

### 2.1.2. L2 Regularization: 0.00125

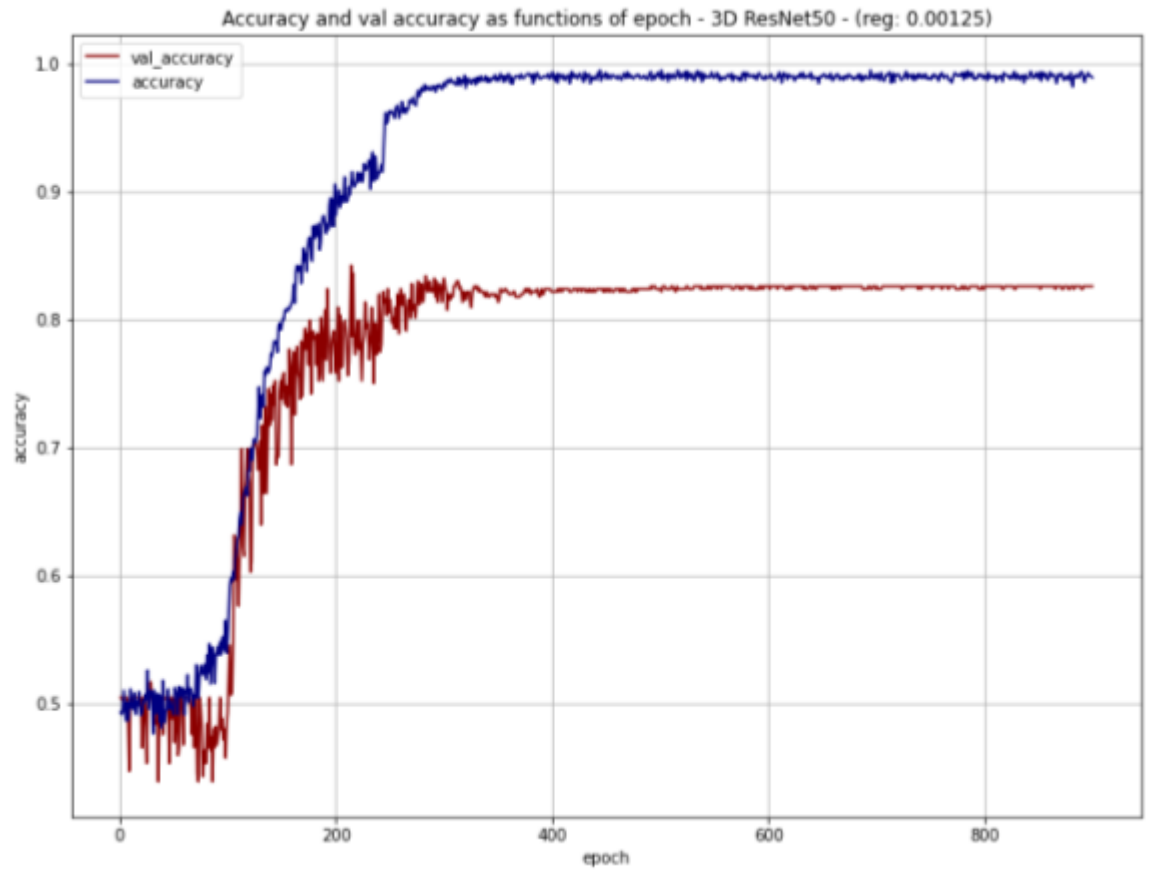

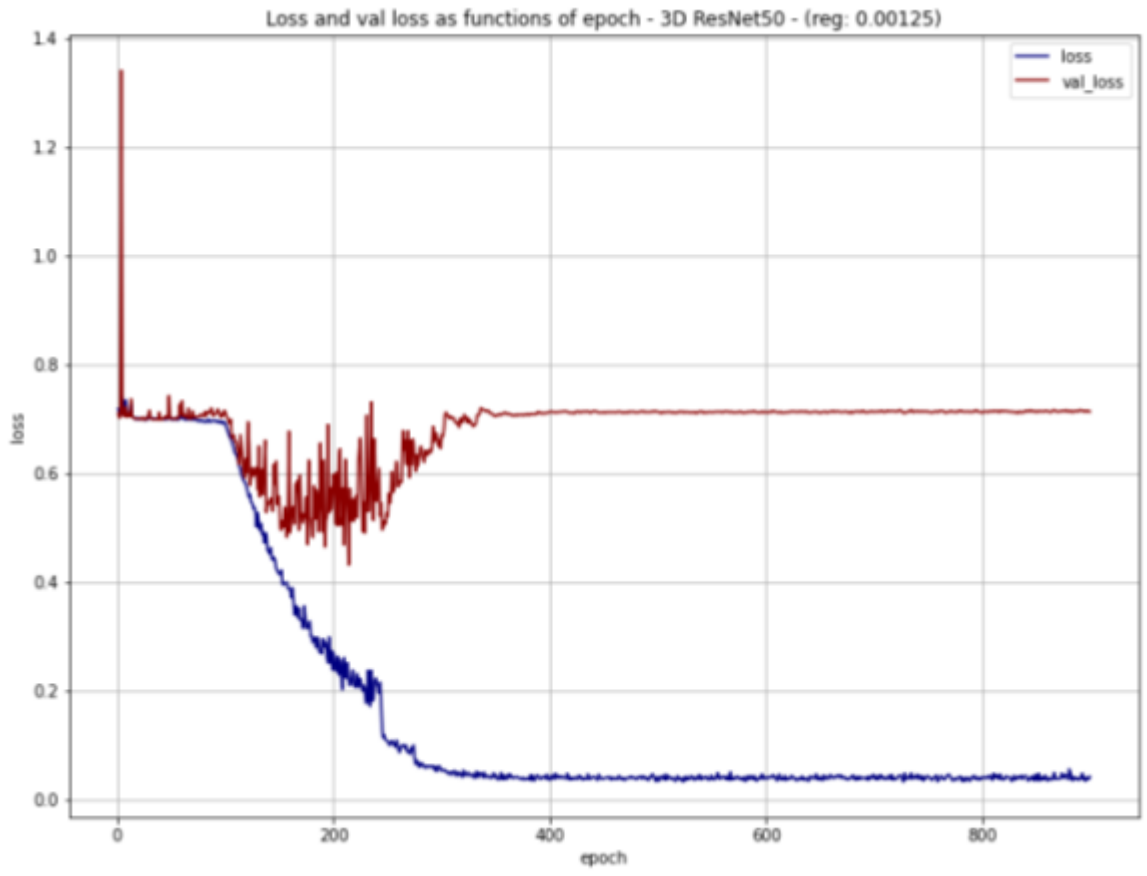

### 2.1.3. L2 Regularization: 0.001

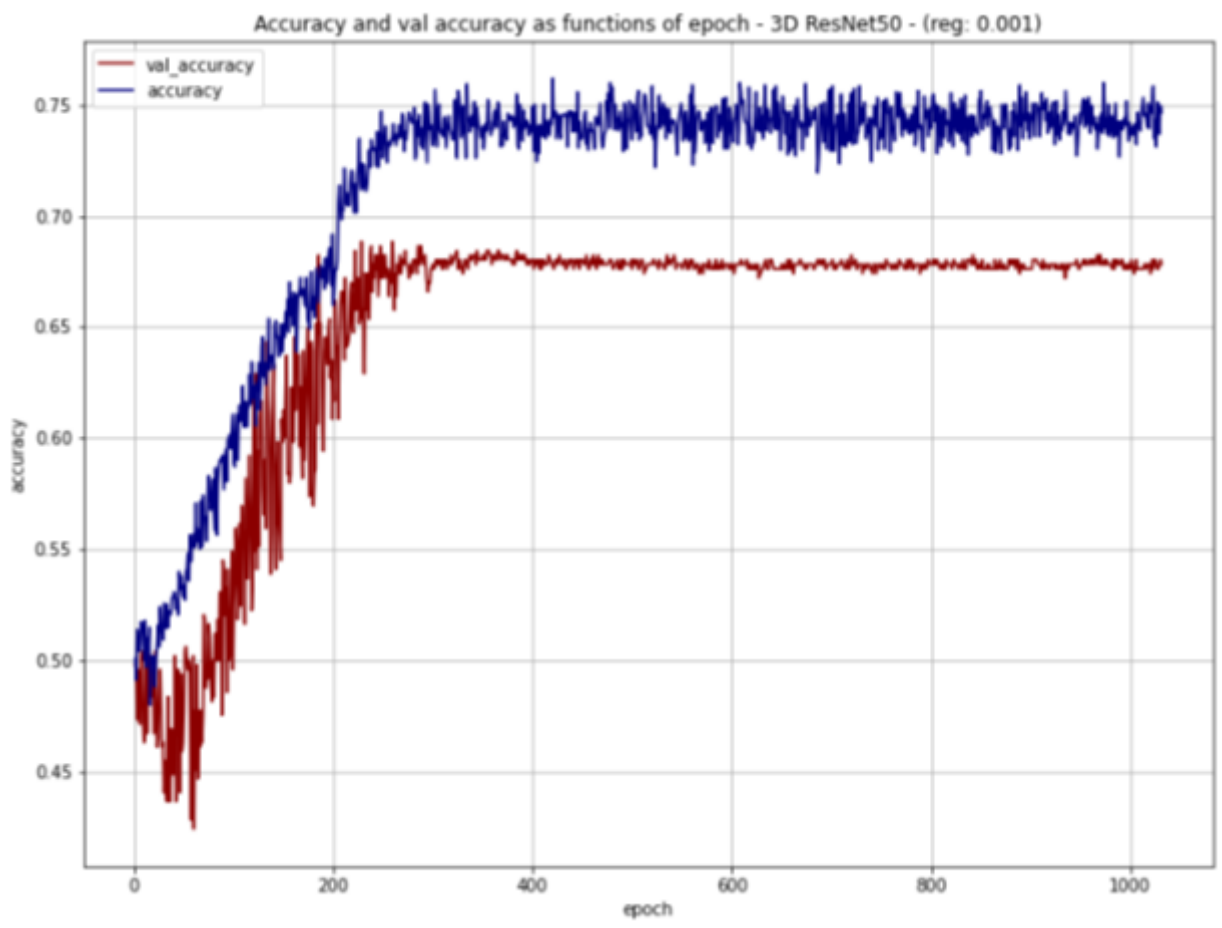

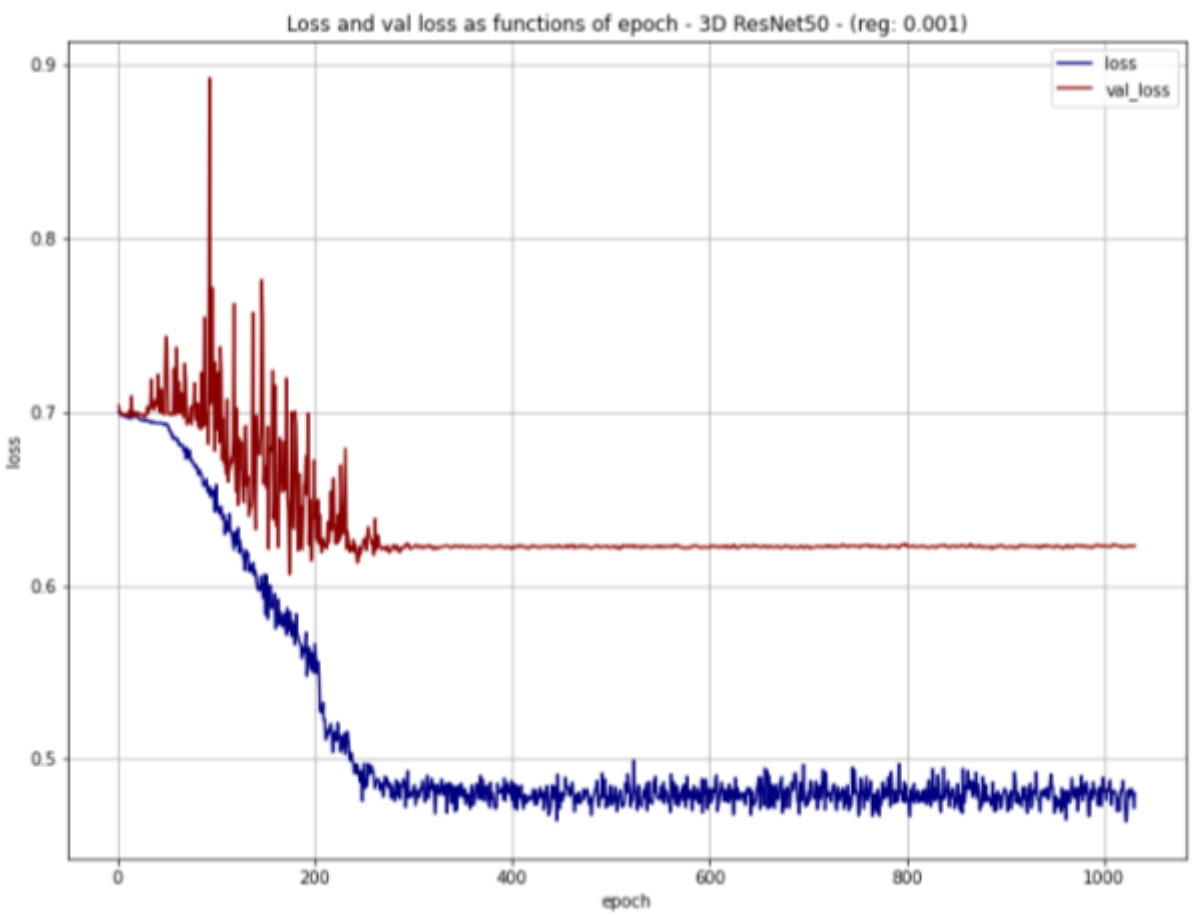

### 2.1.4. L2 Regularization: 0.0009

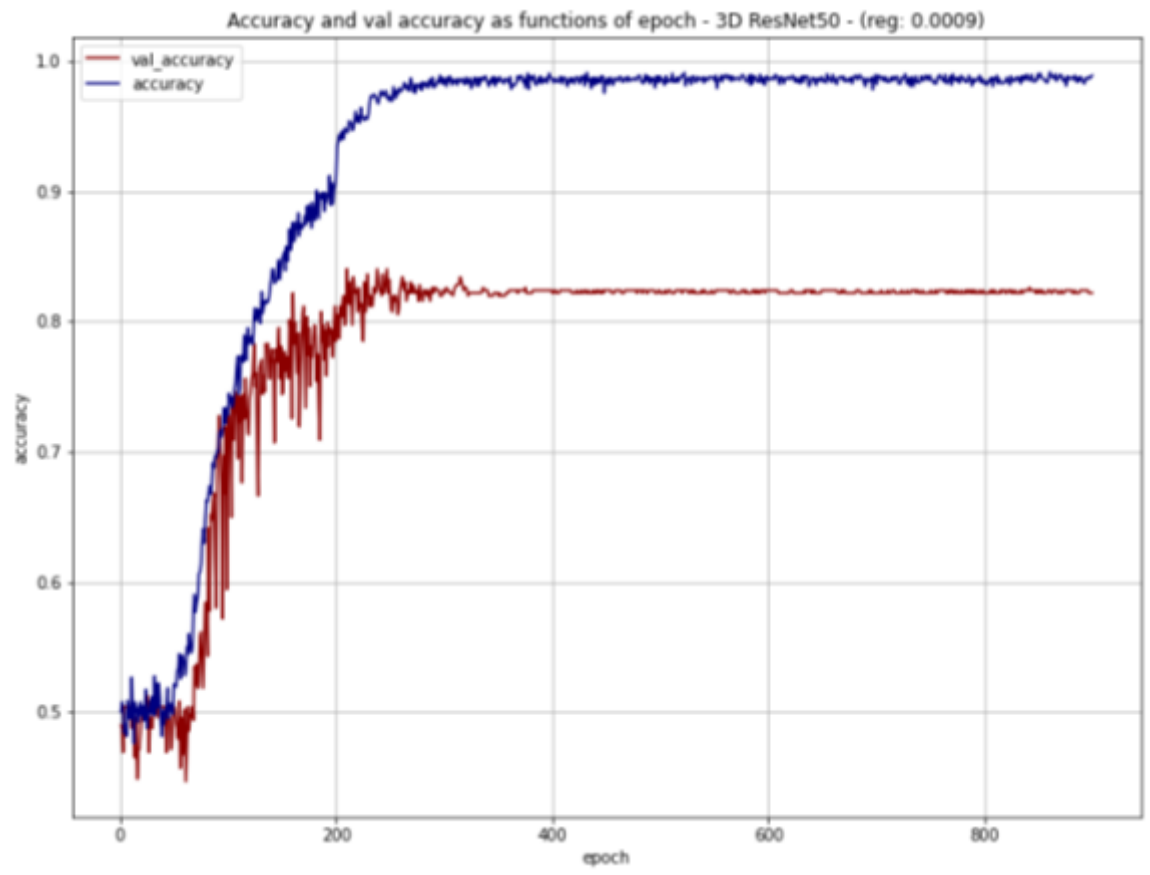

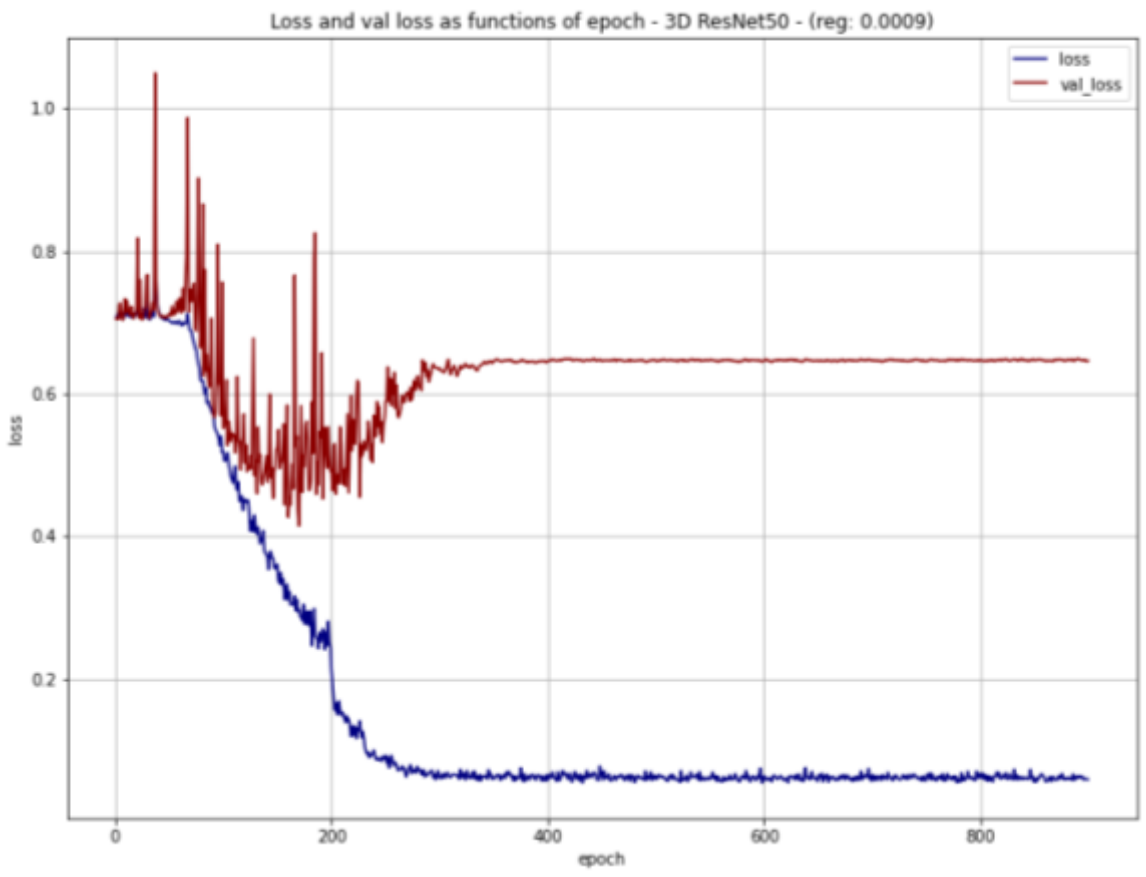

### 2.1.5. L2 Regularization: 0.00075

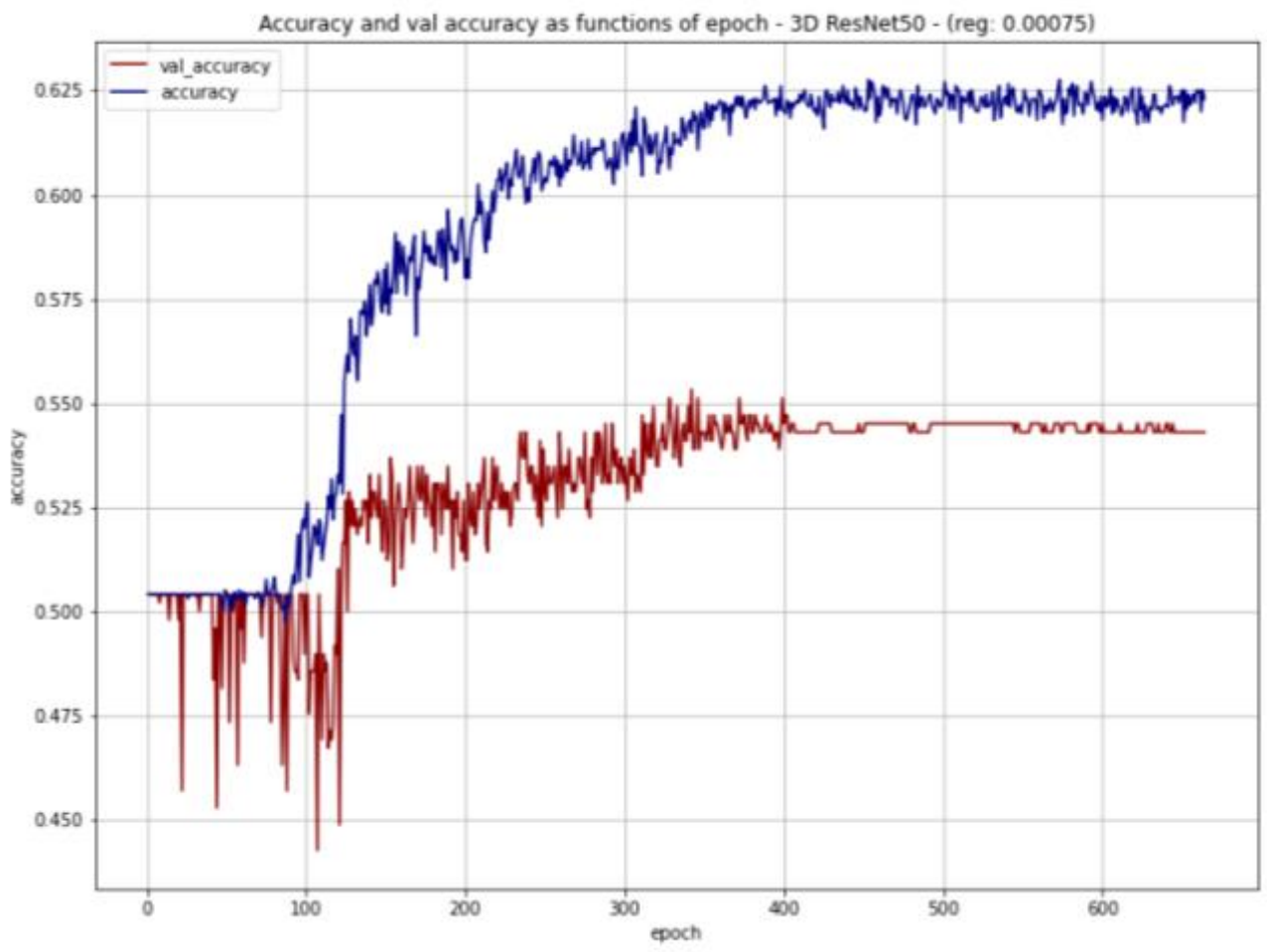

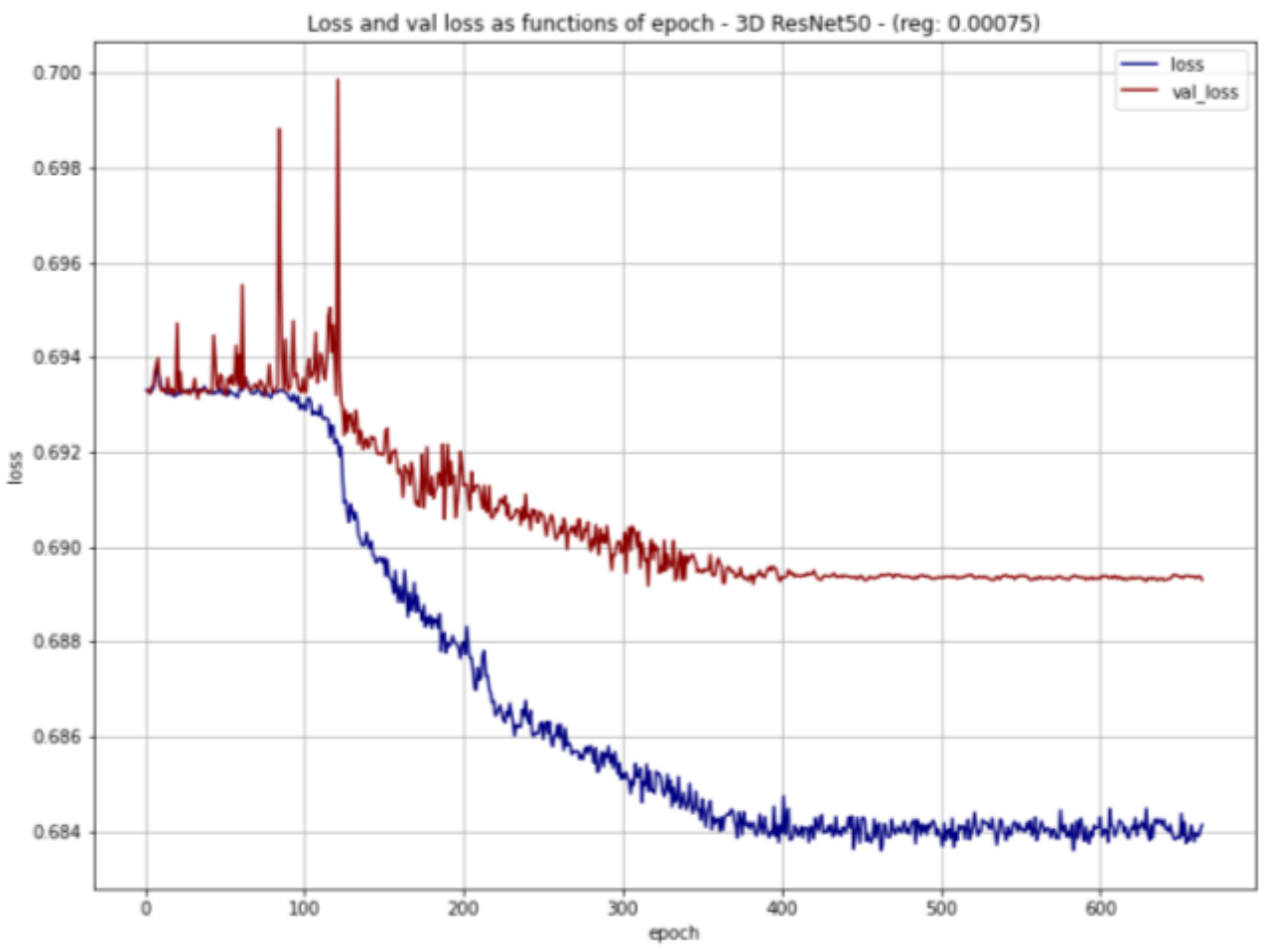

### 2.1.6. L2 Regularization: 0.0006

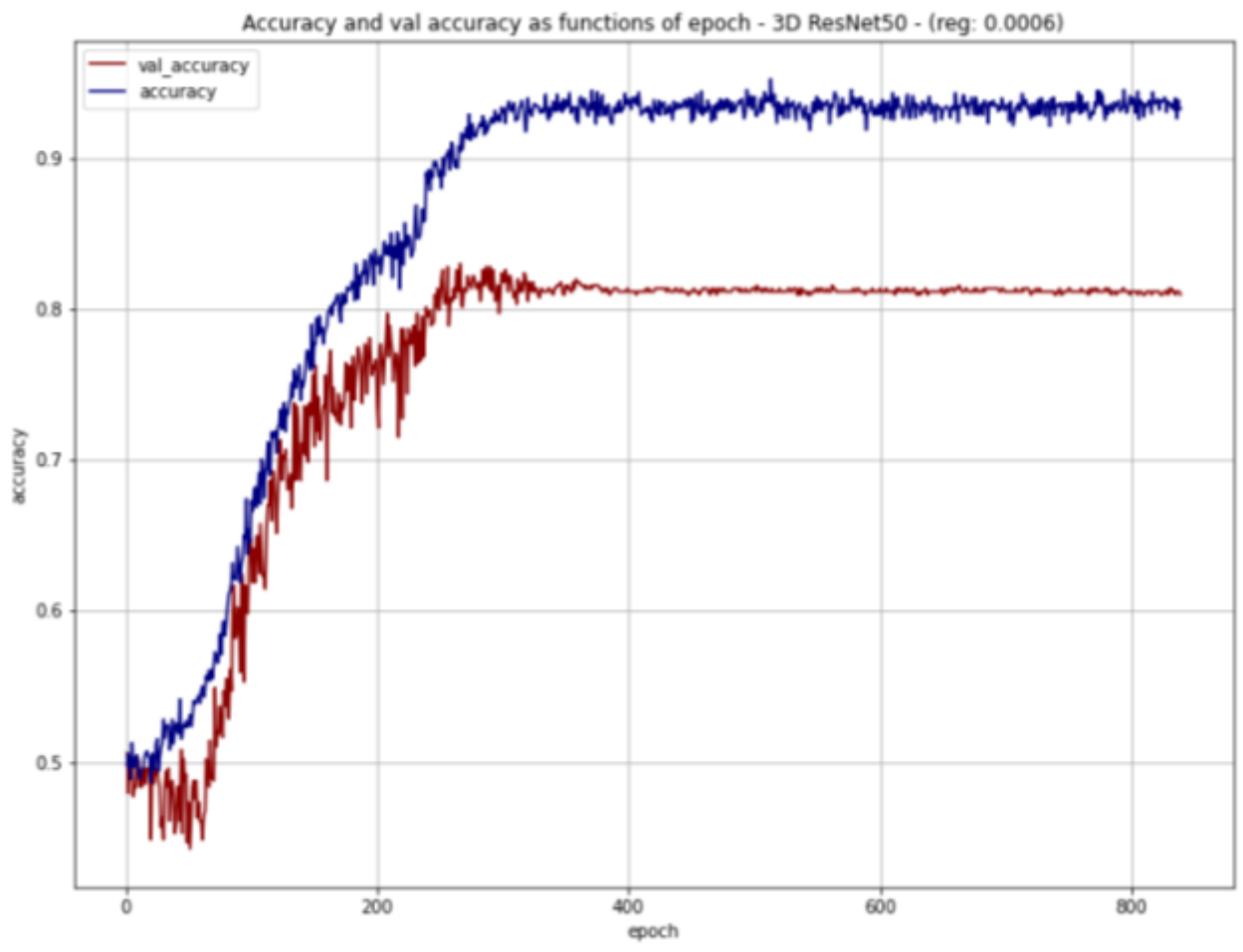

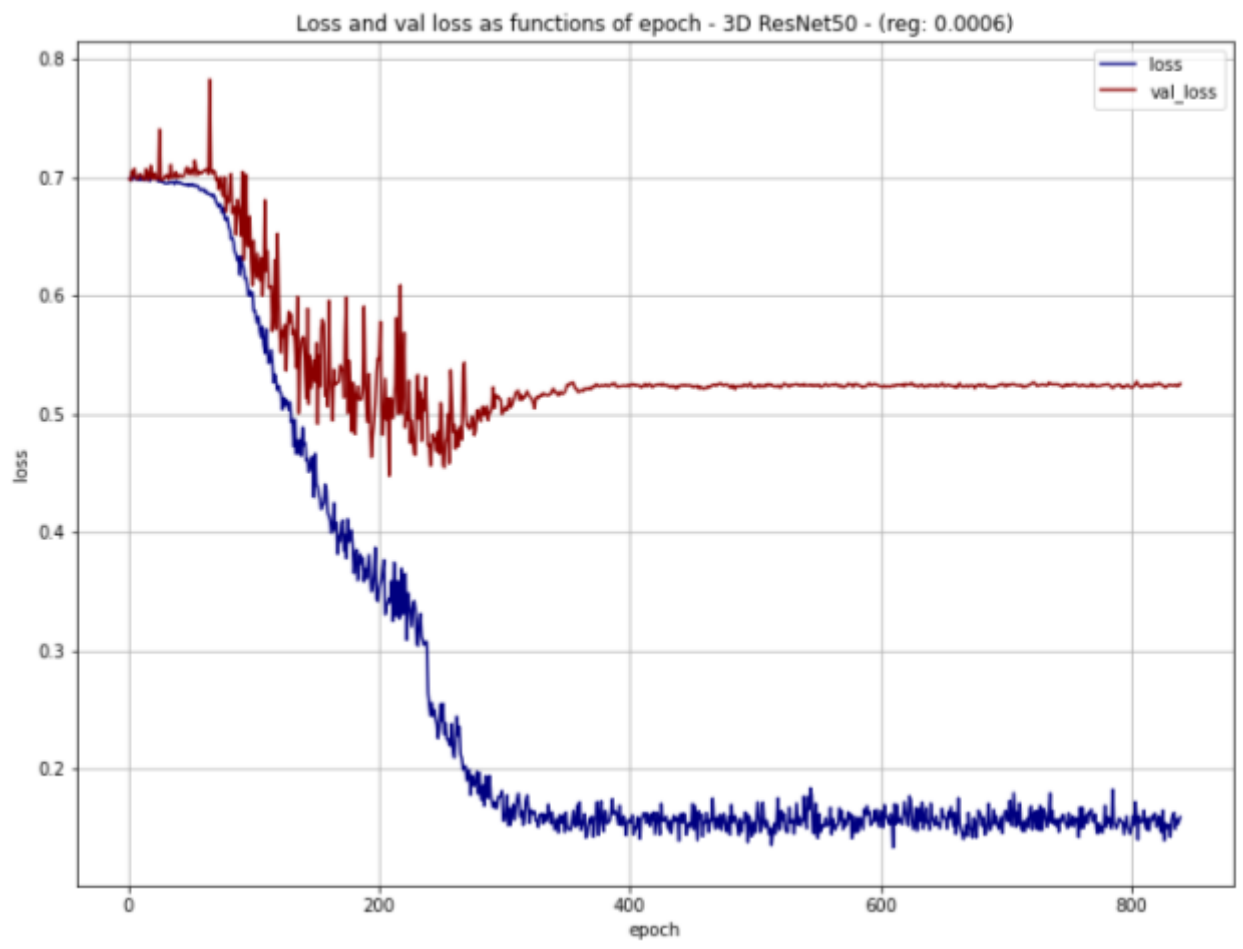

### 2.1.7. L2 Regularization: 0.0005

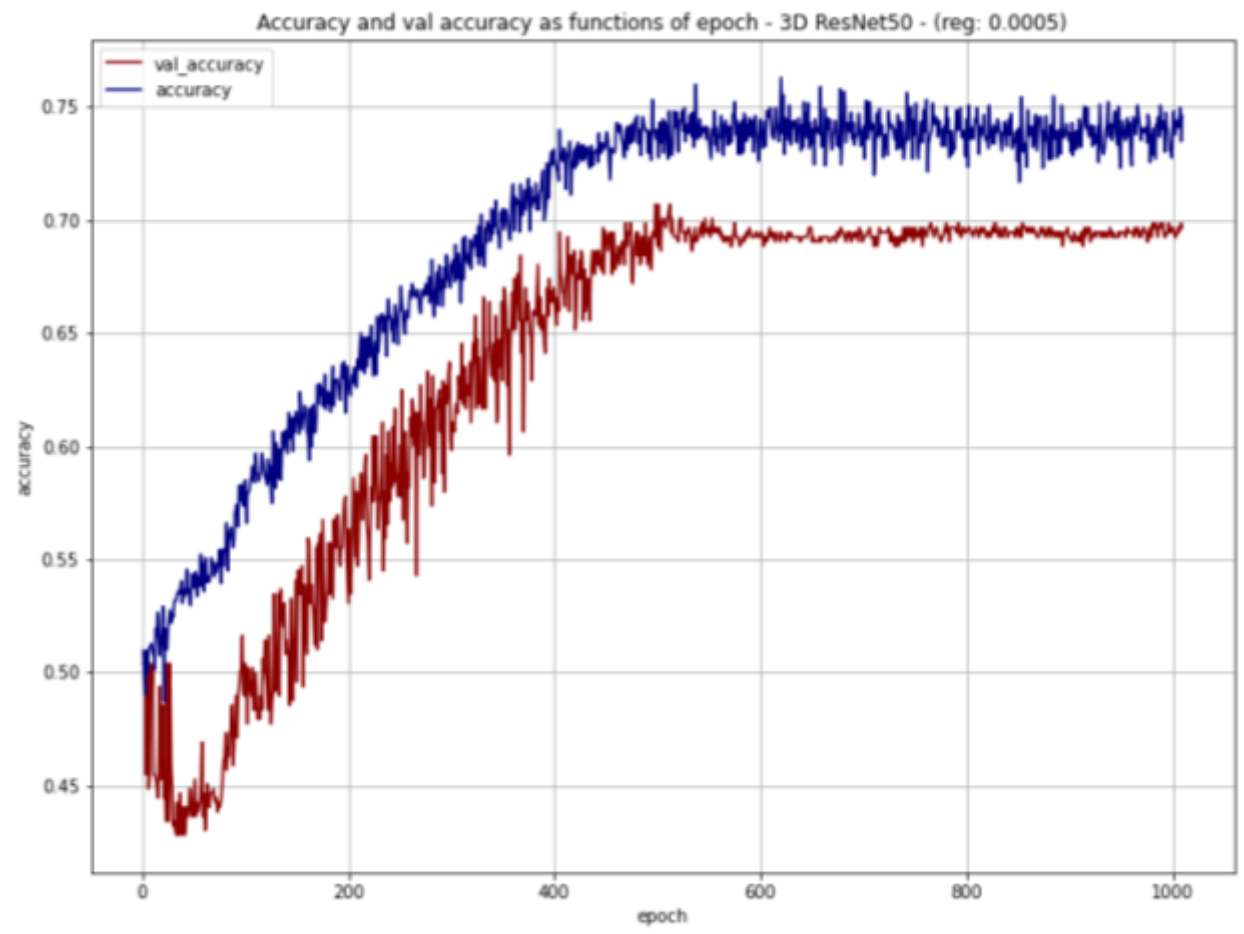

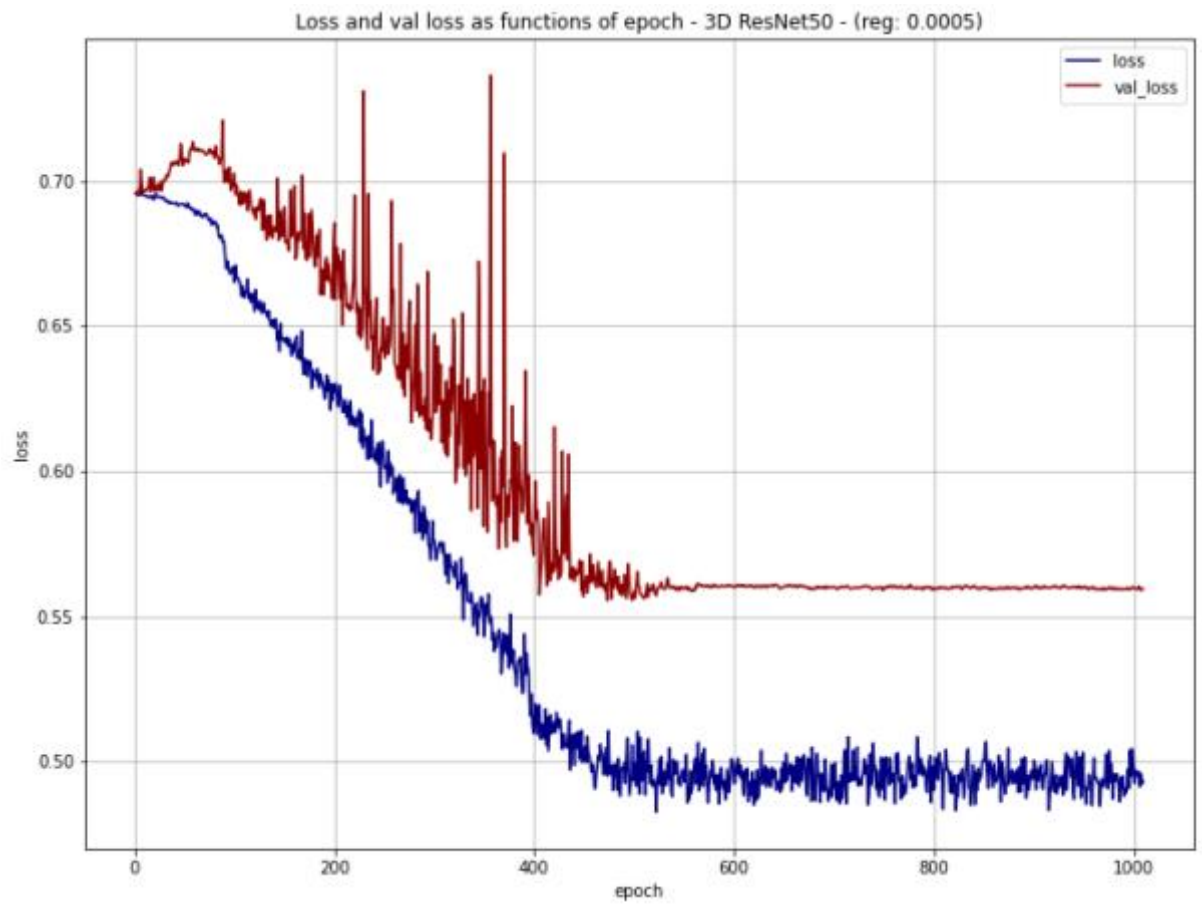

### 2.1.8. L2 Regularization: 0.0002

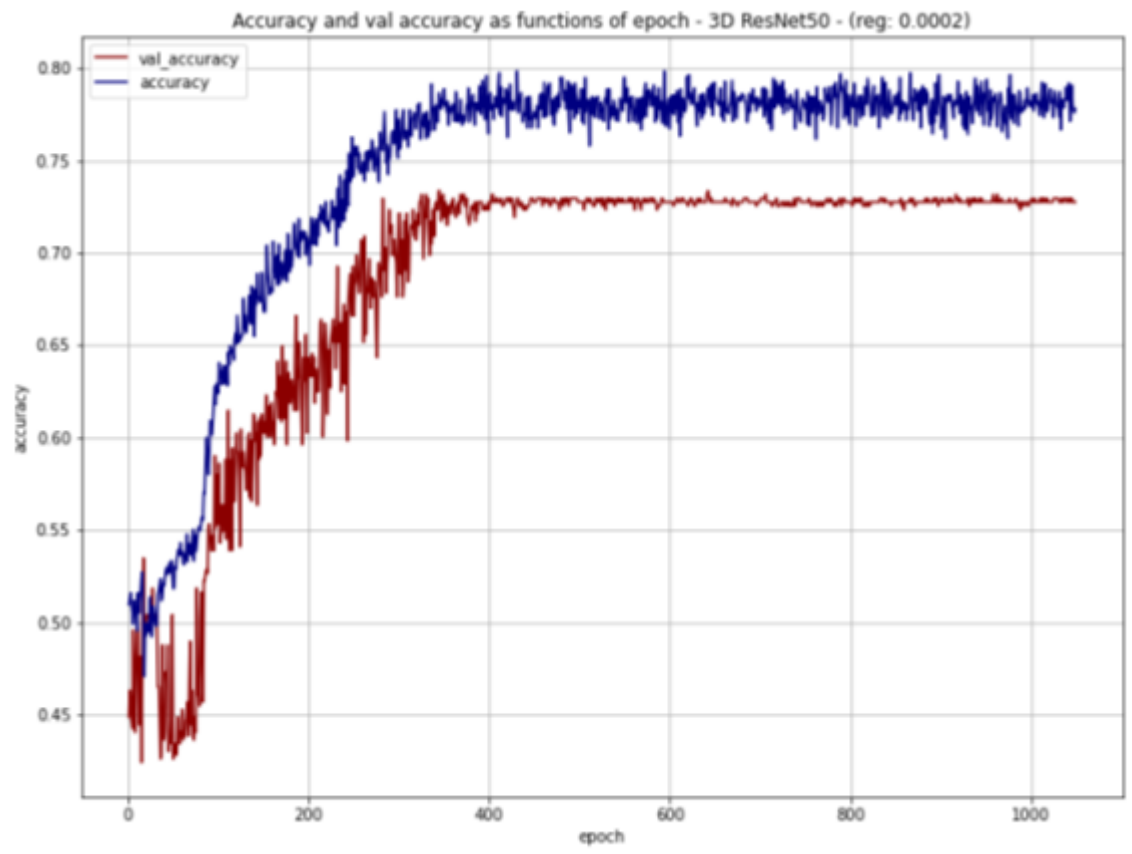

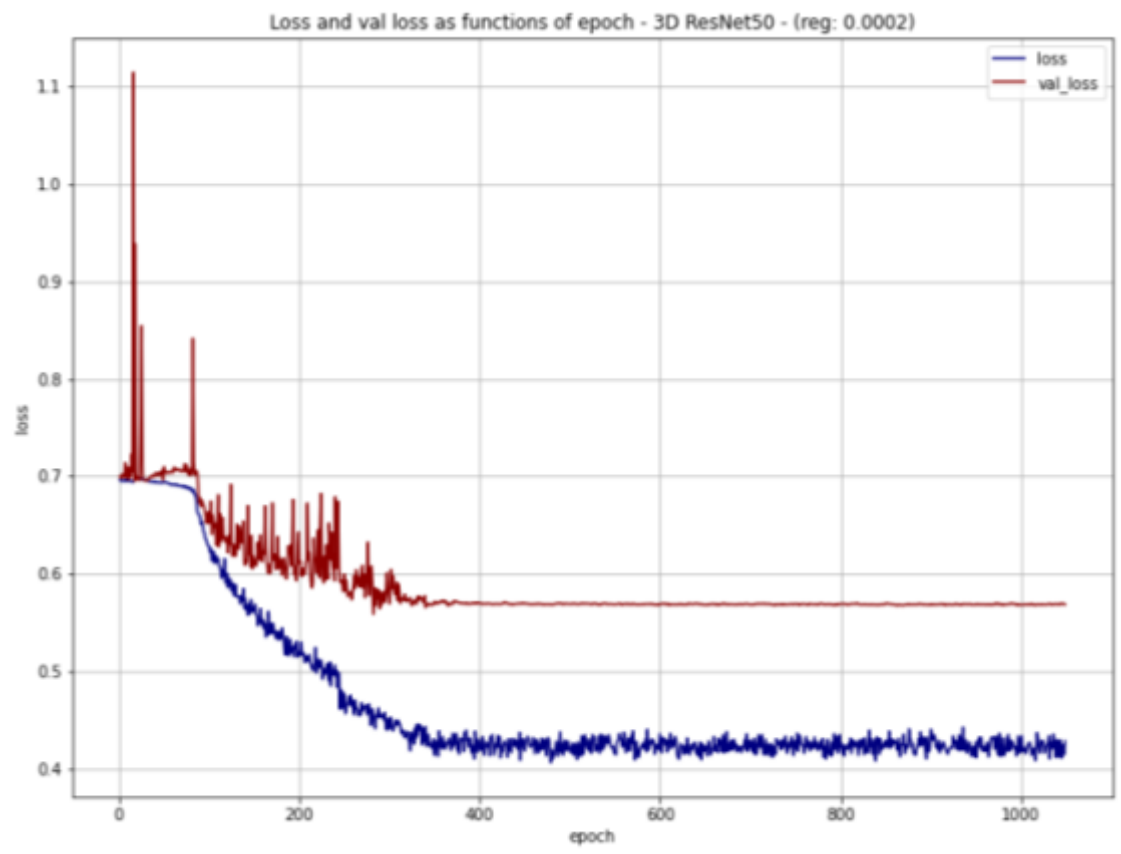

## 2.2. ROC Curves

### 2.2.1. L2 Regularization: 0.0002

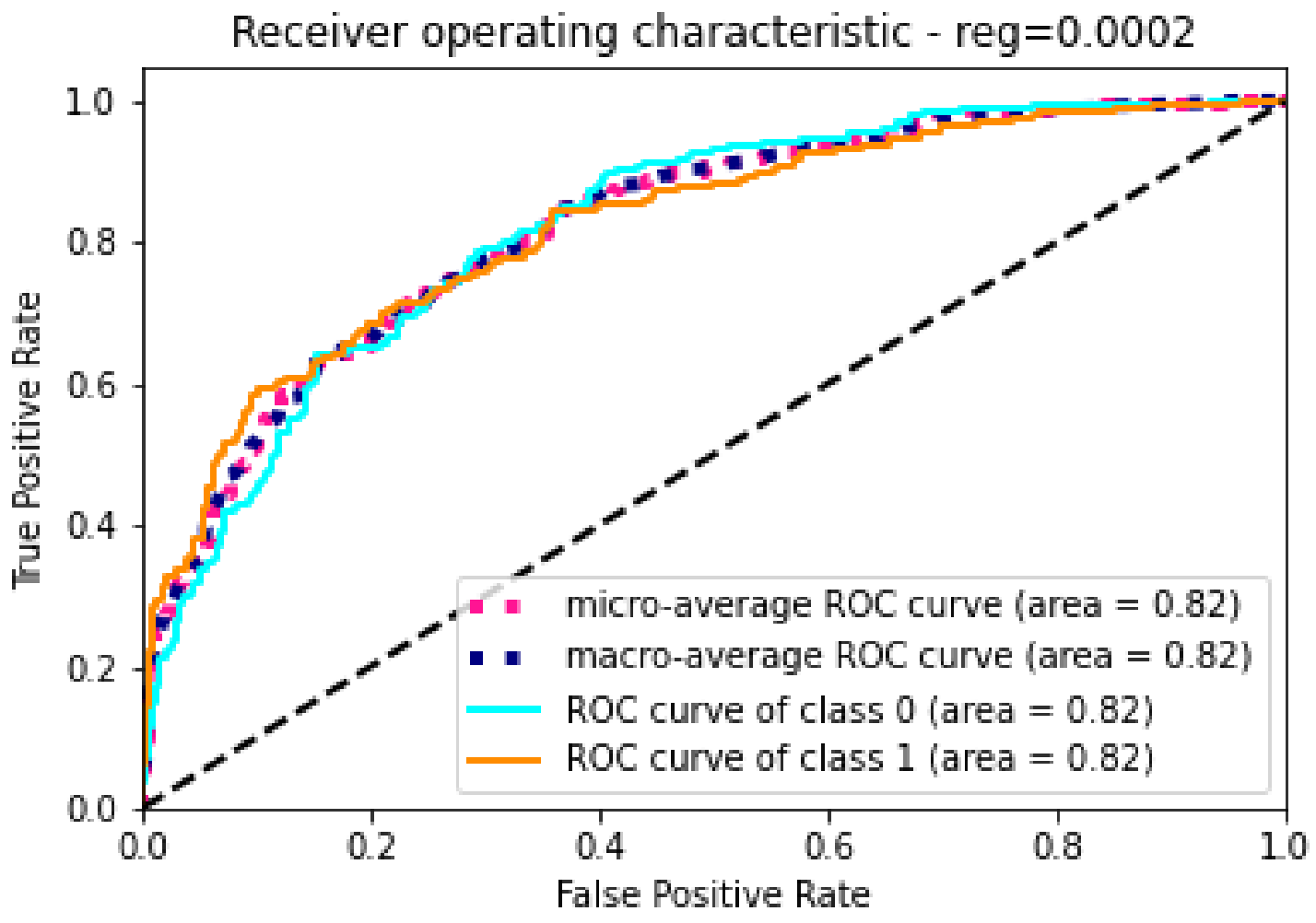

## 2.3. Confusion Matrices, Precision, Recall, F1-scores

### 2.3.1. L2 Regularization: 0.0002

#### 2.3.1.1. Confusion Matrix

| | Neutral (Predicted) | Negative (Predicted) |
|---|---|---|
| Neutral (True) | 250 | 53 |
| Negative (True) | 59 | 248 |

#### 2.3.1.2. Precision, Recall and F1-scores

| | Precision | Recall | F1-score |
|---|---|---|---|
| Class 0: neutral | 0.809 | 0.825 | 0.817 |
| Class 1: negative | 0.824 | 0.808 | 0.816 |

## 2.4. Result Lists

### 2.4.1. L2 Regularization: 0.0002

| Training Instance # | ResNet-50 Accuracy | ResNet-50 Loss |
|---|---|---|
| 0 | 0.754 | 0.546 |
| 1 | 0.803 | 0.545 |
| 2 | 0.820 | 0.555 |
| 3 | 0.754 | 0.491 |
| 4 | 0.790 | 0.495 |
| 5 | 0.770 | 0.505 |
| 6 | 0.770 | 0.556 |
| 7 | 0.738 | 0.496 |
| 8 | 0.885 | 0.483 |
| 9 | 0.721 | 0.553 |